\documentclass[twocolumn]{aastex63}

\usepackage{amsmath}
\usepackage[english]{babel}

\received{August 21, 2019}
\revised{January 18, 2020}
\accepted{February 4, 2020}
\shorttitle{Structure of Brightest Cluster Galaxies and Intracluster Light}
\shortauthors{Kluge et al.}
\submitjournal{ApJS}

\begin{document}

\title{Structure of Brightest Cluster Galaxies and Intracluster Light}

\author[0000-0002-9618-2552]{M. Kluge}
\affil{University-Observatory, Ludwig-Maximilians-University, Scheinerstrasse 1, D-81679 Munich, Germany}
\affil{Max Planck Institute for Extraterrestrial Physics, Giessenbachstrasse, D-85748 Garching, Germany}

\author[0000-0001-6564-9693]{B. Neureiter}
\affil{Max Planck Institute for Extraterrestrial Physics, Giessenbachstrasse, D-85748 Garching, Germany}

\author[0000-0002-5466-3892]{A. Riffeser}
\affil{University-Observatory, Ludwig-Maximilians-University, Scheinerstrasse 1, D-81679 Munich, Germany}

\author[0000-0001-7179-0626]{R. Bender}
\affil{University-Observatory, Ludwig-Maximilians-University, Scheinerstrasse 1, D-81679 Munich, Germany}
\affil{Max Planck Institute for Extraterrestrial Physics, Giessenbachstrasse, D-85748 Garching, Germany}

\author[0000-0002-2152-6277]{C. Goessl}
\affil{University-Observatory, Ludwig-Maximilians-University, Scheinerstrasse 1, D-81679 Munich, Germany}

\author[0000-0003-1008-225X]{U. Hopp}
\affil{University-Observatory, Ludwig-Maximilians-University, Scheinerstrasse 1, D-81679 Munich, Germany}
\affil{Max Planck Institute for Extraterrestrial Physics, Giessenbachstrasse, D-85748 Garching, Germany}

\author{M. Schmidt}
\affil{University-Observatory, Ludwig-Maximilians-University, Scheinerstrasse 1, D-81679 Munich, Germany}

\author{C. Ries}
\affil{University-Observatory, Ludwig-Maximilians-University, Scheinerstrasse 1, D-81679 Munich, Germany}

\author{N. Brosch}
\affil{Tel Aviv University, Tel Aviv, 69978, Israel}

\begin{abstract}

Observations of 170 local ($z\lesssim0.08$) galaxy clusters in the northern hemisphere have been obtained with the Wendelstein Telescope Wide Field Imager (WWFI). We correct for systematic effects such as point-spread function broadening, foreground star contamination, relative bias offsets, and charge persistence. Background inhomogeneities induced by scattered light are reduced down to $\Delta {\rm SB} > 31~g'$ mag arcsec$^{-2}$ by large dithering and subtraction of night-sky flats. Residual background inhomogeneities brighter than ${\rm SB}_{\sigma}< 27.6~g'$ mag arcsec$^{-2}$ caused by galactic cirrus are detected in front of 23\% of the clusters. However, the large field of view allows discrimination between accretion signatures and galactic cirrus. We detect accretion signatures in the form of tidal streams in 22\%, shells in 9.4\%, and multiple nuclei in 47\% of the Brightest Cluster Galaxies (BCGs) and find two BCGs in 7\% of the clusters. We measure semimajor-axis surface brightness profiles of the BCGs and their surrounding Intracluster Light (ICL) down to a limiting surface brightness of ${\rm SB} = 30~g'$ mag arcsec$^{-2}$. The spatial resolution in the inner regions is increased by combining the WWFI light profiles with those that we measured from archival \textit{Hubble Space Telescope} images or deconvolved WWFI images. We find that 71\% of the BCG+ICL systems have surface brightness (SB) profiles that are well described by a single S\'ersic (SS) function, whereas 29\% require a double S\'ersic (DS) function to obtain a good fit. We find that BCGs have scaling relations that differ markedly from those of normal ellipticals, likely due to their indistinguishable embedding in the ICL.

\end{abstract}


\section{Introduction}

Following the first detection of an ``extended mass of luminous intergalactic matter of very low surface brightness" in the Coma cluster \citep{Zwicky1951}, numerous early studies have confirmed that subgroupings of galaxies in clusters ``[...] often share a common outer envelope several hundred kiloparsecs in diameter" (\citealt{Kormendy1974}; also \citealt{Arp1971,Welch1971,Thuan1977}). A similar envelope was discovered to surround the Virgo cluster galaxy M87 (\citealt{Arp1969,DeVaucouleurs1969}).

Today, we know that many galaxy clusters are populated by an outstandingly bright and extended elliptical galaxy near the geometric and kinematic cluster center. They are referred to as brightest cluster galaxies (BCGs). The definition of this galaxy type is similar to the historic definition of cD galaxies (\citealt{Matthews1964,Morgan1965}). cD galaxies form a subset of BCGs that are surrounded by an extended, diffuse stellar envelope. That envelope is part of the ex-situ stellar population that was accreted during mass assembly \citep{Cooper2013,Cooper2015,Pillepich2018}. It is probably mixed with the intracluster light (ICL), which is kinematically unbound from the BCG. In this work, we do not distinguish between stellar envelope, stellar halo, and ICL because they are probably indistinguishable with photometric data alone. \cite{Oegerle2001} classify 20\% of BCGs as cD galaxies. The issue with this subset definition is that the detection of an existing envelope depends on the depth of the observation. Moreover, large samples of BCGs are Gaussian distributed in their brightnesses (\citealt{Postman1995,Hansen2009,Donzelli2011,Lauer2014}), which implies that the transition between cD and non-cD BCGs is continuous. Hence, it makes sense to study BCGs as a generalized class of galaxies.

Contrary to what the name suggests, a BCG is in our adopted definition not necessarily the brightest galaxy in a cluster: it must also lie close to the cluster center as traced by the satellite galaxy distribution or the intracluster medium. Between 20\% and 40\% of central galaxies are not the brightest galaxy in their host clusters (\citealt{Skibba2011,Hoshino2015}). A famous example is M87 in the Virgo cluster. The brightest galaxy is M49, but it is located far off the cluster center. M87 is (in projection) near the X-ray gas emission peak (e.g., \citealt{Kellogg1971}), which is a good tracer for the potential minimum of the cluster. Moreover, the rising velocity dispersion profile of the surrounding planetary nebulae is steeper for M87 \citep{Longobardi2018} than for M49 \citep{Hartke2018}, showing that intracluster planetary nebulae are more centered on M87 than on M49. The velocity dispersion profile of the globular clusters rises toward the outskirts of M87 \citep{Cote2001} but it falls toward the outskirts of M49 \citep{Sharples1998}, showing that intracluster globular clusters are also more centered on M87. All of the arguments above agree that M87 qualifies better as the BCG of the Virgo cluster in our adopted definition.

The currently widely accepted two-phase formation scenario (e.g., \citealt{Contini2014,Contini2018}) states that the BCG formed first by galactic mergers and the ICL was accreted afterward by a mixture of (1) galaxy harassment, that is, high-velocity encounters between satellite galaxies \citep{Moore1996}; (2) tidal stripping induced by effects of dynamical friction against the whole cluster potential (\citealt{Byrd1990,Gnedin2003}); and (3) preprocessing in smaller groups (\citealt{Mihos2004,Rudick2006}). Remnants of these violent processes are predicted by simulations to occur at low surface brightnesses (SBs), mostly below ${\rm SB}\gtrsim29~g'$ mag arcsec$^{-2}$ (\citealt{Rudick2009,Puchwein2010,Harris2017,Mancillas2019}), and are confirmed by observations (e.g., \citealt{Arnaboldi2012,Kormendy2012,Iodice2017,Mihos2017}). We refer to these remnants as accretion signatures.

\begin{figure}
	\centering
	\includegraphics[width=\linewidth]{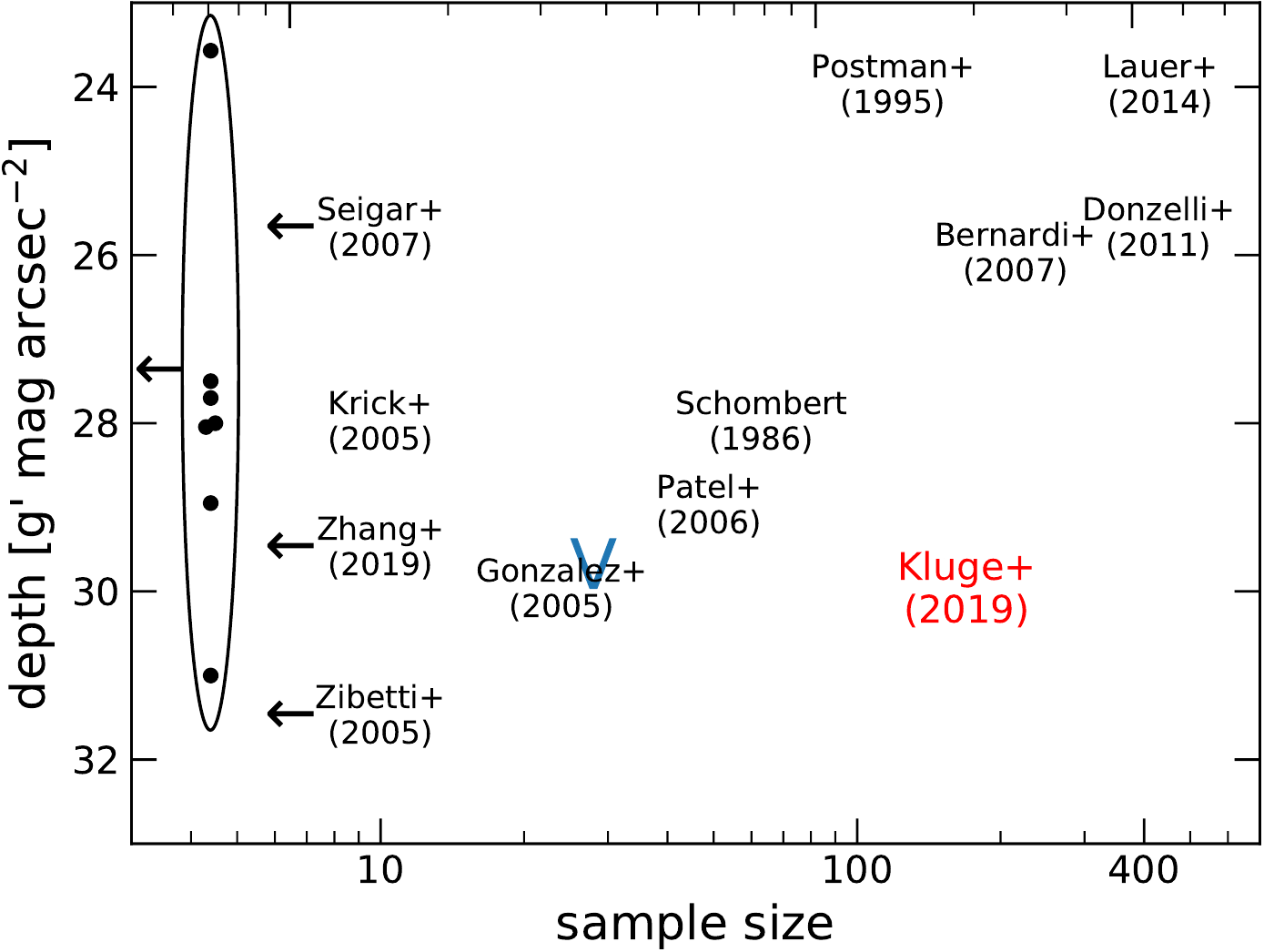}
	\caption{Overview of photometric low-redshift BCG surveys: \cite{Seigar2007,Krick2005,Zhang2019,Zibetti2005,Gonzalez2005a,Patel2006,Schombert1986,Postman1995,Bernardi2007,Lauer2014,Donzelli2011}. The so-far published VST survey of Early-type GAlaxieS (VEGAS) sample is shown by a blue "V" (\citealt{Capaccioli2015,Spavone2017,Spavone2018,Cattapan2019}). The dots embedded in the ellipse represent single- or double-target BCG observations. From top to bottom: \cite{Jorgensen1992,Bender2015,Ferrarese2012,Feldmeier2002,Kormendy2009,Mihos2005,Iodice2016}. The arrows indicate that the sample size is smaller than the label position in the plot. Our survey (red) populates an unexplored parameter space region in  sample size and depth. \label{fig:surveys}}
\end{figure}

The build-up, shape, and substructure of BCG+ICL light profiles, as well as the types and abundances of accretion signatures, are sensitive probes for the dynamical evolution of galaxy clusters (e.g., \citealt{Puchwein2010,Cui2014}). To constrain formation models, especially in the faint outskirts of BCGs, a large sample of BCGs with deep light profiles is needed. Figure \ref{fig:surveys} illustrates that so far, either the studied sample was relatively small (\citealt{Gonzalez2005a,Krick2005,Patel2006,Seigar2007}) or the surface brightness at the transition radius between the two photometric components of double S\'ersic (DS) BCGs is mostly below the limiting magnitude of the surveys (\citealt{Postman1995,Bernardi2007,Donzelli2011,Lauer2014}). In this paper, we present a study that fulfills both criteria. We perform a statistical analysis of the SB profiles, isophotal shape profiles, and structural parameters of BCG+ICLs. An analysis of the correlations of these parameters with host cluster properties and various approaches to separating the ICL from the BCGs is reserved for an forthcoming paper.

\begin{figure*}
	\centering
	\includegraphics[width=\linewidth]{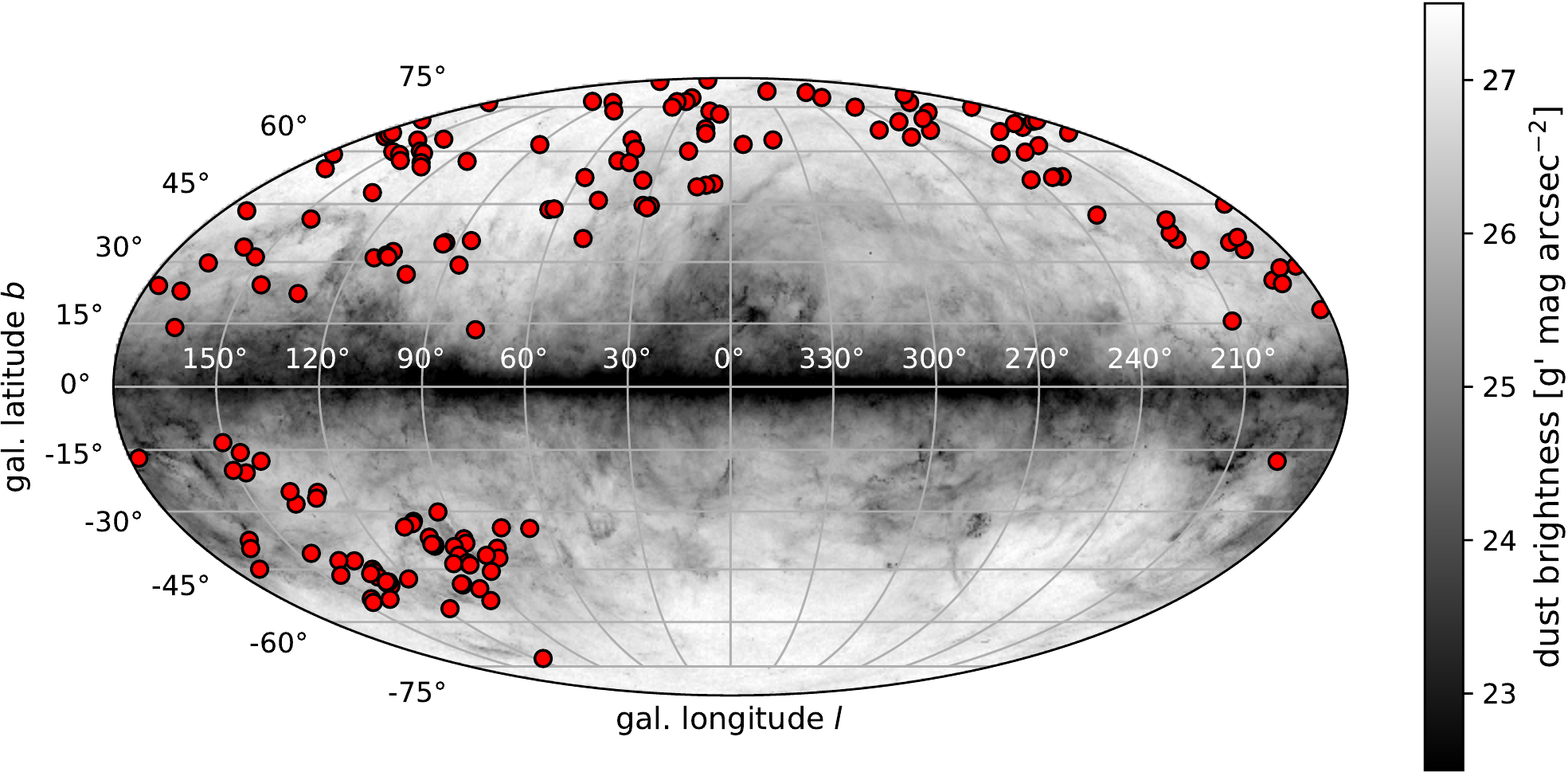}
	\caption{Our sample. The observed galaxy clusters are marked as red points. The background is the far-infrared dust emission map from \cite{Planck2014}. It is scaled to match the emission of the galactic cirrus (see Section \ref{sec:galacticcirrus}).\label{fig:sample}}
\end{figure*}

This paper is organized as follows. In section 2, we present our sample and selection criteria. The data and data reduction are described in section 3. Our methods to measure and combine the surface brightness profiles are described in section 4. Sections 3 and 4 are dedicated to readers who are interested in the image processing techniques necessary for deep imaging. An extensive error analysis is given in section 5. We present our results of accretion signatures, average profiles, and structural parameters in section 6. They are discussed in section 7 and summarized in section 8.

Throughout the paper we assume a flat cosmology with $H_0\,=\,69.6$\,km\,s$^{-1}$ Mpc$^{-1}$ and $\Omega_{\rm m}\,=\,0.286$. Distances and angular scales were calculated using the web tool from \cite{Wright2006}. Virgo infall is not considered. If not stated otherwise, then three types of flux corrections were applied: (1) dust extinction using the maps from \cite{Schlafly2011}, (2) K corrections following \cite{Chilingarian2010} and \cite{Chilingarian2012}, and (3) cosmic $(1+z)^4$ dimming. Magnitudes are always given in the AB system.

\section{Sample selection}

Our sample is based on the Abell--Corwin--Olowin (ACO) catalog \citep{Abell1989}. It contains 4073 rich galaxy clusters, out of which we selected 141 by the following criteria:

\begin{enumerate}
	\item redshift $z \lesssim 0.08$,
	\item galactic latitude $|b| > 13.5\degr$,
	\item decl. $> +5\degr$,
	\item no bright stars nearby.
\end{enumerate}

Regarding the fourth criterion, a stellar brightness limit in the range $5 < g < 9$, where $g$ is the stellar magnitude in the $g$-band, is applied, depending on the projected distance $2.6\degr < d < 0.08\degr$ from the BCG. Additionally, we allow 15 exceptions from the redshift constraint because of preobservational misidentification of the BCG and one exception from the declination constraint: A85 was observed for a different program. The sample is extended with nine clusters from the \cite{Linden2007} catalog, three clusters from the \cite{AWM1977} catalog, and one group from the \cite{MKW1975} catalog. The final sample of 170 BCGs is listed in Table \ref{tab:sample}, and its spatial distribution is shown in Figure \ref{fig:sample}.

\begin{figure}
	\centering
	\includegraphics[width=\linewidth]{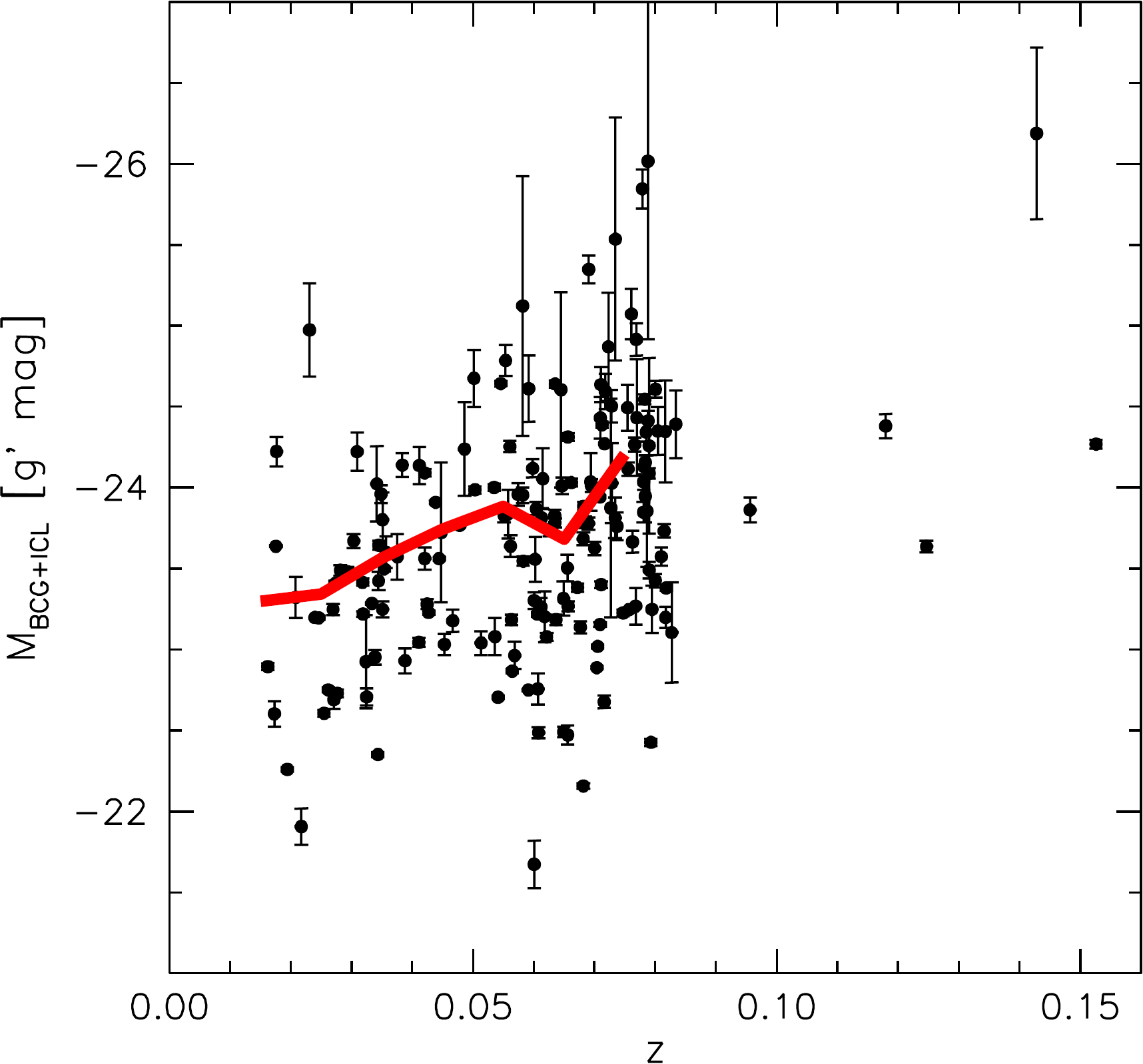}
	\caption{Redshift $z$ of the BCG plotted against the total brightness of the BCG+ICL $M_{\rm BCG+ICL}$ measured in this paper. The red line shows the average brightness in each redshift bin with width $\Delta z = \pm 0.005$. Four outliers with $M_{\rm BCG+ICL} < -27~g'$ mag were neglected because total brightness depends heavily on the extrapolation of the upward-curved light profiles that are unlikely to continue to infinite radius.\\~\\ \label{fig:zvsM}}
\end{figure}

In order to investigate the completeness of our sample, we plot the BCG redshifts against the total BCG+ICL brightness in Figure \ref{fig:zvsM}. A slight Malmquist bias is seen by the upward trend of the average brightness with increasing redshift, shown by the red line.

Furthermore, we compare the completeness of our sample to that of the most comprehensive samples available in the literature, \cite{Lauer2014} and \cite{Donzelli2011}. After applying the same volume-limiting constraints, the overlap of Lauer et al.'s sample on our sample is 90\%. An overall agreement is expected because both Lauer et al.'s and our samples are mainly drawn from the ACO catalogs. Following the same criteria, the overlap with the sample of Donzelli et al. is 89\%, and vice versa 80\%.

The selection of the BCG in each cluster was done manually while inspecting the deep Wendelstein Telescope Wide Field (WWFI) images. We always chose the most extended elliptical galaxy (at the $\sim27~g'$ mag arcsec$^{-2}$ isophote) that is located close to the cluster center, as traced by the X-ray gas or satellite galaxy distribution. It usually coincides with the brightest galaxy in the cluster, but that is not a stringent criterion. Our choices differ in 26 (20\%) out of the 127 clusters that overlap with the \cite{Lauer2014} sample who select the brightest galaxy measured in a metric aperture of 10 h$^{-1}$ kpc radius. That is a consequence of the more cD-like definition of BCGs, that were adopted by us. Out of those 26 galaxies, 15 are marked as the second-brightest galaxy in the Lauer et al. sample.

Private discussion with Tod Lauer and Marc Postman revealed that the choice of the BCG in those clusters is debatable. That is mainly due to (1) the Abell cluster number does not unambiguously identify a cluster in the case of line-of-sight overlap (three cases); (2) there is disagreement on the distance to the cluster center, usually in the case of ongoing mergers of clusters (four cases); (3) the BCG is fainter in the metric aperture but brighter in terms of total luminosity (eight cases); (4) the criteria ``brightest in the metric aperture" and ``most extended" stand in conflict with each other (11 cases); and (5) the criteria "brightest in the metric aperture" and "central" stand in conflict with each other (three cases). The sometimes-occurring conflict between "brightest" and "central" was also pointed out by \cite{Linden2007}.

\startlongtable
\begin{deluxetable*}{llrrcccc}
\tablewidth{700pt}
\tabletypesize{\footnotesize}

\tablecaption{The BCG sample.\label{tab:sample}}
\tablehead{
	\colhead{Cluster} & \colhead{BCG} & \colhead{R.A.} & \colhead{Decl.} & \colhead{$z$} & \colhead{Angular Scale} & \colhead{L14 Selection} & \colhead{\textit{HST} Available}\\
	\colhead{} & \colhead{} & \colhead{(J2000)} & \colhead{(J2000)} & \colhead{} & \colhead{(kpc arcsec$^{-1}$)} & \colhead{} & \colhead{}
}
\colnumbers
\startdata
  A76 & IC 1565 & 00:39:26.3 & +06:44:04 & 0.038372 & 0.765 & \checkmark & \checkmark \\
  A85 & MCG-02-02-086 & 00:41:50.4 & -09:18:11 & 0.055380 & 1.083 & \checkmark & -- \\
  A150 & UGC 716 & 01:09:18.4 & +13:10:09 & 0.061321 & 1.190 & \checkmark & -- \\
  A152 & UGC 727 & 01:10:03.1 & +13:58:42 & 0.058280 & 1.135 & \checkmark & -- \\
  A154 & IC 1634 & 01:11:02.9 & +17:39:47 & 0.069478 & 1.336 & M2& -- \\
  A158 & LEDA 1518776 & 01:11:46.3 & +16:51:29 & 0.064500 & 1.248 & other& -- \\
  A160 & MCG+02-04-010 & 01:12:59.6 & +15:29:29 & 0.043830 & 0.869 & \checkmark & \checkmark \\
  A161 & LEDA 2098391 & 01:15:22.3 & +37:20:24 & 0.076954 & 1.467 & \checkmark & -- \\
  A171 & MCG+03-04-014 & 01:17:17.9 & +16:15:57 & 0.071670 & 1.375 & \checkmark & -- \\
  A174 & 2MASX J01201619+3548272 & 01:20:16.1 & +35:48:27 & 0.078056 & 1.486 & \checkmark & -- \\
  A179 & 2MASX J01223283+1931312 & 01:22:32.8 & +19:31:32 & 0.056194 & 1.097 & M2& -- \\
  A193 & IC 1695 & 01:25:07.6 & +08:41:58 & 0.050171 & 0.987 & \checkmark & \checkmark \\
  A225 & NVSSJ013848+184931 & 01:38:48.9 & +18:49:32 & 0.069375 & 1.334 & \checkmark & -- \\
  A240 & UGC 1191 & 01:42:06.0 & +07:39:54 & 0.062534 & 1.212 & \checkmark & -- \\
  A245 & 2MASX J01435285+0624499 & 01:43:52.8 & +06:24:51 & 0.079310 & 1.508 & other& -- \\
  A257 & 2MASX J01490841+1357474 & 01:49:08.3 & +13:57:48 & 0.070540 & 1.355 & \checkmark & -- \\
  A260 & IC 1733 & 01:50:42.9 & +33:04:56 & 0.035680 & 0.714 & \checkmark & \checkmark \\
  A262 & NGC 708 & 01:52:46.3 & +36:09:07 & 0.016220 & 0.332 & \checkmark & \checkmark \\
  A292 & UGC 1518 & 02:02:18.9 & +19:04:02 & 0.064648 & 1.250 & \checkmark & -- \\
  A347 & NGC 910 & 02:25:26.9 & +41:49:23 & 0.017302 & 0.354 & \checkmark & \checkmark \\
  A376 & UGC 2232 & 02:46:03.9 & +36:54:19 & 0.048610 & 0.958 & \checkmark & \checkmark \\
  A397 & UGC 2413 & 02:56:28.7 & +15:54:58 & 0.034447 & 0.690 & \checkmark & \checkmark \\
  A399 & UGC 2438 & 02:57:53.1 & +13:01:52 & 0.071239 & 1.367 & \checkmark & -- \\
  A400 & NGC 1128 & 02:57:41.6 & +06:01:21 & 0.023980 & 0.487 & \checkmark & \checkmark \\
  A407 & 2MASX J03015146+3550283 & 03:01:51.8 & +35:50:20 & 0.047820 & 0.943 & \checkmark & -- \\
  A426 & NGC 1275 & 03:19:48.1 & +41:30:43 & 0.017560 & 0.359 & -- & \checkmark \\
  A498 & 2MASX J04375071+2112203 & 04:37:50.7 & +21:12:21 & 0.059823 & 1.163 & \checkmark & -- \\
  A505 & UGC 3197 & 04:59:55.6 & +80:10:44 & 0.053504 & 1.048 & \checkmark & -- \\
  A539 & LEDA 17025 & 05:16:37.3 & +06:26:27 & 0.028142 & 0.568 & M2& -- \\
  A553 & 2MASX J06124115+4835445 & 06:12:41.1 & +48:35:45 & 0.069059 & 1.329 & -- & -- \\
  A559 & 2MASX J06395117+6958332 & 06:39:51.0 & +69:58:34 & 0.075700 & 1.445 & -- & -- \\
  A568 & MCG+06-16-019 & 07:07:41.5 & +35:03:32 & 0.081702 & 1.549 & \checkmark & -- \\
  A569 & NGC 2329 & 07:09:08.0 & +48:36:56 & 0.019420 & 0.396 & \checkmark & \checkmark \\
  A582 & 2MASX J07280080+4155074 & 07:28:00.8 & +41:55:08 & 0.060245 & 1.171 & \checkmark & -- \\
  A592 & 2MASX J07424058+0922207 & 07:42:40.6 & +09:22:21 & 0.065651 & 1.268 & other & -- \\
  A595 & MCG+09-13-062 & 07:49:27.2 & +52:02:33 & 0.070938 & 1.362 & M2 & -- \\
  A600 & 2MASX J07563560+6344237 & 07:56:35.5 & +63:44:25 & 0.080997 & 1.537 & \checkmark & -- \\
  A602 & 2MASX J07532661+2921341 & 07:53:26.6 & +29:21:35 & 0.060420 & 1.174 & M2& -- \\
  A607 & SDSSJ075724.71+392106.6 & 07:57:24.7 & +39:21:07 & 0.095620 & 1.784 & -- & -- \\
  A612 & 2MASX J08011329+3440311 & 08:01:13.2 & +34:40:31 & 0.082720 & 1.567 & -- & -- \\
  A634 & UGC 4289 & 08:15:44.8 & +58:19:16 & 0.027090 & 0.548 & \checkmark & \checkmark \\
  A671 & IC 2378 & 08:28:31.6 & +30:25:53 & 0.050320 & 0.990 & \checkmark & \checkmark \\
  A688 & SDSSJ083734.33+154907.6 & 08:37:34.3 & +15:49:08 & 0.152620 & 2.672 & -- & -- \\
  A690 & ICRF J083915.8+285038 & 08:39:15.8 & +28:50:39 & 0.079020 & 1.503 & \checkmark & -- \\
  A695 & 2MASX J08411308+3224596 & 08:41:13.1 & +32:25:00 & 0.071103 & 1.365 & \checkmark & -- \\
  A734 & 2MASX J09003199+1614213 & 09:00:32.0 & +16:14:26 & 0.074717 & 1.428 & -- & -- \\
  A744 & 2MASX J09072049+1639064 & 09:07:20.5 & +16:39:07 & 0.072850 & 1.395 & \checkmark & -- \\
  A757 & 2MASX J09130775+4742307 & 09:13:07.7 & +47:42:31 & 0.051350 & 1.009 & \checkmark & -- \\
  A834 & 2MASX J09413277+6642376 & 09:41:32.7 & +66:42:39 & 0.070910 & 1.361 & \checkmark & -- \\
  A883 & 2MASX J09511516+0529170 & 09:51:15.1 & +05:29:17 & 0.078983 & 1.502 & -- & -- \\
  A999 & MCG+02-27-004 & 10:23:23.7 & +12:50:07 & 0.032490 & 0.653 & \checkmark & \checkmark \\
  A1003 & MCG+08-19-026 & 10:25:01.5 & +47:50:31 & 0.063660 & 1.233 & \checkmark & -- \\
  A1016 & IC 613 & 10:27:07.8 & +11:00:39 & 0.032370 & 0.650 & \checkmark & \checkmark \\
  A1020 & 2MASX J10274949+1026306 & 10:27:49.5 & +10:26:31 & 0.067702 & 1.305 & \checkmark & -- \\
  A1056 & LEDA 2186592 & 10:38:01.7 & +41:46:26 & 0.124670 & 2.251 & -- & -- \\
  A1066 & 2MASX J10393872+0510326 & 10:39:38.7 & +05:10:33 & 0.068170 & 1.313 & \checkmark & -- \\
  A1100 & MCG+04-26-010 & 10:48:45.6 & +22:13:05 & 0.046666 & 0.922 & \checkmark & -- \\
  A1108 & NGC 3405 & 10:49:43.3 & +16:14:20 & 0.021740 & 0.442 & -- & -- \\
  A1142 & IC 664 & 11:00:45.3 & +10:33:12 & 0.033860 & 0.679 & \checkmark & \checkmark \\
  A1155 & 2MASX J11043955+3513477 & 11:04:39.5 & +35:13:49 & 0.076788 & 1.464 & \checkmark & -- \\
  A1173 & 2MASX J11091531+4133412 & 11:09:15.3 & +41:33:42 & 0.076620 & 1.461 & -- & -- \\
  A1177 & NGC 3551 & 11:09:44.4 & +21:45:33 & 0.031830 & 0.640 & \checkmark & \checkmark \\
  A1185 & NGC 3550 & 11:10:38.4 & +28:46:04 & 0.035094 & 0.703 & \checkmark & -- \\
  A1187 & 2MASX J11110955+3935522 & 11:11:09.6 & +39:35:53 & 0.078380 & 1.492 & \checkmark & -- \\
  A1190 & MCG+07-23-031 & 11:11:43.6 & +40:49:15 & 0.078150 & 1.488 & \checkmark & -- \\
  A1203 & 2MASX J11134824+4017083 & 11:13:48.2 & +40:17:09 & 0.075510 & 1.442 & \checkmark & -- \\
  A1213 & 2MASX J11162274+2915086 & 11:16:22.7 & +29:15:09 & 0.045300 & 0.896 & \checkmark & -- \\
  A1218 & 2MASX J11184993+5144291 & 11:18:49.9 & +51:44:30 & 0.079470 & 1.511 & \checkmark & -- \\
  A1228 & UGC 6394 & 11:22:56.4 & +34:06:42 & 0.042710 & 0.847 & other & -- \\
  A1257 & MCG+06-25-069 & 11:26:17.3 & +35:20:25 & 0.034320 & 0.688 & other& -- \\
  A1270 & MCG+09-19-084 & 11:28:41.9 & +54:10:21 & 0.070400 & 1.352 & \checkmark & -- \\
  A1275 & 2MASX J11292709+3638189 & 11:29:27.1 & +36:38:19 & 0.060690 & 1.179 & \checkmark & -- \\
  A1279 & 2MASX J11313927+6714296 & 11:31:39.3 & +67:14:31 & 0.054130 & 1.060 & \checkmark & -- \\
  A1314 & IC 712 & 11:34:49.3 & +49:04:40 & 0.033350 & 0.669 & \checkmark & -- \\
  A1324 & LEDA 2557423 & 11:37:16.3 & +57:06:49 & 0.117960 & 2.146 & -- & -- \\
  A1356 & 2MASX J11422978+1028327 & 11:42:29.8 & +10:28:33 & 0.071612 & 1.374 & \checkmark & -- \\
  A1365 & NVSS J114430+305259 & 11:44:30.5 & +30:53:01 & 0.076260 & 1.455 & \checkmark & -- \\
  A1367 & NGC 3842 & 11:44:02.1 & +19:57:00 & 0.020830 & 0.424 & \checkmark & -- \\
  A1371 & MCG+03-30-100 & 11:45:22.2 & +15:29:44 & 0.068220 & 1.314 & M2& -- \\
  A1400 & 2MASS J11520578+5458171 & 11:52:05.7 & +54:58:18 & 0.060070 & 1.168 & other& -- \\
  A1413 & MCG+04-28-097 & 11:55:18.0 & +23:24:18 & 0.142800 & 2.527 & -- & \checkmark \\
  A1423 & 2MASX J11574738+3342438 & 11:57:47.3 & +33:42:44 & 0.080010 & 1.520 & \checkmark & -- \\
  A1424 & MCG+01-31-003 & 11:57:28.9 & +05:05:21 & 0.080446 & 1.528 & \checkmark & -- \\
  A1435 & MCG+02-31-009 & 12:00:14.3 & +10:41:49 & 0.061220 & 1.189 & -- & -- \\
  A1436 & MCG+09-20-056 & 12:00:13.8 & +56:15:03 & 0.067212 & 1.296 & M2& -- \\
  A1452 & MCG+09-20-071 & 12:03:07.1 & +51:40:31 & 0.065544 & 1.266 & M2& -- \\
  A1507 & NGC 4199A & 12:14:48.6 & +59:54:23 & 0.060070 & 1.168 & \checkmark & -- \\
  A1516 & 2MASX J12185235+0514443 & 12:18:52.3 & +05:14:45 & 0.078342 & 1.491 & -- & -- \\
  A1526 & 2MASX J12214375+1345168 & 12:21:43.8 & +13:45:17 & 0.081740 & 1.550 & -- & -- \\
  A1534 & MCG+10-18-041 & 12:24:42.7 & +61:28:15 & 0.070010 & 1.345 & \checkmark & -- \\
  A1569 & 2MASX J12362580+1632181 & 12:36:25.7 & +16:32:19 & 0.068464 & 1.318 & other& -- \\
  A1589 & MCG+03-32-083 & 12:41:17.4 & +18:34:29 & 0.071800 & 1.377 & \checkmark & -- \\
  A1610 & IC 822 & 12:47:45.5 & +30:04:39 & 0.060622 & 1.178 & \checkmark & -- \\
  A1656 & NGC 4874 & 12:59:35.7 & +27:57:34 & 0.023100 & 0.469 & M2 & \checkmark \\
  A1668 & IC 4130 & 13:03:46.6 & +19:16:18 & 0.063510 & 1.230 & \checkmark & -- \\
  A1691 & MCG+07-27-039 & 13:11:08.6 & +39:13:37 & 0.072300 & 1.386 & \checkmark & -- \\
  A1749 & IC 4269 & 13:29:21.0 & +37:37:23 & 0.055785 & 1.090 & \checkmark & -- \\
  A1767 & MCG+10-19-096 & 13:36:08.3 & +59:12:24 & 0.071062 & 1.364 & \checkmark & -- \\
  A1775 & MCG+05-32-063 & 13:41:49.1 & +26:22:25 & 0.075460 & 1.441 & \checkmark & \checkmark \\
  A1795 & MCG+05-33-005 & 13:48:52.5 & +26:35:35 & 0.063550 & 1.231 & \checkmark & \checkmark \\
  A1800 & UGC 8738 & 13:49:23.5 & +28:06:27 & 0.078288 & 1.490 & \checkmark & -- \\
  A1809 & 2MASX J13530637+0508586 & 13:53:06.4 & +05:09:00 & 0.078850 & 1.500 & \checkmark & -- \\
  A1812 & 2MASX J13522099+3730370 & 13:52:21.0 & +37:30:38 & 0.061810 & 1.199 & -- & -- \\
  A1825 & UGC 8888 & 13:58:00.4 & +20:37:57 & 0.062135 & 1.205 & \checkmark & -- \\
  A1828 & 2MASX J13581472+1820457 & 13:58:14.7 & +18:20:47 & 0.064913 & 1.255 & \checkmark & -- \\
  A1831 & MCG+05-33-033 & 13:59:15.1 & +27:58:35 & 0.076110 & 1.452 & \checkmark & -- \\
  A1890 & NGC 5539 & 14:17:37.7 & +08:10:47 & 0.058180 & 1.134 & \checkmark & -- \\
  A1899 & MCG+03-37-008 & 14:21:41.7 & +17:45:09 & 0.056445 & 1.102 & \checkmark & -- \\
  A1904 & MCG+08-26-024 & 14:22:10.2 & +48:34:15 & 0.070980 & 1.363 & \checkmark & -- \\
  A1913 & 2MASX J14263943+1641139 & 14:26:39.4 & +16:41:15 & 0.053610 & 1.050 & other& -- \\
  A1982 & MCG+05-35-020 & 14:51:14.4 & +30:41:33 & 0.056322 & 1.100 & \checkmark & -- \\
  A1983 & MCG+03-38-044 & 14:52:55.3 & +16:42:11 & 0.044764 & 0.886 & M2 & \checkmark \\
  A2022 & MCG+05-36-002 & 15:04:15.9 & +28:29:48 & 0.058189 & 1.134 & \checkmark & -- \\
  A2029 & IC 1101 & 15:10:56.1 & +05:44:42 & 0.077900 & 1.484 & \checkmark & \checkmark \\
  A2052 & UGC 9799 & 15:16:44.5 & +07:01:18 & 0.034470 & 0.691 & \checkmark & \checkmark \\
  A2061 & 2MASX J15212054+3040154 & 15:21:20.6 & +30:40:16 & 0.078820 & 1.499 & \checkmark & -- \\
  A2063 & MCG+02-39-020 & 15:23:05.3 & +08:36:34 & 0.034170 & 0.685 & \checkmark & -- \\
  A2065 & MCG+05-36-020 & 15:22:24.0 & +27:42:52 & 0.069020 & 1.328 & \checkmark & -- \\
  A2107 & UGC 9958 & 15:39:39.0 & +21:46:58 & 0.042060 & 0.835 & \checkmark & -- \\
  A2122 & UGC 10012 & 15:44:59.0 & +36:06:35 & 0.066210 & 1.278 & \checkmark & -- \\
  A2147 & UGC 10143 & 16:02:17.0 & +15:58:29 & 0.035380 & 0.708 & \checkmark & \checkmark \\
  A2151 & NGC 6041 & 16:04:35.8 & +17:43:18 & 0.035100 & 0.703 & \checkmark & -- \\
  A2152 & MCG+03-41-095 & 16:05:29.2 & +16:26:10 & 0.044440 & 0.880 & \checkmark & -- \\
  A2162 & NGC 6086 & 16:12:35.5 & +29:29:06 & 0.031900 & 0.641 & \checkmark & -- \\
  A2197 & NGC 6173 & 16:29:44.9 & +40:48:42 & 0.029380 & 0.592 & \checkmark & \checkmark \\
  A2199 & NGC 6166 & 16:28:39.1 & +39:33:11 & 0.030920 & 0.622 & \checkmark & \checkmark \\
  A2247 & UGC 10638 & 16:50:58.6 & +81:34:30 & 0.038809 & 0.774 & M2& -- \\
  A2248 & 2MASX J16573834+7703462 & 16:57:38.4 & +77:03:46 & 0.065641 & 1.268 & M2& -- \\
  A2255 & 2MASX IJ1712287+640338 & 17:12:28.8 & +64:03:39 & 0.073440 & 1.406 & -- & -- \\
  A2256 & UGC 10726 & 17:04:27.1 & +78:38:26 & 0.059170 & 1.152 & \checkmark & -- \\
  A2271 & MCG+13-12-022 & 17:18:16.6 & +78:01:07 & 0.057439 & 1.120 & \checkmark & -- \\
  A2293 & 2MASX J18012131+5739016 & 18:01:21.3 & +57:39:02 & 0.073396 & 1.405 & M2& -- \\
  A2308 & 2MASX J18340865+7057188 & 18:34:08.5 & +70:57:20 & 0.083424 & 1.579 & \checkmark & -- \\
  A2319 & MCG+07-40-004 & 19:21:10.0 & +43:56:45 & 0.054588 & 1.068 & -- & -- \\
  A2388 & LEDA 140981 & 21:53:39.3 & +08:15:10 & 0.061500 & 1.194 & \checkmark & -- \\
  A2469 & - & 22:40:34.3 & +12:17:56 & 0.065600 & 1.267 & other& -- \\
  A2495 & MCG+02-58-021 & 22:50:19.7 & +10:54:13 & 0.080060 & 1.521 & \checkmark & \checkmark \\
  A2506 & NGC 7432 & 22:58:01.9 & +13:08:05 & 0.025464 & 0.516 & -- & -- \\
  A2513 & NGC 7436 & 22:57:57.5 & +26:09:01 & 0.024600 & 0.499 & -- & -- \\
  A2516 & 2MASX J23001449+1835027 & 23:00:14.5 & +18:35:03 & 0.081825 & 1.551 & -- & -- \\
  A2524 & 2MASX J23031792+1740232 & 23:02:55.8 & +17:45:01 & 0.081490 & 1.546 & \checkmark & -- \\
  A2558 & 2MASX J23124349+1021435 & 23:12:43.5 & +10:21:44 & 0.064900 & 1.255 & \checkmark & -- \\
  A2572 & NGC 7597 & 23:18:30.2 & +18:41:21 & 0.037540 & 0.749 & other& -- \\
  A2589 & NGC 7647 & 23:23:57.4 & +16:46:38 & 0.041170 & 0.818 & \checkmark& \checkmark \\
  A2593 & NGC 7649 & 23:24:20.0 & +14:38:50 & 0.042042 & 0.835 & \checkmark & \checkmark \\
  A2618 & 2MASX J23340547+2259000 & 23:34:05.5 & +22:59:00 & 0.072813 & 1.395 & \checkmark & -- \\
  A2622 & 2MASX J23350151+2722203 & 23:35:01.5 & +27:22:21 & 0.063441 & 1.229 & \checkmark & -- \\
  A2625 & 2MASX J23374932+2048340 & 23:37:49.3 & +20:48:34 & 0.059118 & 1.151 & \checkmark & -- \\
  A2626 & IC 5338 & 23:36:30.6 & +21:08:51 & 0.055108 & 1.078 & \checkmark & \checkmark \\
  A2630 & 2MASX J23380105+1554022 & 23:38:01.0 & +15:54:02 & 0.068170 & 1.313 & other & -- \\
  A2634 & NGC 7720 & 23:38:29.4 & +27:01:54 & 0.030350 & 0.611 & \checkmark & \checkmark \\
  A2637 & 2MASXI J2338533+212752 & 23:38:53.3 & +21:27:53 & 0.073702 & 1.410 & \checkmark & -- \\
  A2657 & 2MASX J23445742+0911349 & 23:44:57.4 & +09:11:36 & 0.041081 & 0.817 & M2 & \checkmark \\
  A2665 & MCG+01-60-039 & 23:50:50.5 & +06:08:59 & 0.056100 & 1.096 & \checkmark & -- \\
  A2666 & NGC 7768 & 23:50:58.5 & +27:08:51 & 0.026955 & 0.545 & \checkmark & \checkmark \\
  A2675 & 2MASX J23554260+1120355 & 23:55:42.6 & +11:20:36 & 0.076893 & 1.466 & \checkmark & -- \\
  A2678 & 2MASX J23554532+1139135 & 23:55:45.3 & +11:39:14 & 0.078125 & 1.487 & M2& -- \\
  AWM1  & NGC 2804 & 09:16:50.0 & +20:11:55 & 0.027670 & 0.559 & -- & -- \\
  AWM5 & NGC 6269 & 16:57:58.1 & +27:51:16 & 0.034891 & 0.699 & -- & -- \\
  AWM7 & NGC 1129 & 02:54:25.2 & +41:34:37 & 0.017639 & 0.361 & -- & \checkmark \\
  L2027 & LEDA 1479941 & 00:43:11.8 & +15:16:03 & 0.078650 & 1.497 & -- & -- \\
  L2030 & NGC 7237 & 22:14:46.9 & +13:50:28 & 0.026180 & 0.530 & -- & -- \\
  L2069 & 2MASX J01072180+1416240 & 01:07:21.8 & +14:16:24 & 0.078512 & 1.494 & -- & -- \\
  L2093 & 2MASX J01092719+1415359 & 01:09:27.2 & +14:15:37 & 0.060780 & 1.181 & -- & -- \\
  L2211 & NGC 7651 & 23:24:26.0 & +13:58:21 & 0.042460 & 0.843 & -- & -- \\
  L3009 & 2MASX J09204890+4039516 & 09:20:48.8 & +40:39:52 & 0.072690 & 1.393 & -- & -- \\
  L3055 & 2MASX J07464283+3059493 & 07:46:42.9 & +30:59:50 & 0.056850 & 1.109 & -- & -- \\
  L3152 & NGC 6338 & 17:15:22.9 & +57:24:41 & 0.027300 & 0.552 & -- & \checkmark \\
  L3186 & 2MASXJ 17153003+6439511 & 17:15:30.0 & +64:39:52 & 0.079040 & 1.503 & -- & -- \\
  MKW4 & NGC 4104 & 12:06:39.0 & +28:10:29 & 0.028605 & 0.577 & -- & -- \\
\enddata
\tablecomments{BCG sample. Cluster names beginning with ``AWM", ``L", and ``MKW"  are taken from the \cite{AWM1977}, \cite{Linden2007}, and \cite{Morgan1965} catalogs, respectively. A comparison to the BCG selection by \cite{Lauer2014} (hereafter L14), is given in column (7). The items stand for agreement (\checkmark), our BCG choice is the second-ranked galaxy L14 (M2), or the cluster is not present in L14 (--) and our choice is neither the first- nor the second-ranked galaxy in L14. The last column states whether \textit{Hubble Space Telescope} images were used to increase the spatial resolution of the inner light profiles.}
\end{deluxetable*}


\begin{figure*}[t]
	\includegraphics[width=\linewidth]{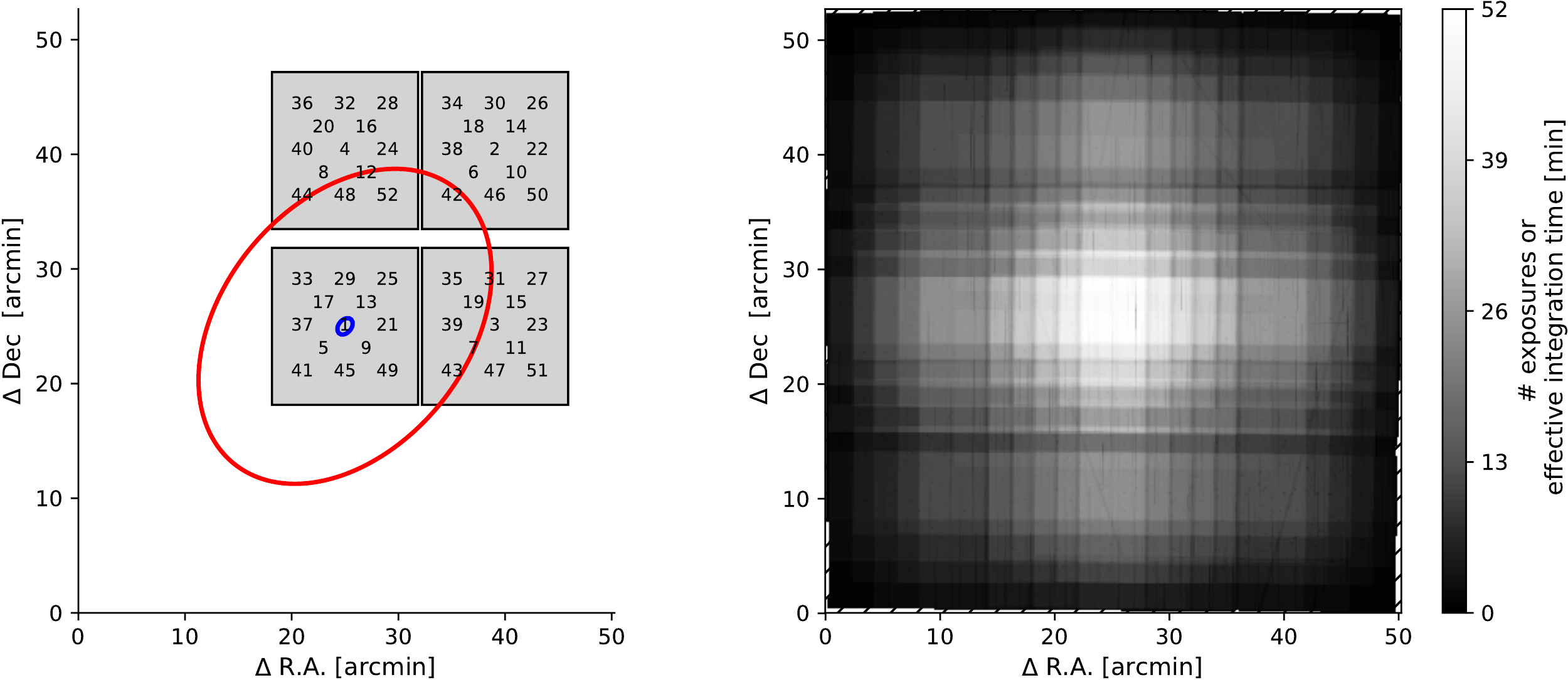}
	\caption{Left: illustration of the dither pattern. The four CCDs are represented by gray squares. The illustrated pointing corresponds to the first element of the dither pattern. The position of the BCG on the detectors is indicated by the number $i$ for each dither element $i$. Blue and red ellipses correspond to the isophotes with ${\rm SB}=30~g'$s mag arcsec$^{-2}$ for the apparently smallest BCG, A2630 (semimajor axis radius $a=50\arcsec$), and the apparently largest BCG, A1367 ($a=955\arcsec$), respectively, that were observed with this dither pattern. Right: stacked weight file of A600. The spatially dependent number of exposures that were added are color coded.\label{fig:ditherpattern}}
\end{figure*}

\section{Data}\label{sec:data}

The observations have been carried out with the 2m Fraunhofer telescope at the Wendelstein Observatory. It is located in the Bavarian Alps, 70 km southeast from Munich, Germany. The telescope is a modern Alt-Az instrument that has been in operation since late 2013. We have utilized the Wendelstein Wide Field Imager (WWFI; \citealt{Kosyra2014}) for our survey as its first light instrument. Its optical components are designed to minimize ghost intensities \citep{Hopp2014}, which qualifies the setup well for a deep imaging study. 

The field of view with $27.6\arcmin\times 28.9\arcmin$ in combination with the large dither pattern is wide enough to cover the ICL down to an SB of 30 $g'$ mag arcsec$^{-2}$ while still providing sufficient sky coverage (see Figure \ref{fig:ditherpattern}, left panel) to model the background accurately. That corresponds to a median semimajor-axis radius of $a=350\pm128$ kpc for our sample.

\begin{figure*}
	\includegraphics[width=\linewidth]{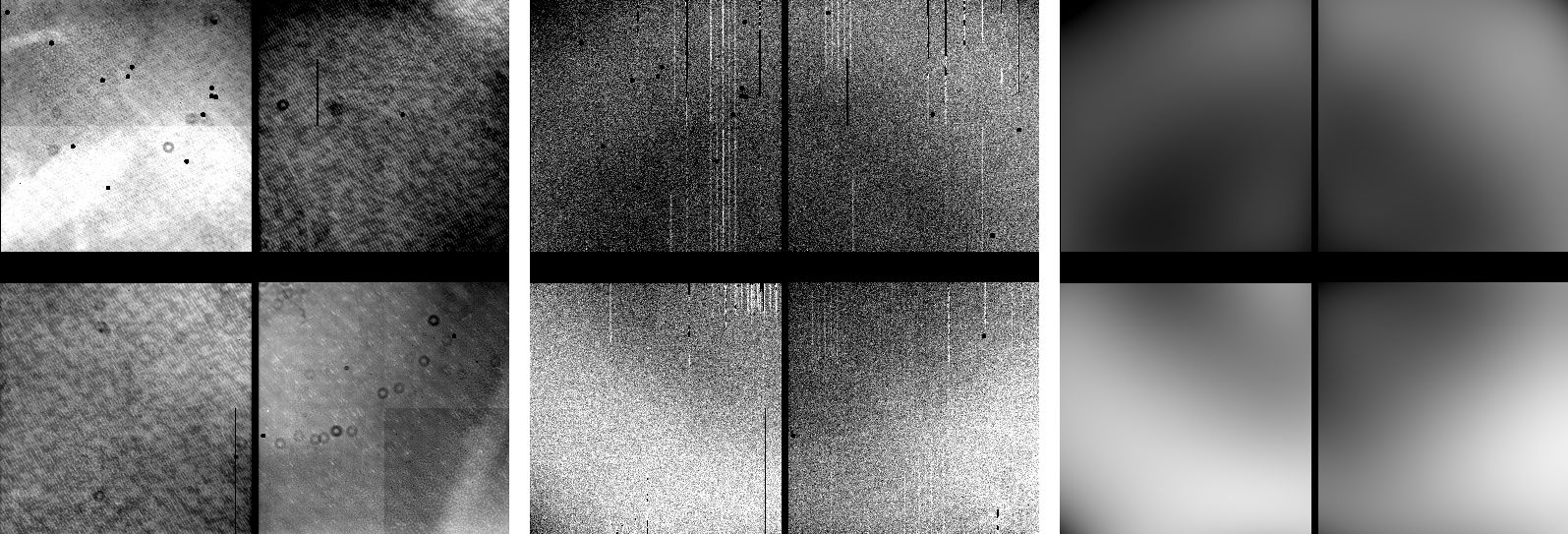}
	\caption{Left: example master flat in the $g'$-band. Variations are on the 3\% level. Middle panel: example night-sky flat, also in the $g'$-band. Variations are on the 2\% level. Charge persistence stripes are visible as vertical white lines. Right panel: fit of the night-sky flat with 2D fourth-order polynomials for each of the four CCDs.\label{fig:flats}}
\end{figure*}

The camera consists of four 4096$\times$4109 pixel sized e2v CCD detectors installed in a camera by Spectral Instruments. The detectors are aligned in a 2$\times$2 mosaic (see Figure \ref{fig:flats}). On the sky, the gaps between the CCDs are 98\arcsec~in the north--south direction and 22\arcsec~in the east--west direction. A large 52-step dither pattern is chosen for the observations to fill up the gaps and provide sufficient sky coverage. It is illustrated in Figure \ref{fig:ditherpattern}. For the first four exposures, the BCG is centered on each CCD, then shifted by two arcminutes in the R.A. or decl. direction before repeating the four exposures off-center. That procedure is repeated 13 times whereby the shifting direction changes for each step. In other words, the dither pattern is a 13-step spiral where each step is repeated on all four CCDs. This strategy allows us to model any time-stable background pattern accurately, which is especially important near the location of the BCG. The total integration time per target is 52 minutes and is split into 60 s single exposures.

We have chosen the $g'$-band for all observations because the night-sky brightness is more stable in that filter band compared to redder bands, due to the absence of strong emission lines. The fact that optical reflections have lower intensities is also important.

The pixel scale of 0.2\arcsec/pixel oversamples the seeing-limited data. The typical seeing of FWHM $\simeq1.2\pm0.2\arcsec$ allows us to resolve the inner cores of BCGs after deconvolving the central image regions. If available, we use high-resolution \textit{Hubble Space Telescope} imaging data downloaded from the Hubble Legacy Archive (https://hla.stsci.edu) to measure the central light profiles and combine them with the profiles measured from wide-field WWFI data.

As the main interest of this work is the faint outskirts of BCGs, the observing constraints were prioritized more on dark and photometric conditions than on excellent seeing. Hence, the median seeing for our survey is worse than the median site seeing of 0.8\arcsec~reported by \cite{Hopp2008}.

\subsection{Data Reduction}\label{sec:datareduction}

The data reduction pipeline was specifically developed and assembled for the WWFI. It includes bias subtraction, flat-fielding, charge persistence, bad pixel and cosmic ray masking, photometric zero-point calibration, background subtraction, bright star removal, resampling, and coadding. The photometric zero-points are calibrated using Pan-STARRS1 DR2 catalogs \citep{Flewelling2020} and provide SB profiles consistent with the Sloan Digital Sky Survey (SDSS). A comparison of 10 BCG SB profiles with those measured from SDSS DR12 image data shows that the SB profiles agree within 0.02 mag arcsec$^{-2}$ before point-spread function (PSF) debroadening correction. Dark current is negligible at the low CCD operating temperature of --115\degr C. Detailed descriptions of all important aspects regarding deep surface photometry follow in the next subsections.

\subsubsection{Bias}

Bias exposures show a chess field--like pattern. Each of the 16 readout amplifiers places a unique bias offset on the corresponding data region. A time-stable vertical line pattern is hidden beneath these offsets. To get rid of this line pattern, we subtract a master-bias image that was created by averaging all the bias images taken in the relevant month. The offsets are subtracted beforehand. Cosmic rays are removed with the tool {\tt cosmicfits} \citep{Goessl2002}.

The values of the offsets themselves are not stable over time. They fluctuate mildly on minute time-scales. We measure $\sim0.1$ ADU residuals even after the clipped average of the corresponding overscan regions had been subtracted. The origin of this effect could be a heating up of the readout electronics, which is correlated to the number of charges being read out. An alternative explanation is based on electromagnetic interferences from a nearby transmitting antenna. In order to eliminate the varying offsets from the science images, we match the background fluxes along 30-pixel-wide adjacent stripes along the borders of each quadrant to the average value of these stripes. This is done for each CCD independently. Regions affected by charge persistence are masked beforehand in order to dismiss contaminated pixels (see Section \ref{sec:cpmasking}).

\subsubsection{Flat-fielding}\label{sec:flatfielding}

We correct for large-scale illumination inhomogeneities and small-scale patterns like dust using calibration images that were taken each night during twilight. These twilight flats are flux-aligned with fourth-order polynomials to each other and then combined into a master flat (Figure \ref{fig:flats}, left panel). Every bias-subtracted science image is divided by this master flat. However, large-scale residuals on a 2\% level remain (Figure \ref{fig:flats}, middle panel). That is common for wide-field imagers. The residual patterns are almost (but not perfectly!) point-symmetric around the center and stable for one pointing in one night. We have identified three properties of this pattern that point toward a color effect as its origin:
(1) the closer to dark time the flats are taken, the weaker is the pattern structure;
(2) it is weaker in narrowband filters; (3) the quotient of two exposures that were taken while first a green and second a red LED illuminated the inner dome shows a similar pattern with $\sim$2\% large-scale variation.

We conclude that the bluer color of the twilight sky compared to the redder night sky, in combination with color-dependent stray light originating inside the optical path, lead to inhomogeneities in the flat-fielding process. No improvement in flatness was accomplished by using dome flats instead of twilight flats. Even though the chosen lamp produced redder light than the twilight sky, the difficulty of illuminating a large inner dome surface homogeneously from a short distance limits the possibility of achieving perfect flatness.

The multiplicative scaling of the flat-field pattern correlates positively with the night-sky brightness. Color changes toward a bluer night sky that are due to airglow, city lights, or during lunar twilight invoke an inversion of the pattern. We factor in that behavior by scaling night-sky flats (NSFs) accordingly (see Equation (\ref{eq:nsfscaling}) in Section \ref{sec:bgsub}).

\subsubsection{Charge Persistence Masking}\label{sec:cpmasking}

Bright foreground stars from the Galaxy are unavoidable in all observed fields, especially due to the wide field of the WWFI camera. The extreme numbers of photons arriving from these stars trigger a tremendous release of free electrons into the valence band of the CCD detector. A small fraction of them gets trapped inside defects in the silicon lattice. These trapped electrons are then released over time where the release rate follows a power law $\dot{N}\propto t^{-1}$. That process can last for hours, depending on the severity of saturation. After the trapped electrons are released, they are stored inside the pixels' potential wells until readout.

When the electrons are being shifted toward the readout register as part of the readout process, they temporarily affect the pixel values along their path. More precisely, the electron bulk loses a fraction in lattice defects of the pixels along the readout direction. Many of these secondarily trapped electrons are released over the first 2 $\mu$s, which is the length of a readout step. As a result, a saturation stripe appears in the same exposure where saturation happened, but opposite to the shifting direction. Another charge persistence stripe appears in subsequent exposures in the shifting direction because the damaged pixels slowly release the remaining trapped electrons (see Figure \ref{fig:flats}, middle panel). Over time, these artificial signals contaminate an increasing fraction of the total field of view because of the dithering strategy.

Our masking strategy is to store the locations where stars saturated and check the corresponding stripes' background flux relative to the $\pm15$-pixel-wide areas alongside them. The charge persistence stripe is being masked when the contaminating signal is higher than the local background plus 0.2 times the rms background scatter. The factor 0.2 was empirically determined to minimize false-positive detections. The location information of a positive detection is forwarded to the subsequent images until the stripe is no longer detectable.

\subsubsection{Bright Star Removal}\label{sec:brightstarremoval}

Bright foreground stars have to be removed from the images for two reasons. Some of the PSFs' extended wings (see Figure \ref{fig:psfmod} and, e.g., \cite{Kormendy1973}) overlap in projection with the targeted intracluster light and they furthermore complicate the background modeling. We follow the strategy from \cite{Slater2009} to model and subtract the $\sim$100 brightest stars in the observed fields. Their approach has been successfully applied for the Burrell Schmidt Deep Virgo Survey \citep{Mihos2017}. \cite{Duc2015} and later \cite{Karabal2017} performed a similar correction for the {\tt MATLAS} survey data, but with time-consuming manual modeling.

We split the cleaning procedure into two steps. First, we subtract the circular PSF light profile from every star, and second, we model and subtract the out-of-focus reflections, which are location dependent. The circularly symmetric light profile for a zeroth magnitude star is shown in Figure \ref{fig:psfmod}. It spans $\sim$14\arcmin~in radius and $\sim19~g'$ mag arcsec$^{-2}$ dynamic range in surface brightness. The blue line is a \cite{Moffat1969} fit to the core and depends on the seeing. The outer components are time-stable because they are due to the optics. Surprisingly, they can be modeled by three $r^{1/4}$ profiles. The outermost $r^{1/4}$ component is extrapolated to the edge of the field of view. We are mostly interested in removing the wings accurately. A single PSF model is therefore sufficient for all observations. The analytic SB profile shown by the red line in Figure \ref{fig:psfmod} is used to model all stars that are listed in the Tycho-2 catalog \citep{Tycho2000} and located inside of a circular aperture with radius $r < 1.3\degr$ around the center of the field. Stellar magnitudes are converted from the Tycho $B_T$ and $V_T$ to the Johnson $B_J$ and $V_J$ filter system using Equation (1.3.20) from \cite{Tycho1997}:

\begin{figure}
	\includegraphics[width=\linewidth]{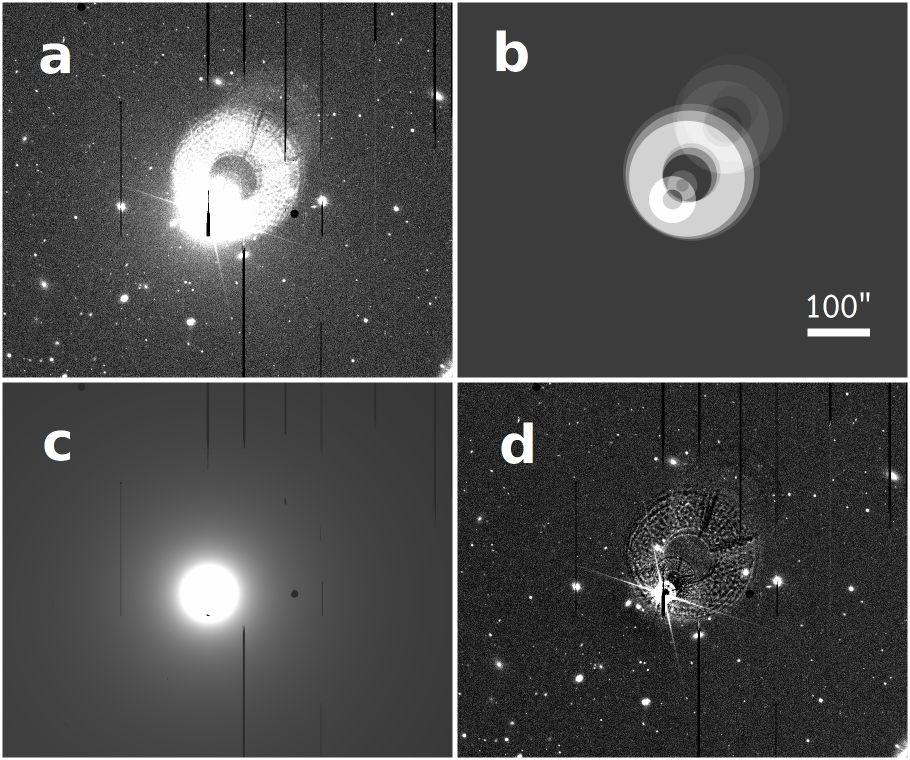}\\
	\includegraphics[width=\linewidth]{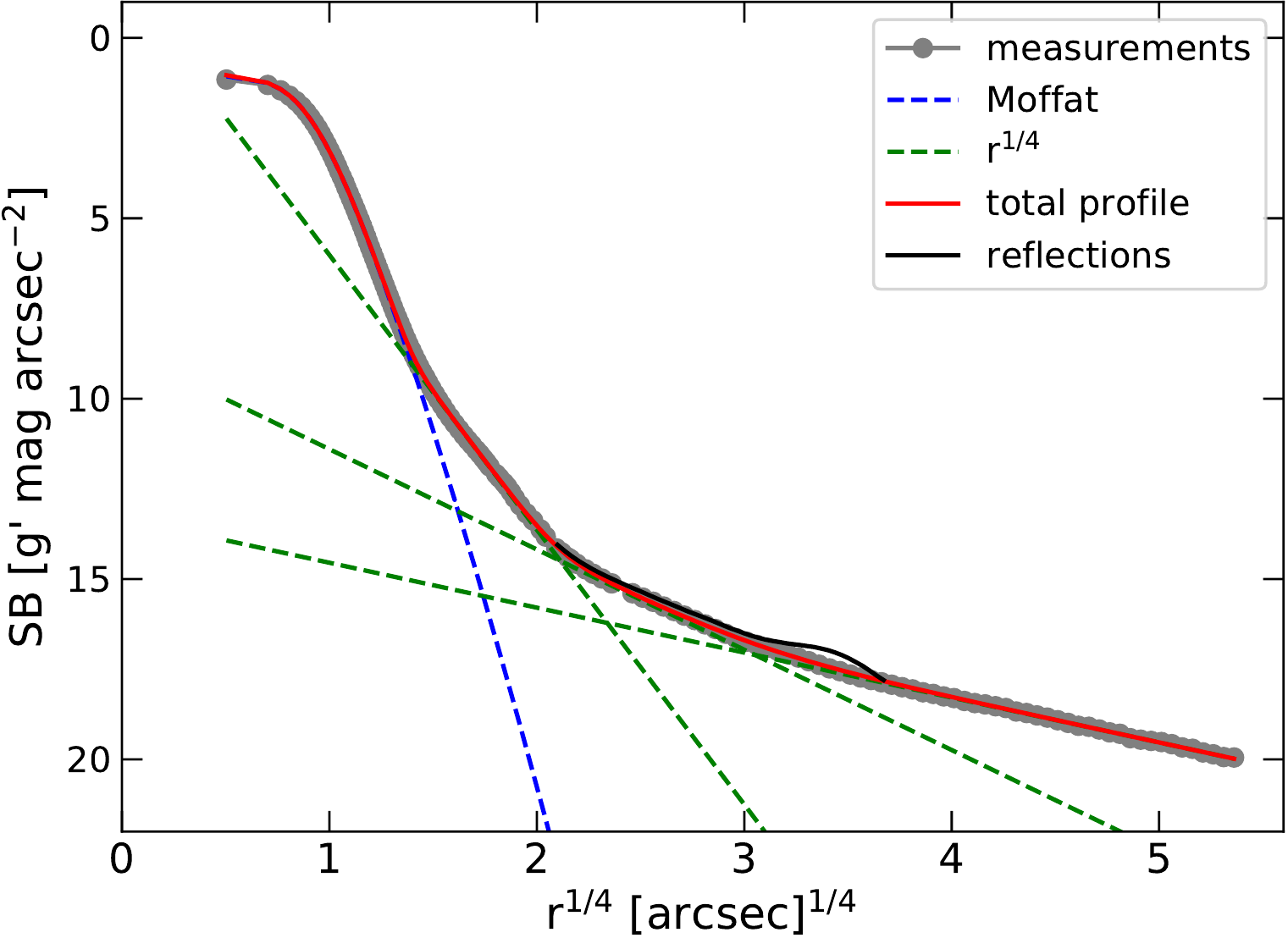}
	\caption{Top panel: (a) example cutout of a bright star, (b) model of the ghosts, (c) model of the point symmetric component of the PSF, (d) residual after subtracting both models. Bottom panel: SB profile of a zeroth magnitude star. The multicomponent fit to the data points is shown as a red line and it is used for the modeling. The FWHM of the Moffat fit (blue dashed line) to the central region is \mbox{FWHM $=1.05\arcsec$}. The outer three $r^{1/4}$ components (green dashed lines) are formed by the optical elements in the telescope. The contribution from the reflections is plotted separately as the black line.\label{fig:psfmod}}
\end{figure}

\begin{align}
V_J &= V_T - 0.09 (B - V )_T,\\
(B - V)_J &= 0.85 (B - V )_T
\end{align}

and are then converted to SDSS $g$-band magnitudes using the equation derived by \cite{Jester2005}:

\begin{equation}
g = V_J + 0.60(B_J-V_J) - 0.12.
\end{equation}

Our photometric zero-points are calibrated to the Pan-STARRS photometric filter system. The difference between SDSS $g$ and Pan-STARRS $g'$ magnitudes (e.g., $g-g'=0.09$ for the Sun; \citep{Willmer2018}) is not relevant here because the reference $g'$-band PSF SB profile is assigned its $g$-band catalog magnitude. It is scaled for different stars according to their $g$-band magnitudes. However, the uncertainties of the color transformations propagate an error to the individual scaling of the model stars. Further relevant effects are intrinsic stellar variabilities and the uncertainty of the preliminary photometric zero-point calibration at that intermediate step of the data reduction. We want to minimize the average residuals of the brightest stars. That is achieved by introducing an empirically determined scaling factor $f^i_{\rm scaling}$ for all stars in each cluster pointing $i$. Our manual choices vary in $0.9 \lesssim f^i_{\rm scaling} \lesssim 1.1$.

\begin{table}
	\centering
	\begin{tabular}{c|c|c|c}
		j & $s$ & $r_{\rm outer}$ (arcsec) & SB ($g'$ mag arcsec$^{-2}$)\\
		\hline
		1 & 0.04371 & 37.4 & 17.03 \\
		2 & 0.08649 & 23.8 & 17.98 \\
		3 & 0.10602 & 92.6 & 17.28 \\
		4 & 0.11811 & 102.6 & 17.40 \\
		5 & 0.12555 & 109.6 & 18.22 \\
		6 & 0.26040 & 74.0 & 19.12 \\
		7 & 0.28365 & 87.6 & 19.12 \\
		8 & 0.31713 & 86.4 & 20.04 \\
	\end{tabular}
	\caption{Reflection properties for a zeroth magnitude star. The relative shift $s$, given in column (2), is defined in Equation (\ref{eq:ringscaling}) as the center offset of a ring with respect to the source divided by its distance from the optical axis. The outer radius of each reflection ring $j$ is given in column (3). The inner radius is always $r_{\rm inner} = 0.424 r_{\rm outer}$. The surface brightness normalized to a 0-th magnitude star of each ring is given in column (4).\label{tab:rings}}
\end{table}

Reflections are considered separately. They are particularly prominent in wide-field imagers, due to the need for multiple corrector optics in order to correct for field distortions. They arise from light that is reflected at various surfaces during its path through the telescope system. These surfaces are the front and back sides of filter glasses, corrector lenses, and the CCD entrance window. The longer path lengths result in out-of-focus duplicates next to bright light sources, so-called ghosts (Figure \ref{fig:psfmod}, top panel). For the WWFI $g'$-band, we calculate that 1.78\% of the PSF's light is redistributed into these ghosts, which is consistent with the findings of \cite{Hopp2014}. We identify eight rings with parameters listed in Table \ref{tab:rings}. The rings are only concentric if the light source is located on the optical axis, that is, close to the field center. They are shifted radially outward in any other case. The relative shift $s$ of ring $j$ is in good approximation linearly dependent on the distance of the star at position $\boldsymbol{r}_{\rm star}$ from the optical axis at position $\boldsymbol{r}_{\rm oa}$:

\begin{equation} \label{eq:ringscaling}
\boldsymbol{r}_j = \boldsymbol{r}_{\rm star} + s \cdot (\boldsymbol{r}_{\rm star} - \boldsymbol{r}_{\rm oa}^{q_i})
\end{equation}
with
\begin{align}
\boldsymbol{r}_{\rm oa}^{q_1}[\rm px] &= (4011,4162), \nonumber \\
\boldsymbol{r}_{\rm oa}^{q_2}[\rm px] &= (4007,-433), \nonumber \\
\boldsymbol{r}_{\rm oa}^{q_3}[\rm px] &= (-195,-443), \nonumber \\
\boldsymbol{r}_{\rm oa}^{q_4}[\rm px] &= (-195,4159),
\end{align}

being the position of the optical axis in the coordinate system of each CCD $q_i$. The central coordinates of the rings are $\boldsymbol{r}_j$. The outer radii $r_{\rm outer}$ are tabulated in Table \ref{tab:rings}. The inner radii are always $0.424r_{\rm outer}$, corresponding to the shaded area of the support for the secondary mirror. The surface brightness of the ring $j$ is $SB_j + g'$ for a star with a $g'$-band magnitude $g'$. The values for ${\rm SB}_j$ are given in Table \ref{tab:rings} and are estimated by scaling the brightness of each ring model independently so that the total residual after subtraction is minimal.

\subsubsection{Background Subtraction}\label{sec:bgsub}

After flat-fielding, large-scale variations in the background pattern are apparent on a 2\% level (see Figure \ref{fig:flats}, middle panel). That corresponds to a surface brightness of ${\rm SB}\sim26~g'$ mag arcsec$^{-2}$. In order to measure accurate SBs at the 30 $g'$ mag arcsec$^{-2}$ level, the background has to be flat on the same level. The necessary calibration has to be performed on the individual images because the dithering between observations would otherwise result in sharp-edged jumps in the background pattern of the coadded mosaic.

The delicacy for every background subtraction method lies in the risk of accidentally subtracting the incompletely masked ICL, which mimics an artificial background pattern. The easiest method to follow would be simple surface polynomial or surface spline fitting (e.g., \citealt{Capaccioli2015}) of the source-masked images. We have discarded this approach because of its severe risk of subtracting part of the ICL. This method is furthermore fairly unstable when the polynomials or splines are unconstrained, due to large masks. This can lead to overshooting, especially when an edge of the image is masked. A detailed explanation of our masking procedure is given in Section \ref{sec:masks}.

We apply the more robust method of subtracting an average model of the background pattern, a so-called night-sky flat (NSF). This is only possible because the background pattern is to zeroth order time-stable (see Section \ref{sec:flatfielding}). A separate NSF is created for every visit, that is, for each target in each night. The major advantage of this method is that the background pattern is known at and around the masked BCG. That is because masked regions in individual images are filled up in the NSF by averaging numerous dithered exposures. Moreover, the issue of incomplete masks is reduced because only a small number of images are contaminated by undetected PSF or galaxy halos at a specific, fixed image location, again thanks to the large dither pattern.

The NSF can either be created from separate sky pointings (\citealt{Iodice2017,Spavone2017}) or from the target pointings themselves. The first option is safer because the risk of including part of the ICL in the NSF is eliminated. The necessary observing strategy is, on the other hand, twice as time-consuming. We take advantage of the $\sim 4\times$ larger field of view compared to the extent to which we measure ICL. An optimized dithering strategy (see Section \ref{sec:data}) ensures that the background can be determined from the target exposures themselves while maximizing the exposure time on-target and minimizing the contribution of the incompletely masked ICL on the NSFs.

The PSF-subtracted (see Section \ref{sec:brightstarremoval}) and source-masked single images are normalized and averaged to an NSF. The normalization is necessary because fluctuations in the sky brightness on a 2\% level are common between exposures, and the normalization is allowed since the background pattern is to first order multiplicative because of its origin in flat-fielding residuals (see Section \ref{sec:flatfielding}).

A number of undetected charge persistence stripes usually become visible in the deep NSFs (see Figure \ref{fig:flats}, middle panel, and Section \ref{sec:cpmasking}). We therefore fit two-dimensional fourth-order polynomials to each CCD region in order to improve the NSF smoothness (see Figure \ref{fig:flats}, right panel). The smoothed NSF is then rescaled in flux back to the individual images $i$ from which it was created. The scaling formula is

\begin{align}
{\rm NSF}_i(x,y) = {\rm NSF}(x,y) \times a_i + b_i. \label{eq:nsfscaling}
\end{align}

We allow for an additive $b_i$ and multiplicative $a_i$ scaling. Two fitting parameters are needed to account for the gradual flipping of an outward-increasing to an outward-decreasing brightness of the background pattern as the night-sky color becomes bluer (see Section \ref{sec:flatfielding}).

Every NSF is scaled slightly too high because incomplete masks are more present in the individual images than in the averaged NSF. That leads to a small, negative background constant on the order of negative $\sim30~g'$ mag arcsec$^{-2}$ that remains in the coadded mosaics (see Section \ref{sec:mockbcg}). This constant is later determined as the value to which the linear light profiles converge at large radii (see e.g., \citealt{Spavone2017}).

\subsection{Source Masking} \label{sec:masks}

\begin{figure*}
	\centering
	\includegraphics[width=0.445\linewidth]{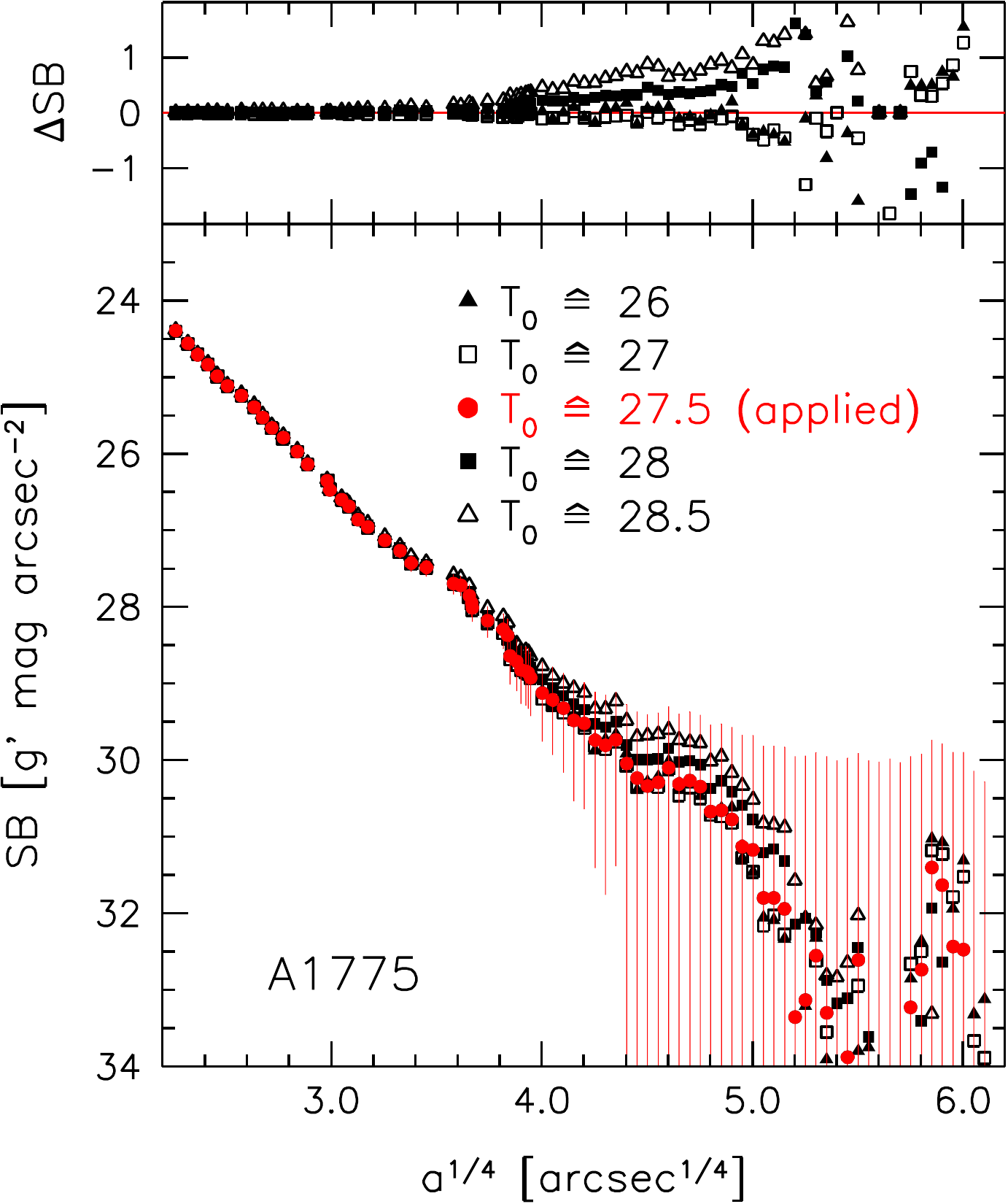} ~~~~~~~~~~~
	\includegraphics[width=0.46\linewidth]{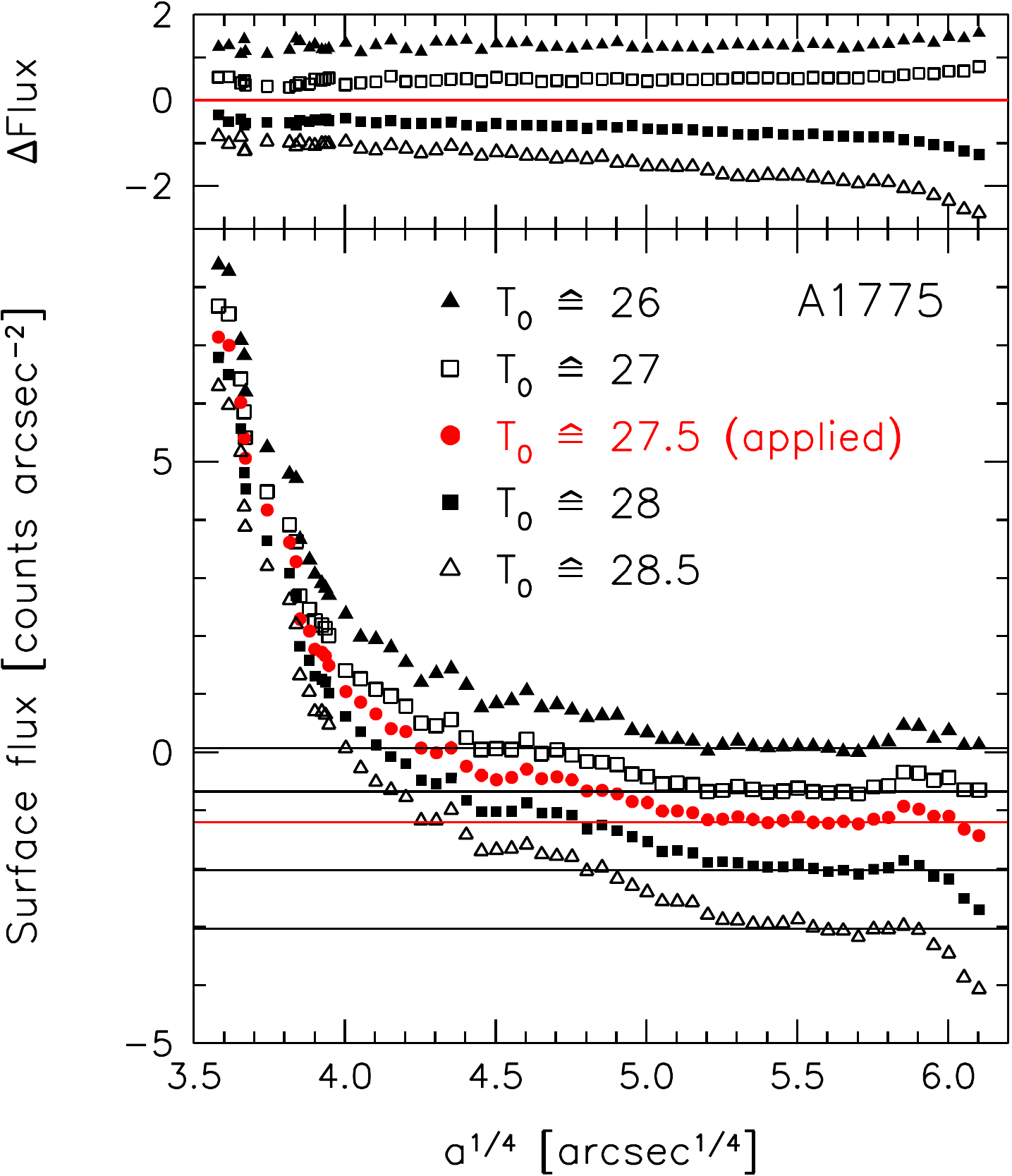}
	\caption{Surface brightness (left) and surface flux (right) profiles of the BCG in A1775 measured for different masking thresholds $T_0$ (Equation (\ref{eq:T02})) converted to units of $g'$ mag arcsec$^{-2}$. The residual differences from the measurements done with the applied masking threshold (red) are shown in the top panels. The residual background choices are shown for each masking threshold as the horizontal lines in the right panel.\label{fig:maskthres}}
\end{figure*}

Two situations require source masking: (1) before averaging images to create NSFs and (2) before measuring the BCG/ICL light profiles. Both situations require different masking techniques, but both resulting masks need to be as complete and on large scales as homogeneous as possible. Tools that are being used by other authors include the IRAF task {\tt objmasks} \citep{Mihos2017} or {\tt ExAM} (\citealt{Huang2011,Spavone2017}), which is based on SExtractor \citep{Bertin1996} catalogs.

We have developed our own technique specifically to tackle the requirement of homogeneity. Our adopted, large dither pattern causes a spatially varying signal-to-noise ratio of $\Delta ({\rm S/N}) > 2$. That is a severe problem for the choice of masking thresholds:

\begin{enumerate}
\item A constant signal masking threshold leads to a more frequent masking of noise peaks as false-positive detections in the low S/N regions.
\item A constant S/N masking threshold leads to fewer detections of sources in the low-S/N regions.
\end{enumerate}

While choice (1) results in a decrease of the average flux value in the low-S/N regions, choice (2) evokes the opposite. Both options therefore introduce a significant bias in the NSF scaling and isophotal flux measurements. A compromise between the two options, that is, a spatially dependent scaling of the masking threshold $T(x,y)$ by the square root of the local background rms scatter rms$(x,y)$, results in satisfyingly homogeneous masks for a low average masking threshold $T_0$:

\begin{equation}
T(x,y) \ge T_0 \cdot \left(\frac{\sqrt{{\rm rms}(x,y)}}{{\rm median}(\sqrt{{\rm rms}(x,y)})}\right). \label{eq:maskthresh1}
\end{equation}

The information of the local background noise rms$(x,y)$ is stored in the stacked weight maps generated by {\tt SWarp} (see Figure \ref{fig:ditherpattern}, right panel, for an example).

We now explain our choices for the average masking thresholds for each of the two scenarios that were mentioned in the beginning of this subsection.

\subsubsection{Masks for Background Modeling}

The basis for this type of mask is a roughly background-subtracted, coadded image. This is in our case a mosaic that was created after subtracting second-order 2D polynomials from the masked single exposures. We mention here that this procedure is iterative. Since there can be no celestial sources with sizes smaller than the seeing, we smooth the stack using a 2D Gaussian filter with standard deviation $\sigma = 11$ px in order to avoid mask fragmentation. All pixels get masked that have greater values than the locally calculated threshold $T(x,y)$, where $T_0$ in Equation (\ref{eq:maskthresh1}) corresponds to a surface brightness of 27.5 $g'$ mag arcsec$^{-2}$. The regions around the BCG and around bright stars are conservatively enlarged by hand to a size where we expect the $29\sim30~g'$ mag arcsec$^{-2}$ isophote to be. We again stress here that ICL residuals at these levels are damped in the NSF by averaging the widely dithered single exposures. That is confirmed by the recovery of a mock BCG-SB profile down to ${\rm SB}>30~g'$ mag arcsec$^{-2}$ as presented in Section \ref{sec:mockbcg}.

\subsubsection{Masks for Light Profile Measurements}

Before measuring the flux from a BCG+ICL, we have to mask all other sources except for the BCG+ICL itself. Our approach to this problem is to subtract a model for the BCG+ICL before creating the mask. We exploit the fact that the BCG+ICL system is usually the largest object in the field of view and has the shallowest SB profile gradient. Thus, it can be approximately modeled by a medium-scale background fitting method. That model is created by performing a bicubic spline interpolation to a grid of points that was generated by calculating the clipped median in (51 $\times$ 51) pixel square apertures around the corresponding locations. After subtracting this model from the stack, we generate and combine one mask for the small and one mask for the medium-sized sources. The stack is smoothed with a 2D Gaussian filter with $\sigma = 5$ px (which is the typical seeing) for the first mask and $\sigma = 21$ px (which is about half of the background interpolation step size) for the second mask. All pixels are masked that have values greater than

\begin{equation}
T(x,y) \ge T_0 \cdot {\rm rms}(x,y) \cdot \left(\frac{\sqrt{{\rm rms}(x,y)}}{{\rm median}(\sqrt{{\rm rms}(x,y)})}\right)^{-1}, \label{eq:T02}
\end{equation}

\noindent where $T_0=0.15$ is given here in units of the local S/N per pixel. We emphasize here that the threshold is extremely low because of the preceding smoothing. Also note that the scaling term is now inverted because the threshold is expressed differently. The chosen threshold $T_0=0.15$ corresponds on average to a surface brightness of $27~\widehat{\lesssim}~T_0~\widehat{\lesssim}$ 27.5 $g'$ mag arcsec$^{-2}$ (see Figure \ref{fig:maskthres}, red label). We decided to fix the masking threshold this time in units of (scaled) S/N because it provides a more consistent mask homogeneity between stacks of different integration times. In practice, fainter average thresholds result to zeroth order in a higher residual background constant (see the horizontal lines in Figure \ref{fig:maskthres}, right panel). That is because the overall distribution of background galaxies is largely homogeneous on the spatial scales of the outermost isophotes. This constant is determined in any case during the measurement of the light profile and thus introduces no bias. A first-order effect of a too-faint masking threshold is a downward bending of the outer surface flux profile (see the slope of the residuals in Figure \ref{fig:maskthres}, right panel). That is due to the outward-decreasing signal-to-noise ratio, as explained in the beginning of this subsection. The fainter the threshold, the more sensitive the mask homogeneity becomes toward spatially varying S/N ratios. The effect is reduced by $\sim50\%$ by the spatial scaling of the threshold $T_0$ (Equation (\ref{eq:T02})) but not fully eliminated. The downward bending also biases the background constant choice to too-low values. Both effects combined result in too-bright SB data points in $3.6<(a[{\rm arcsec}])^{1/4}<5.5$ (empty triangles and filled squares in Figure \ref{fig:maskthres}, left panel). The same panel also shows that the SB profiles derived with masking thresholds $26~\widehat{\leq}~T_0~\widehat{\leq}$ $27.5$ are consistent with each other. The explained effects are less important for shallower thresholds because fewer pixels are affected. The optimal threshold is therefore the faintest one that produces a surface flux profile that is still consistent with those derived with shallower thresholds. For the case of A1775, we find the optimal threshold to be $T_0=0.15~\widehat{=}~27.5~g'$ mag arcsec$^{-2}$.

The masks are expanded by convolving them with a circular kernel with radius $r=9$ px for the first mask and $r=11$ px for the second mask so that no light around small and medium-sized sources leaks visibly out of the expanded masks. The spline interpolation produces artifacts in the central areas of the BCG. We unmask and remask these regions by hand. Finally, the regions around bright and extended sources excluding the BCG+ICL are conservatively expanded by hand to a size where we expect the $29\sim30~g'$ mag arcsec$^{-2}$ isophotes to be. The average masked fraction in the final masks is $33\pm5\%$.

\subsection{Astrometry, Resampling, and Stacking}\label{sec:stacking}

The astrometric solutions were calculated with {\tt SCAMP} \citep{Bertin2006}. The resampling and coadding of the calibrated images is performed with {\tt SWarp} \citep{Bertin2010}. The individual images are weighted by their inverse background rms scatter squared to obtain an optimal S/N for extended sources.

\newpage

\section{Surface Brightness Profiles and Isophotal Shape Parameters} \label{sec:composite}

\subsection{Fitting Ellipses to the Isophotes}\label{sec:ellfitn}

Azimuthally averaged surface brightness (SB) profiles of all BCGs are measured by fitting ellipses to the galaxies' isophotes with the code {\tt ellfitn} from \cite{Bender1987}. All ellipses have five degrees of freedom: semi major axis radius $a$, ellipticity $\epsilon = 1 - b/a$ where $b$ is the semi-minor axis radius, central coordinates $X_0$ and $Y_0$, and position angle PA.

Deviations $\Delta r_i$ of the $i$th isophote from a perfect ellipse are expanded in a Fourier series,
\begin{equation}
\Delta r_i = \sum\limits_{k=3}^{19}[a_k\cos(k\theta_i)+b_k\sin(k\theta_i)],\label{eq:fourier}
\end{equation}
where $\theta$ is the eccentric anomaly.

The routine breaks down usually around ${\rm SB} \sim 27~g'$ mag arcsec$^{-2}$ where the low-SB halos of satellite galaxies deform the ICL isophotes on the one hand, but too conservative masking on the other hand prevents the routine from finding enough sampling points for the ellipse fitting. In order to estimate the light profiles beyond that SB, we fix the isophotal shape parameters $\epsilon$, PA, $X_0$, and $Y_0$ for all ellipses that are larger than the one where the scatter in these parameters increases significantly. The semimajor-axis radius for that ellipse is on average $207\pm141$ kpc with a median of 178 kpc. No isophotal parameters besides the flux are determined beyond this radius. The fluxes along all elliptical isophotes in the extended WWFI profiles are then determined by the method described in Section \ref{sec:fluxmeasurement}.

Systems with strong overlap between the BCG and satellite galaxies (e.g., A1656) are handled by mirroring parts of the uncontaminated side of the BCG on the contaminated side before measuring the isophotal shapes.

\subsection{Isophotal Flux Measurement} \label{sec:fluxmeasurement}

\begin{figure}
	\includegraphics[width=\linewidth]{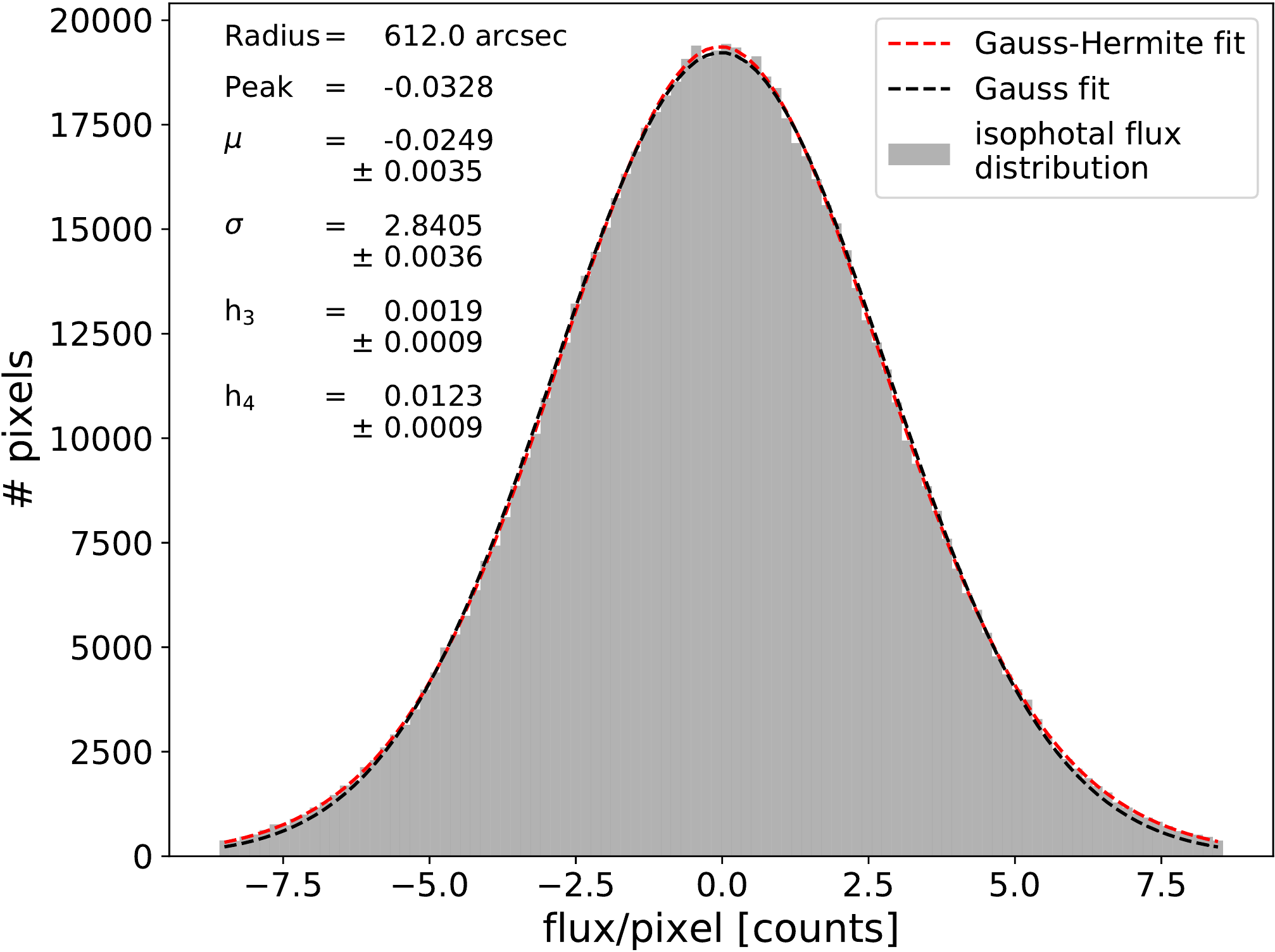}
	\caption{Pixel histogram of an example isophote with ${\rm SB}=30~g'$ mag arcsec$^{-2}$. The flux units are calibrated for a pixel scale of 0.2\arcsec/px and a photometric zero-point of $ZP=30~g'$ mag. The Gaussian fit is overplotted as a black dashed line, and the Gauss--Hermite fit that includes two higher orders $h_3$ and $h_4$ is overplotted as a red dashed line. The value $\mu$ is used to the calculate the SB of the isophote. The negative background constant is not yet subtracted here.\label{fig:fluxmeasurement}}
\end{figure}

The tools for the flux measurements are developed by us on the basis of Python scripts. They use the output isophotal shape tables that are provided by {\tt ellfitn}. The flux along an isophote is measured in an elliptical annulus centered around that isophote. The annulus has a thickness of the average separation between two consecutive isophotes, calculated in $r^{1/4}$ units and evaluated at the isophotal radius. In other words, the annuli are not overlapping but all together cover the full area. We measure the isophotal flux by fitting a Gaussian with two higher-order moments \citep{vanderMarel1993} to the pixel histogram (see Figure \ref{fig:fluxmeasurement}). The distribution is $\kappa-\sigma$ clipped on both sides at three times the standard deviation. The third and fourth Gauss--Hermite moments are given by

\begin{align}
f(F) = A \exp(-0.5F^2) \times [1 &+ h_3 (c_1 F + c_3 F^3)\\
& + h_4 (c_0 + c_2 F^2 + c_4 F^4], \nonumber
\end{align}
where $F=(x-\mu)/\sigma$ with $\mu$ being the mean and $\sigma$ being the standard deviation of the standard Gaussian. The normalization coefficient is $A$ and the other coefficients are given as \mbox{$c_0=\sqrt{6}/4$}, \mbox{$c_1=-\sqrt{3}$}, \mbox{$c_2=-\sqrt{6}$}, \mbox{$c_3=2\sqrt{3}/3$}, and \mbox{$c_4=\sqrt{6}/3$}. We use $\mu$ as the final value for the flux measurement.

The wings of the distribution are larger than what would be estimated from a simple Gaussian fit. Incompletely masked stellar halos, galactic outskirts, or cirrus introduce an asymmetry of the distribution toward more positive values, which we describe by the $h_3$ component. Noisier than average images are weighted less during coaddition. The result on the pixel histogram is similar to adding a second Gaussian component of low amplitude but with larger standard deviation to the high-S/N data. We quantify that behavior with the symmetric $h_4$ component. 

The systematic errors in the light profile of the mock galaxy (see Section \ref{sec:mockbcg}) were smallest when using the mean of the higher order Gaussian $\mu$ as a robust estimator for the flux. We therefore calculate all SBs from this parameter.

A residual, negative background constant remains in every coadded mosaic (see Section \ref{sec:bgsub}). We estimate this constant as the value to which the linear light profile converges at large radii. An example is shown in Section \ref{sec:mockbcg}. That constant is subtracted from all flux data points before these are converted into magnitudes.

\subsection{Composite SB Profiles} \label{sec:merging}

To improve the spatial resolution of the inner light profiles, we deconvolve the innermost $80\times80\arcsec$ of our WWFI data using the {\tt MIDAS} task {\tt deconvolve/image}. The task uses 40 iterations of the Richardson--\cite{Lucy1974} algorithm. If available (see Table \ref{tab:sample}), we use instead archival \textit{Hubble Space Telescope (HST)} data in the filter band that is closest to the $g'$-band. The background constant is poorly calibrated in the \textit{HST} imaging data. We vary it manually until the inferred SB profile agrees best and over the largest radial interval with the WWFI-determined SB profile. The photometric zero-point of the \textit{HST} data is also adjusted in the same way.

\begin{figure}
	\includegraphics[width=\linewidth]{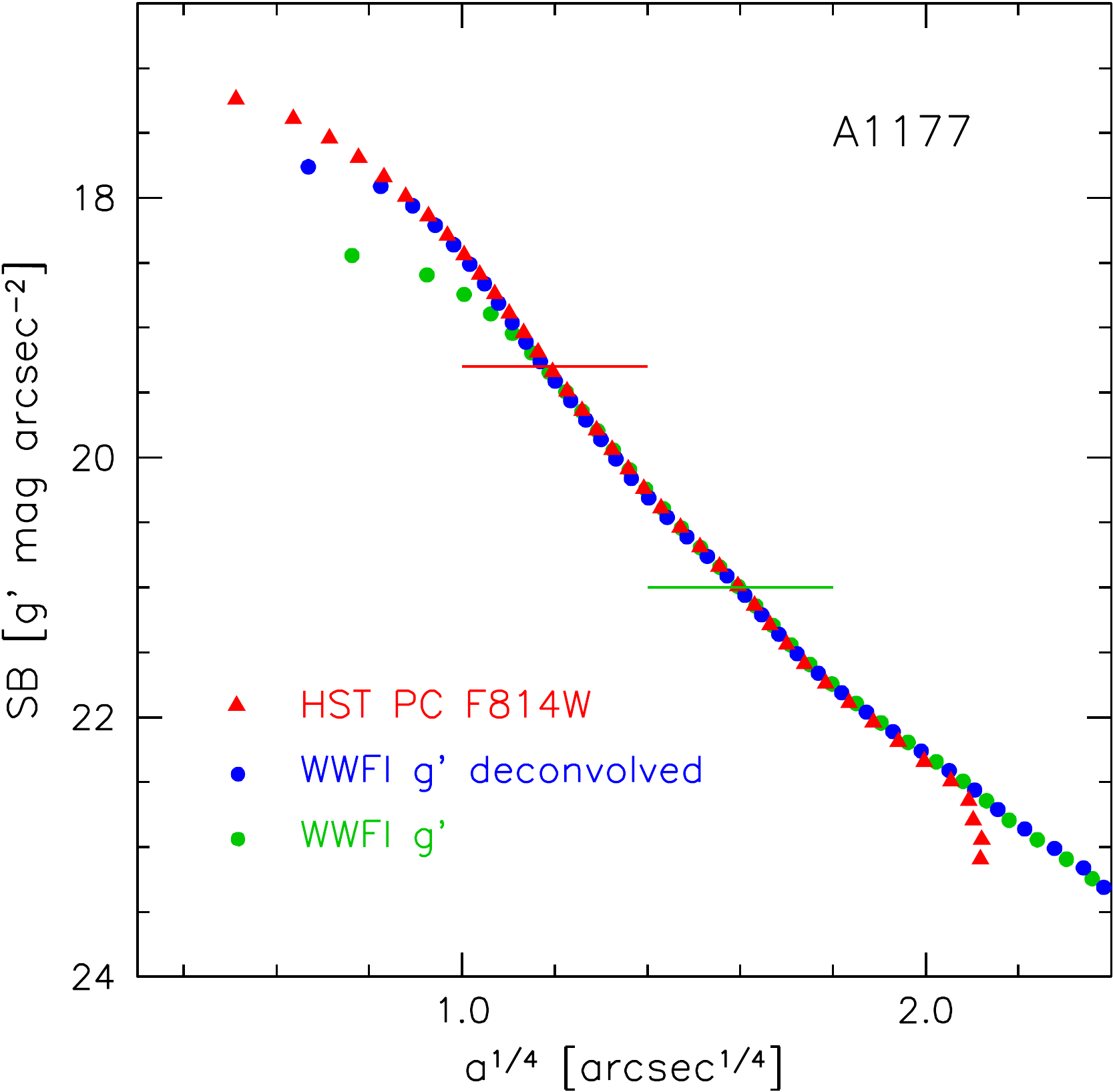}
	\caption{Inner light profile of NGC 3551 in A1177. The green and blue data points are obtained from WWFI data before and after deconvolution, respectively. The Nyquist sampling limit is reached at $a^{1/4} = 0.8\arcsec^{1/4}~\widehat{=}~0.4\arcsec=a$. The red data points are from archival \textit{HST} data. The transition region between \textit{HST} and nondeconvolved WWFI data is in between the two horizontal lines.\label{fig:seeing}}
\end{figure}

A transition region is defined for each light profile where both the \textit{HST} or deconvolved WWFI profile and the extended WWFI profile overlap well (horizontal lines in Figure \ref{fig:seeing}). Both profiles are merged in this transition region by weighted averaging of the data points.

The merging and replacing of the inner data points are done for all isophotal shape parameters.

\newpage

\subsection{S\'ersic Fits} \label{sec:sersicfits}

The \cite{Sersic1968} function is an empirical description for SB profiles of elliptical galaxies. It fits the semimajor-axis profile shapes of elliptical and spheroidal galaxies overwhelmingly well over a large radial range that, for most galaxies, covers $93-99$\% of the total galaxies' light \citep{Kormendy2009}. It was first used to fit SB profiles of BCGs by \cite{Graham1996} who demonstrated its superiority to the hitherto preferentially used de Vaucouleurs ($n=4$) profile because the S\'ersic indices of BCGs are usually $n>4$. The S\'ersic function is given by

\begin{equation}
{\rm SB}(a)={\rm SB}_{\rm e} + c(n)\cdot \left[\left(\frac{a}{a_{\rm e}}\right)^{\frac{1}{n}}-1\right], \label{eq:ss}
\end{equation}

where $a_{\rm e}$ is the effective radius, that is, the semimajor-axis radius of the isophote that encloses one-half of the galaxies' total light. The effective surface brightness ${\rm SB}_{\rm e} = {\rm SB}(a_{\rm e})$ is the SB at radius $a_{\rm e}$. Half of each galaxy's total light is below this SB. It is not to be confused with $<SB_{\rm e}>$, the average surface brightness inside of $a_{\rm e}$, which is often used in the literature and significantly brighter than ${\rm SB}_{\rm e}$. The normalization constant $c(n)=2.5\times(0.868n - 0.142)$ ensures that $a_{\rm e}$ encloses half of the total light. Finally, the S\'ersic index $n$ controls the outer upward bend of the profile. Higher S\'ersic indices hint at a more dominant halo.

If the curvature becomes too strong, then $n$ diverges. For instance, the SB profile of the L3009 BCG ($n=77\pm111$) has a curvature close to that critical value. The power-law slope of the SB profile for divergent $n$ is +5. Stronger curvatures cannot be fitted by a single S\'ersic function. We then extend the fitting formula by a second S\'ersic function SB$_2$ to account for an outer light excess above the inner S\'ersic function SB$_1$:

\begin{equation}
{\rm SB}(a)=-2.5\log_{10}(10^{-0.4 {\rm SB}_1(a)} + 10^{-0.4 {\rm SB}_2(a)}). \label{eq:ds}
\end{equation}

The radius where both S\'ersic profiles cross, that is, where their SBs are equal, is referred to as transition radius $r_{\times}$. The SB at that point is the transition surface brightness ${\rm SB}_{\times}$.

The outer component is sometimes interpreted as the ICL, which is thereby assumed to be photometrically distinct. The (non)justification of this interpretation will be discussed in a forthcoming paper. The BCGs whose light profile can be fitted well enough by only one S\'ersic function are referred to as {\it single S\'ersic BCGs} (SS BCGs), and the BCGs that need two additive S\'ersic functions are referred to as {\it double S\'ersic BCGs} (DS BCGs).

An alternative explanation for the origin of some DS profile shapes could be due to a central poststarburst stellar population that formed after a wet merger, as it is often seen in extra-light ellipticals (e.g., \citealt{Faber1997,Kormendy1999,Kormendy2009,Kormendy2013}). The origin of the DS shape would then be unrelated to the ICL phenomenon. Those BCGs have small DS transition radii relative to their effective radii $r_{\times} < 0.1 r_{\rm e}$ and small DS transition SBs of ${\rm SB}_{\times} < 23~g'$ mag arcsec$^{-2}$ \citep{Hopkins2009,Kormendy2009}. We neglect those inner regions for the fits.

The composite SB profiles including the S\'ersic fits are shown in Appendix \ref{sec:sbprofiles}, and the best-fit parameters are presented in Section \ref{sec:strucparams}. They are corrected for the broadening effects of the PSF wings (see Section \ref{sec:psfeffects}). The fits are performed using the {\tt python scipy} routine {\tt curve\_fit}, which is based on a nonlinear least squares method using the Levenberg--Marquardt algorithm. As pointed out by many authors (e.g., \citealt{Seigar2007,Kormendy2009,Spavone2017}), a simple $\chi^2$ minimization based on measurement uncertainties is not optimal. The reasons are as follows: (1) the S\'ersic model is empirical and does not describe the SB profile shapes perfectly, especially not the intrinsic ``wiggles"; (2) the brightest SB data points have negligible uncertainties compared to the faintest, outermost ones, which would therefore render the outermost data points useless; (3) errors are strongly correlated, and (4) symmetric uncertainties in the background constant are asymmetric in magnitude units. The SB profiles in Appendix \ref{sec:sbprofiles} show that the scatter and systematic deviations from the S\'ersic fits increase at faint SB levels. Therefore, we want to lower their weight but not neglect them for the fits. To achieve that, we minimize the function $\chi^2 = \sum_i ({\rm SB}_i - {\rm SB}^{\rm fit}_i) / \Delta {\rm SB}_i$ where SB$_i$ is the $i$th SB data point, ${\rm SB}^{\rm fit}_i$ is the value from the fit and $\Delta {\rm SB}_i$ are the uncertainties in SB that depend on SB$_i$ itself. The latter do not represent the measurement uncertainties but still increase toward fainter SBs. We use a combination of two uncertainties. One is the background uncertainty of $\Delta {\rm BG} = \pm1$ count arcsec$^{-2}$, given for a photometric zero-point of $ZP=30~g'$ mag (see Section \ref{sec:mockbcg}). Since the linear error bars are asymmetric in magnitude units, we mirror the upper error bars downward. We also added quadratically a systematic uncertainty of 0.18 $g'$ mag arcsec$^{-2}$, which is on the order of stronger intrinsic ``wiggles" in the SB profiles. We get

\begin{equation}
	\Delta {\rm SB} = {\rm SB} + 2.5\log( 10^{-0.4({\rm SB}-{\rm ZP})}+\Delta {\rm BG})+{\rm ZP} + 0.18.
\end{equation}

The errors of the best-fit parameters are estimated using Monte Carlo simulations. They are on the same order of magnitude as the uncertainties due to profile cropping (see Section \ref{sec:integrationcrop}).

The cores below a median major-axis radius of $a = 0.86\pm0.26\arcsec$ are excluded from the fits. The (usually) missing light has negligible influence on the structural parameters.

\subsection{2D Profile Integration} \label{sec:integrationeps}

We calculate the total flux $F_{\rm tot}$ and half-light parameters $r_{\rm e}$ and SB$_{\rm e}$ of the galaxies by integrating the 2D light profiles numerically while considering the radially varying ellipticities. The SB and ellipticity profiles are spline-interpolated and then evaluated on a grid with equidistant step sizes $\Delta a[\arcsec]^{1/4} = 0.001$. The ellipticities below (beyond) the first (last) measured data point are kept fixed. The SBs fainter than the last measured data point or below the limiting magnitude of our survey, ${\rm SB}_{\rm lim} = 30~g'$ mag arcsec$^{-2}$, are replaced by the single or double S\'ersic fit. The two outer limits to which we integrate the light profiles are ${\rm SB} = 30~g'$ mag arcsec$^{-2}$ and effectively infinity.

The step sizes $a_{i+1} - a_i$ are smaller than the scales on which the flux $F$ and ellipticity $\epsilon$ change significantly. In that limit holds the approximation

\begin{align}
F_{\rm tot} & \simeq \frac{1}{2} (F_{\rm tot}^{\rm lower} + F_{\rm tot}^{\rm upper}),
\end{align}

where

\begin{align}
F_{\rm tot}^{\rm upper} & = \sum_i F_{i+1} \cdot \pi (a_{i+1}^2(1-\epsilon_i) - a_i^2(1-\epsilon_i)),\\
F_{\rm tot}^{\rm lower} & = \sum_i F_i     \cdot \pi (a_{i+1}^2(1-\epsilon_i) - a_i^2(1-\epsilon_i)).
\end{align}

The effective radius $r_{\rm e}$ is the semimajor-axis radius of the isophote that encircles one-half of the galaxy's integrated flux $F_{\rm tot}/2$. The effective surface brightness SB$_{\rm e}$ is the SB at that isophote.

\section{Error Analysis and Correction for Systematic Effects}

\subsection{Background Subtraction} \label{sec:mockbcg}

\begin{figure*}
	\centering
	\includegraphics[width=0.445\linewidth]{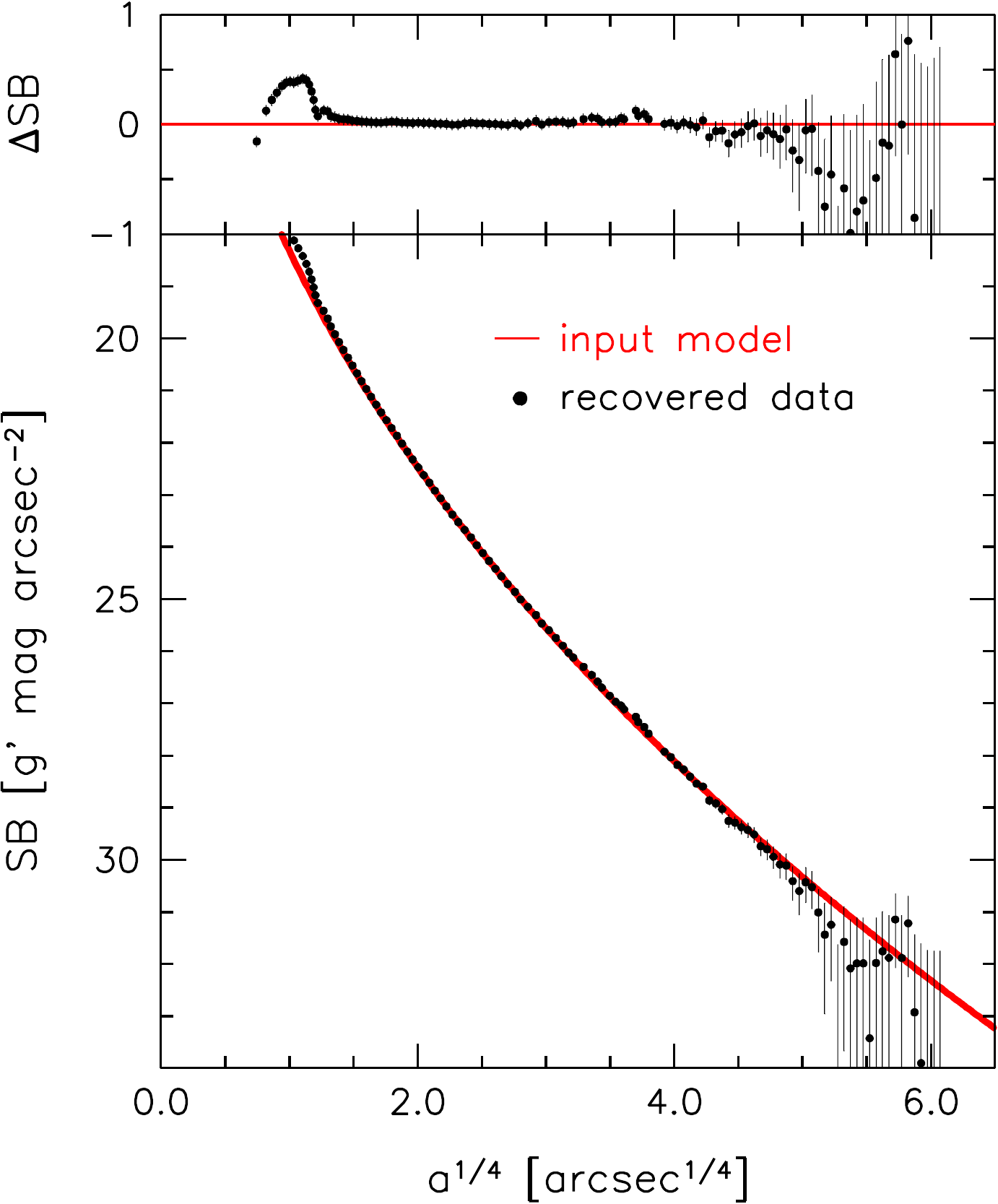} ~~~~~~~~~~~
	\includegraphics[width=0.46\linewidth]{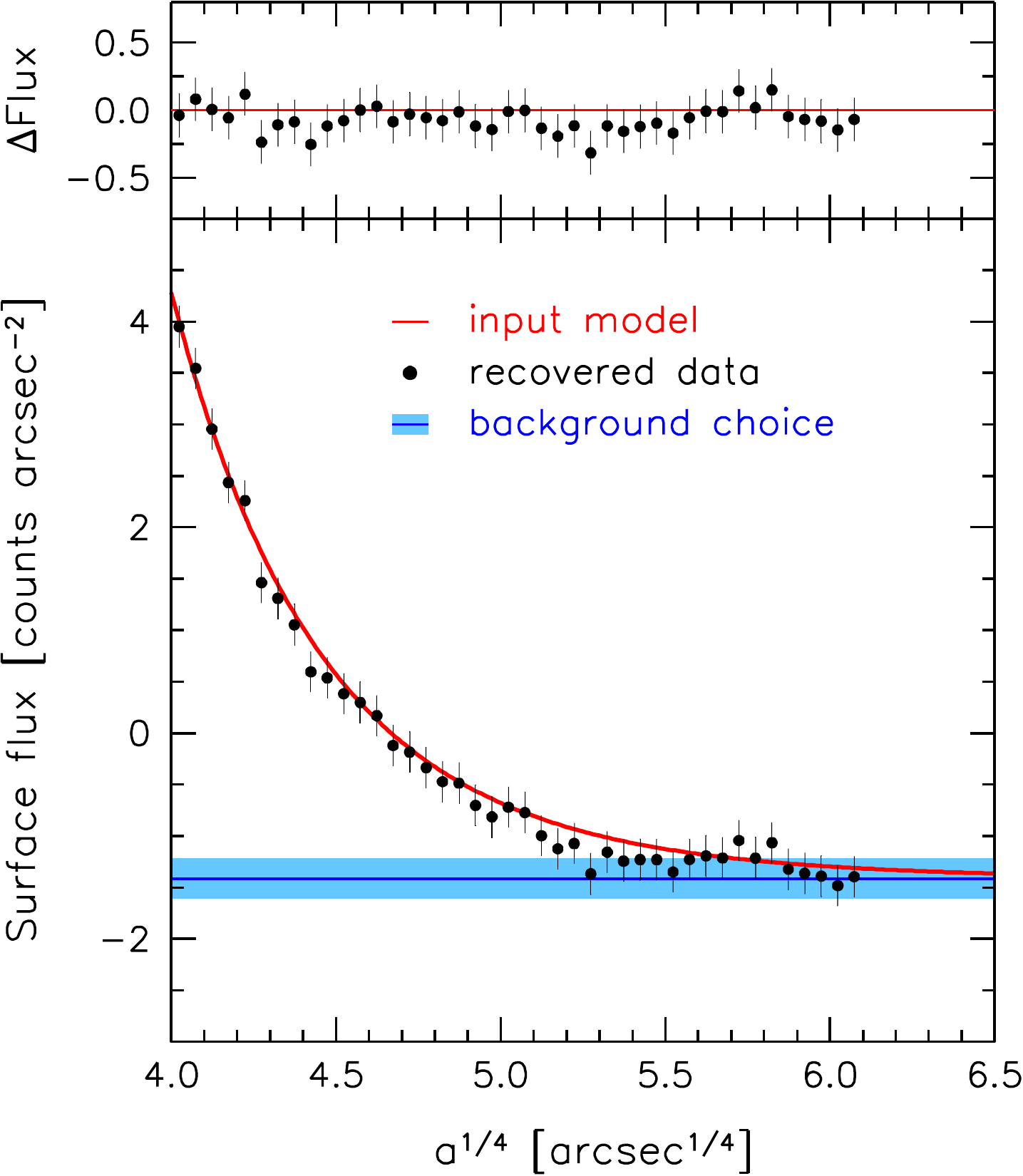}
	\caption{Recovered SB profile of a mock BCG that was inserted into raw data of a sky pointing. The regular data reduction including background subtraction was performed after that. The profile of the input model is plotted as a red line. The error bars are defined by the subjective uncertainty (blue shades) of the residual background constant (blue line).\label{fig:mockbcg}}
\end{figure*}

The extended and faint nature of the ICL makes it susceptible to being subtracted in the progress of background subtraction. We examine the magnitude of this effect with the help of mock data. An empty sky region was observed with the same strategy as the galaxy clusters. Then we insert a mock BCG with a perfect S\'ersic light profile ($r_{\rm e}=100\arcsec$, ${\rm SB}_{\rm e}=26~g'$ mag arcsec$^{-2}$, $n=9$) into the raw data and reduce the data. The deviation of the measured light profile from the input profile provides a measure of the errors that we introduce by background subtraction and masking.

\begin{figure*}
	\centering
	\includegraphics[width=0.49\linewidth]{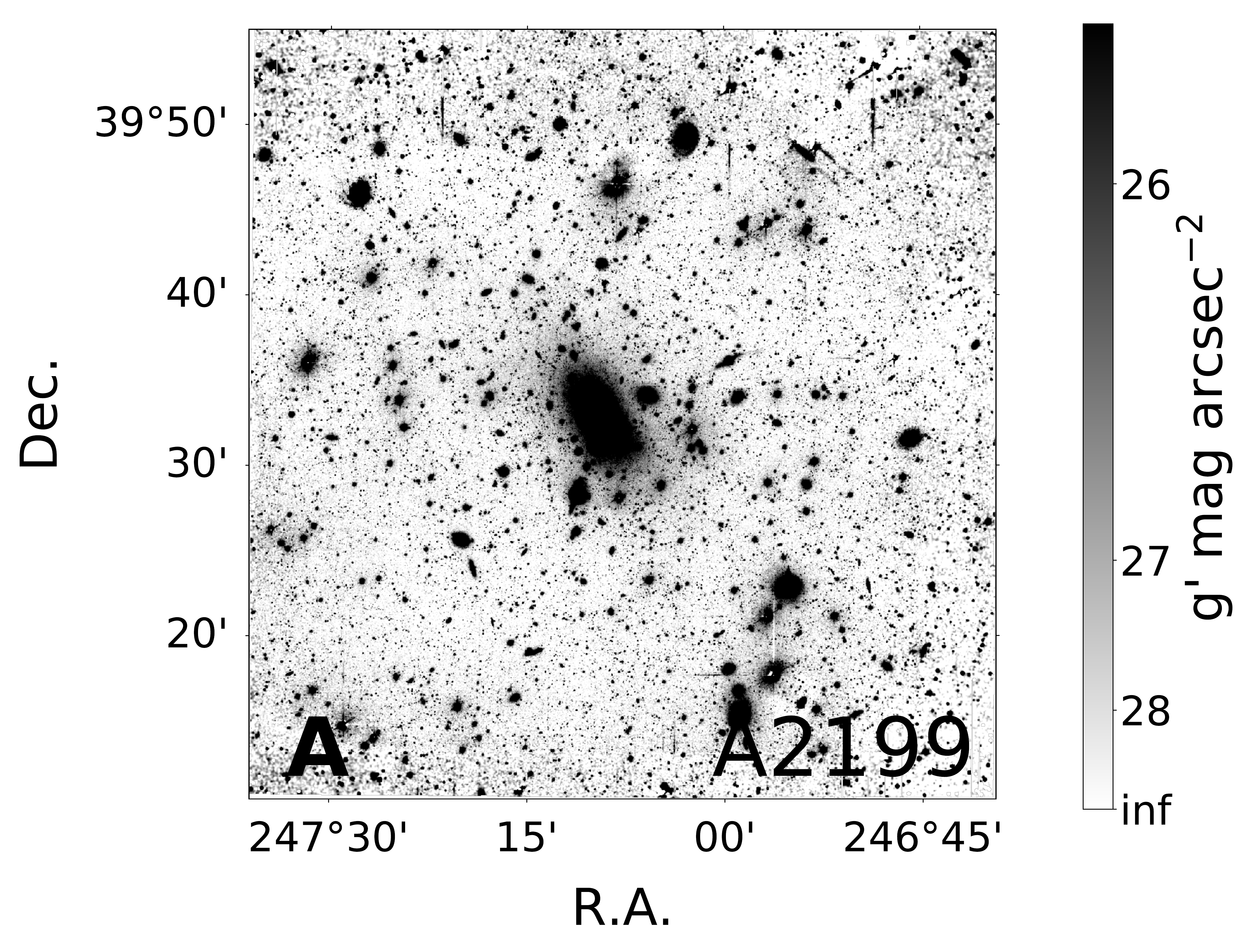}~~~
	\includegraphics[width=0.49\linewidth]{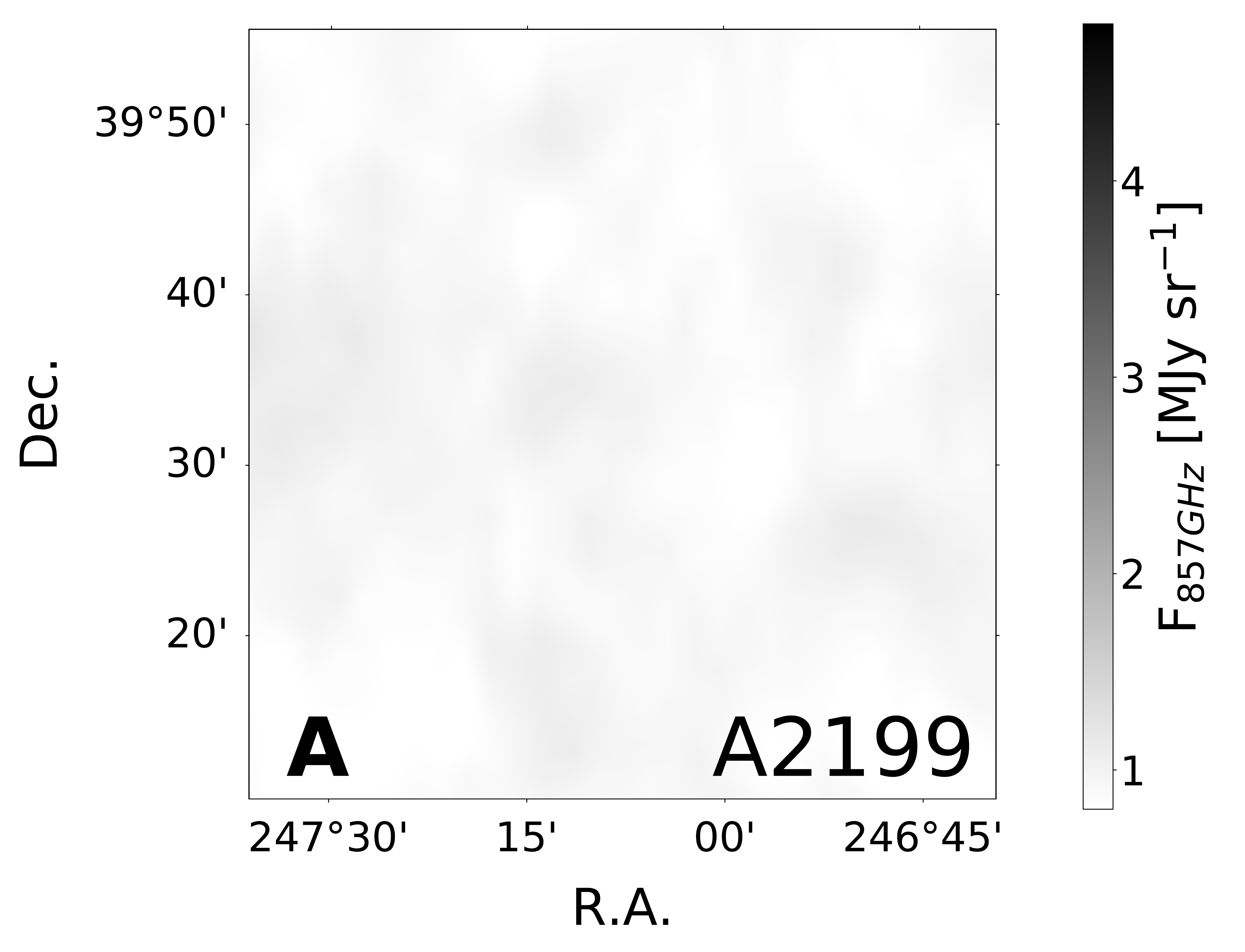}\\
	\includegraphics[width=0.49\linewidth]{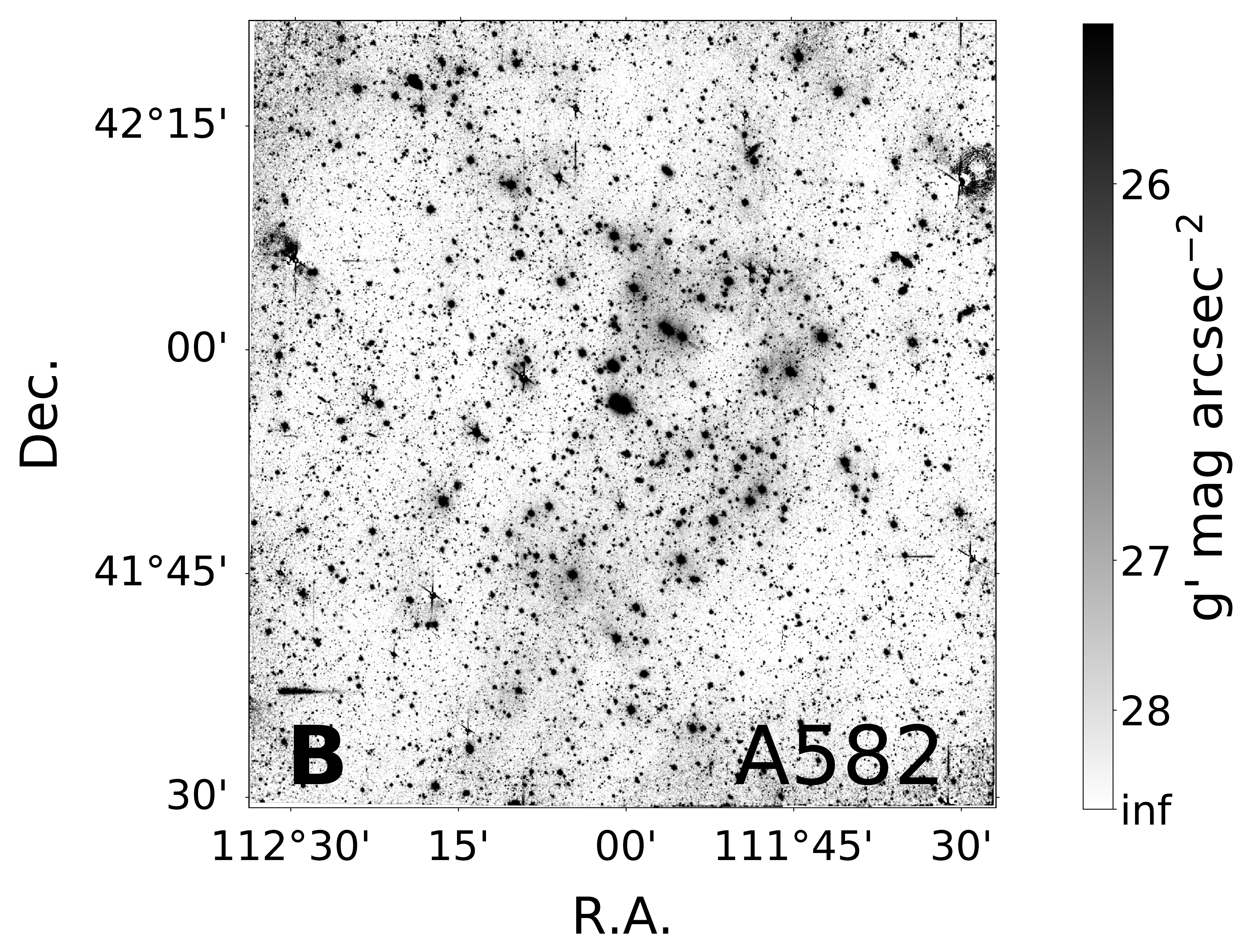}~~~
	\includegraphics[width=0.49\linewidth]{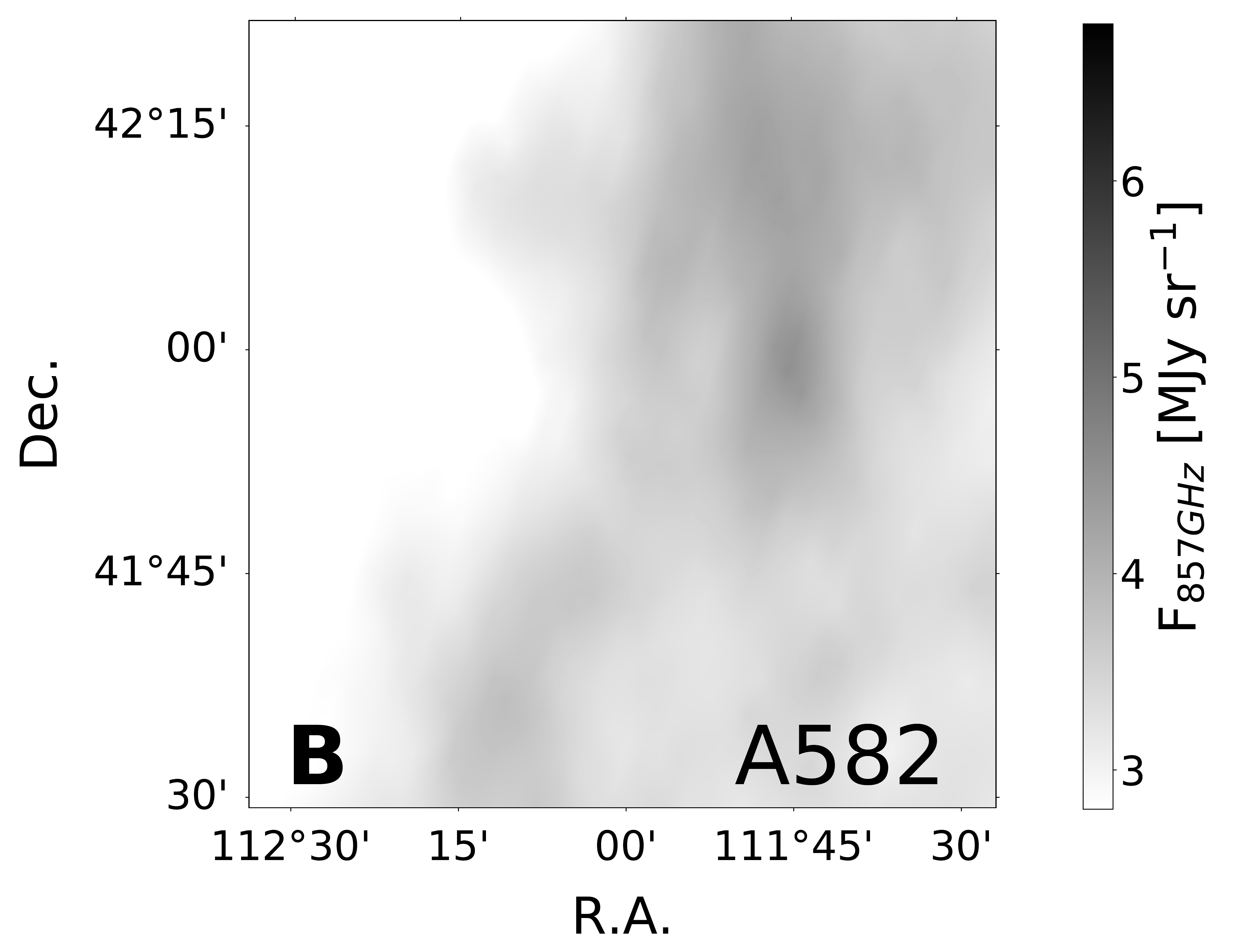}\\
	\includegraphics[width=0.49\linewidth]{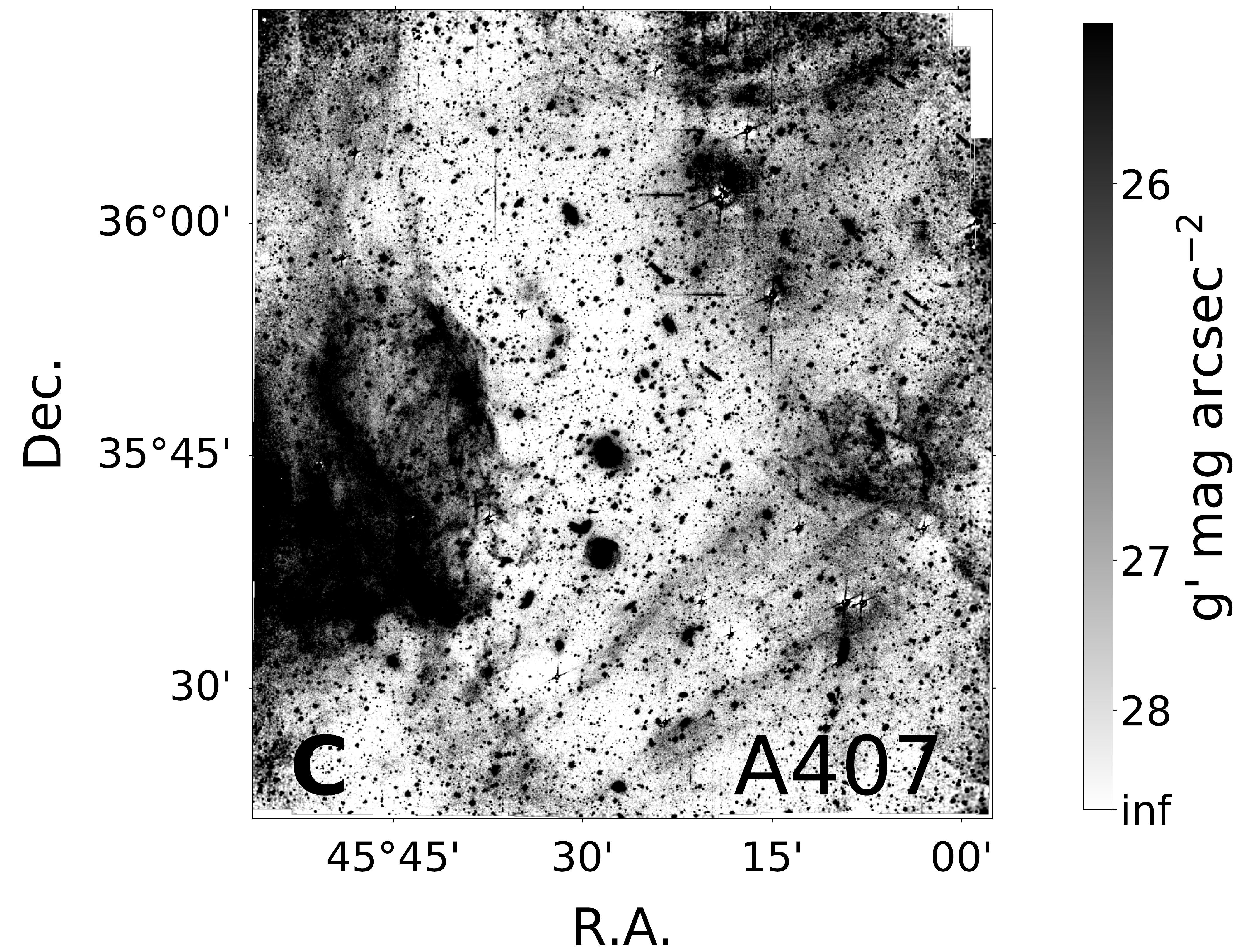}~~~
	\includegraphics[width=0.49\linewidth]{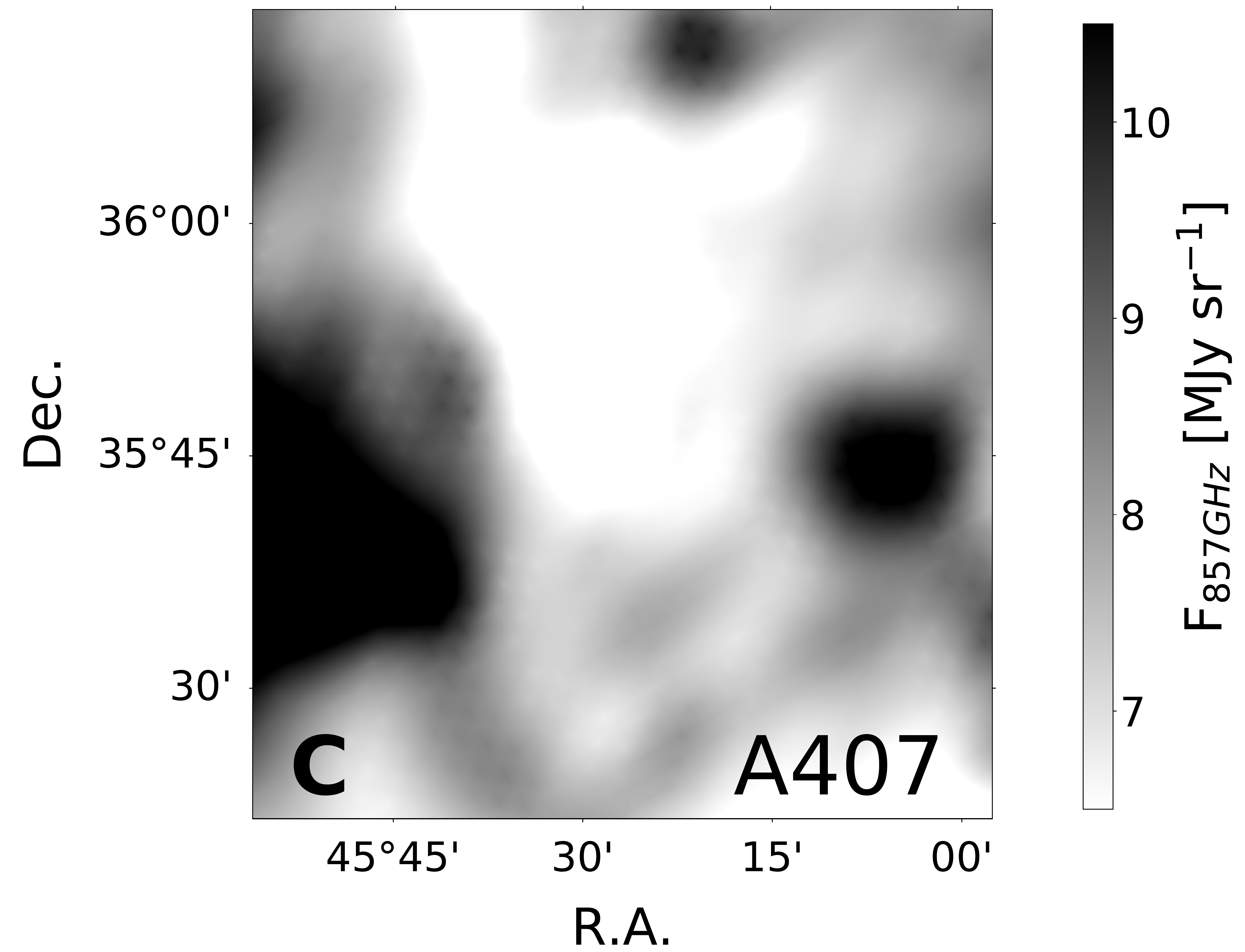}
	\caption{Left panels from top to bottom: three examples for contamination by Galactic cirrus across the fields of view of A2199, A582, and A407. All image cutouts are centered on the BCG. Contamination categories increase from A to C downward. Right panels: far-infrared 857 GHz maps of the same regions from \cite{Planck2014}. Note that the background constant and gradient in the optical images were modeled and subtracted by the night-sky flat procedure (see Section \ref{sec:bgsub}).\label{fig:cirrus}}
\end{figure*}

In Figure \ref{fig:mockbcg} we show that the light profile of the mock galaxy is well preserved down to ${\rm SB} = 31~g'$ mag arcsec$^{-2}$. The main source of error is the choice of the residual background constant. It is always negative because of the flux-scaling of the NSF to the incompletely masked individual exposures (see Section \ref{sec:bgsub}). We conservatively estimate it to be $\pm1$ count arcsec$^{-2}$, based on the outermost surface flux profile flatnesses of the worst 75$^{\rm th}$ percentile of all BCGs in our sample. There is a tendency to choose a too-high value because of the finite field of view. The uncertainty corresponds to a limiting surface brightness of ${\rm SB}_{\rm lim} = 30~g'$ mag arcsec$^{-2}$. That estimation is in agreement with the comparison of the SB profiles measured from WWFI data and larger field-of-view data from the 40cm Wendelstein Telescope and 70cm Jay Baum Rich Telescope (JBRT; Section \ref{sec:comparison}). The effect of choosing a too-high background constant is a drop in the outermost SB data points. That error only concerns surface brightnesses that are below our limiting magnitude ${\rm SB}_{\rm lim} = 30~g'$ mag arcsec$^{-2}$.

\subsection{Galactic Cirrus}\label{sec:galacticcirrus}

Foreground dust in the Galaxy fundamentally limits the depth of optical imaging data (e.g., \citealt{Miville2016} and references therein). It reflects the integrated stellar light of the Galaxy and becomes visible as filamentary structures that are easily misinterpreted as stellar streams. The dust emits at far-infrared and radio wavelengths and is thus easy to identify as not of extragalactic origin (\citealt{Duc2015,Besla2016}).

We estimate the cirrus flux in our observations by scaling the 857 GHz (350 $\mu$m) far-infrared emission maps published by \cite{Planck2014} so that the overall variations in dust flux match the ones of the galactic cirrus in our most strongly contaminated cluster A407 (see Figure \ref{fig:cirrus}):

\begin{align}
F_{\rm cirrus}^{\rm g'}~{\rm [counts]} \approx 0.5 F_{\rm 857 GHz}~{\rm [MJy~sr^{-1}]},
\end{align}

where the units on the left-hand side are calibrated to a photometric zero-point of $ZP=30~g'$ mag and a pixel scale of 0.2\arcsec/pixel. We match the variations in flux and not the absolute flux because the average background was already subtracted from the WWFI stacks during data reduction. The residual cirrus is visible down to a surface brightness of ${\rm SB}\sim 28~g'$ mag arcsec$^{-2}$, to which level we mask it by hand. Hidden cirrus below this SB level can evoke a systematic scatter in the outermost data points of the light profiles.

We define three categories of increasing cirrus contamination: A (invisible in the optical images), B (weak contamination), and C (strong contamination; see Figure \ref{fig:cirrus}). Not the total dust flux but its large-scale variations have the strongest influence on the light profiles. We estimate these variations as the standard deviation $\sigma$ of the dust surface flux in binned, $15\times15$ px sized thumbnails of the one-square-degree fields of view. The thresholds are expressed as surface brightness variations SB$_\sigma$ in units of $g'$ mag arcsec$^{-2}$:

\begin{align}
{\rm Category~A:} && 27.6 < & ~{\rm SB}_\sigma        \nonumber\\
{\rm Category~B:} && 26.9 < & ~{\rm SB}_\sigma < 27.6 \nonumber\\
{\rm Category~C:} &&        & ~{\rm SB}_\sigma < 26.9.
\end{align}

The cirrus-contamination category of each cluster is labeled on the image cutouts in Appendix \ref{sec:screenshots}. In our sample, 131 clusters (77\%) belong to category A, 28 clusters (16.5\%) to category B, and 11 clusters (6.5\%) to category C.

That strength of contamination is reduced (1) by manual masking, (2) by applying a robust estimator on the pixel histogram (see Section \ref{sec:fluxmeasurement}), (3) because the flux is averaged along the large isophotes in the low-SB galactic outskirts, and (4) because large-scale variations, that is, a gradient across the field of view, are included in the NSFs and subtracted.

An all-sky map of the scaled far-infrared map is shown in Figure \ref{fig:sample}.

\subsection{PSF Effects}\label{sec:psfeffects}

Seeing has a distorting influence on the light profiles: central galaxy light is redistributed toward larger radii. This effect manifests itself as (1) a flattening and circularization in the inner few arcseconds and (2) brighter SB in the range $1 \lesssim r \lesssim 4$ PSF FWHM. The effect is of the order of the core sizes of the local BCGs that we aim to resolve. As explained in Section \ref{sec:composite}, we therefore replace the central part of the light profiles by the ones that we either measured from archival (undeconvolved) \textit{Hubble Space Telescope} imaging data or deconvolved WWFI data. As shown in Figure \ref{fig:seeing}, the deconvolved profiles are accurate to almost Nyquist sampling quality, that is, $\gtrsim 0.4\arcsec$~resolution. The cores with sizes of order 1\arcsec~are therefore real and not resolution artifacts.

\begin{figure}
	\includegraphics[width=\linewidth]{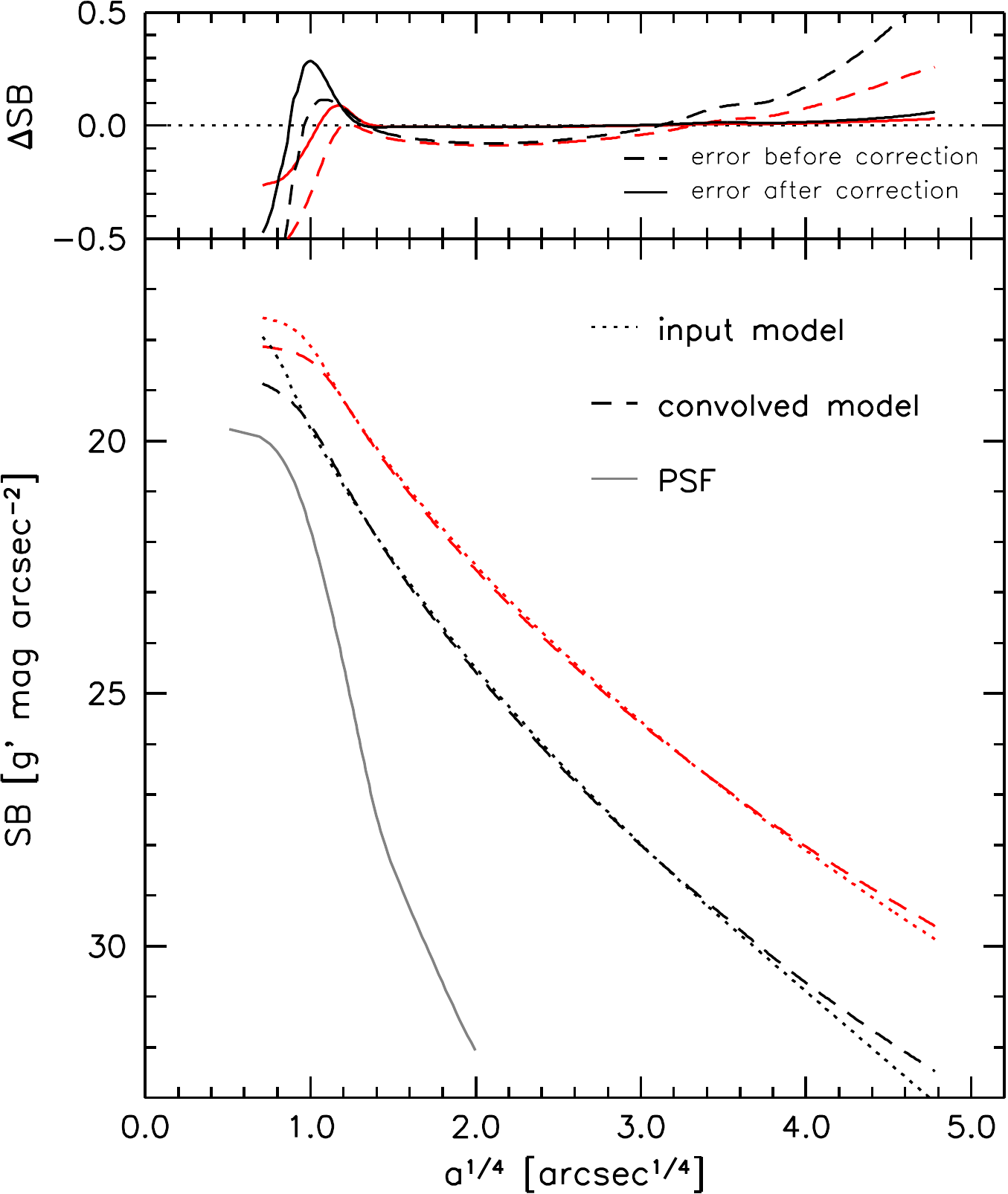}
	\caption{Effect of PSF broadening on SB profiles of two different mock BCGs (red and black). The dotted lines show the original SB profiles, and the dashed lines show the profiles after convolution with the PSF. The dashed and continuous lines in the top panel show the residuals before and after correction, respectively.\label{fig:deco}}
\end{figure}

Not only the PSF's center but also its outer wings distort the galaxy light profiles (see \citealt{Duc2015} and references therein or \citealt{Trujillo2016}). The PSF's wings and reflections from the BCG's center overlap with the ICL. In other words, light is redistributed from the center to the outskirts. We refer to this effect as PSF broadening. It becomes a problem when the galaxy's center is bright compared to its outer halo. Figure \ref{fig:deco} shows the severity of this effect. SB profiles of two (red and black) representative model BCGs are plotted as dotted lines. The SB profiles after 2D convolution with the PSF are overplotted as dashed lines. The systematic error due to PSF broadening is $0.1 > \Delta {\rm SB} > 0.5~g'$ mag arcsec$^{-2}$ and increasing with galaxy size.

We now describe our correction method for the broadening effect. The accurate approach would be to deconvolve the imaging data prior to the SB profile measurement. However, this is computationally challenging considering the large kernel size of $\sim2000\times2000$ pixels and Richardson--Lucy deconvolution \citep{Lucy1974} being an iterative procedure. We use a computationally faster method that is based on the approximation that the amount of scattered light is small (1.78\%, see Section \ref{sec:brightstarremoval}) compared to the total light. Under these circumstances, a secondary convolution $i*t$ by image processing results in similar light scattering even after the primary convolution $i=r*t$ by the telescope optics is already inherent to the images. That is quantified by

\begin{align}
i &= r*t,\\
r  &\approx i - (i*t - i), \label{eq:deco}
\end{align}

where $r$ is the unknown intrinsic light distribution, $t$ is the kernel, i.e. the PSF, and $i$ is the image data after primary convolution. We apply a 2D convolution\footnote{We also experimented with a 1D convolution using the {\tt python} package {\tt scipy.signal.convolve}. A simple test applied to a slice along the major axis of a BCG and PSF light profiles results in a stronger broadening effect than for the 2D convolution and, hence, erroneous results.} to images that were regenerated from the isophotal shape parameters. The scattered light is reconstructed by subtracting the twice-convolved image $i*t$ from the primary convolved image $i$. Then, by subtracting this scattered light from the primary convolved image, the intrinsic light distribution is recovered (see Equation (\ref{eq:deco})).

The deviation of the corrected from the original (intrinsic) light profiles of the mock BCGs is shown in Figure \ref{fig:deco} (top panel) as continuous lines. The SB data points at radii $a>4\arcsec$ agree well with the input model. The inner regions are badly recovered because the small-influence approximation fails there. However, these regions of the profiles are replaced by those derived from \textit{HST} or deconvolved WWFI data (see Section \ref{sec:merging}).

Each SB profile of the real BCGs is corrected individually. The median correction for the structural parameters that are determined by direct integration of the light profiles (i.e., independent of the S\'ersic fits) is

\begin{align}
r_{\rm e,30}^{\rm corrected}/r_{\rm e,30}^{\rm uncorrected} &= 0.94 \pm 0.03,\\
SB_{\rm e,30}^{\rm corrected} - SB_{\rm e,30}^{\rm uncorrected} &= -0.11 \pm 0.05,\\
M_{\rm tot,30}^{\rm corrected} - M_{\rm tot,30}^{\rm uncorrected} &= 0.03 \pm 0.02,
\end{align}

where the index "30" indicates that the parameters were determined by integrating the light profiles out to the isophote with ${\rm SB} = 30~g'$ mag arcsec$^{-2}$ (for details, see Section \ref{sec:integrationeps}).
As expected, only a small influence on the integrated brightness is found. The integration aperture is sufficiently large so that the redistribution of the light is close to negligible. The effective radii are increased and the effective surface brightnesses are dimmed by the broadening effect.

\begin{figure*}
	\centering
	\includegraphics[width=0.46\linewidth]{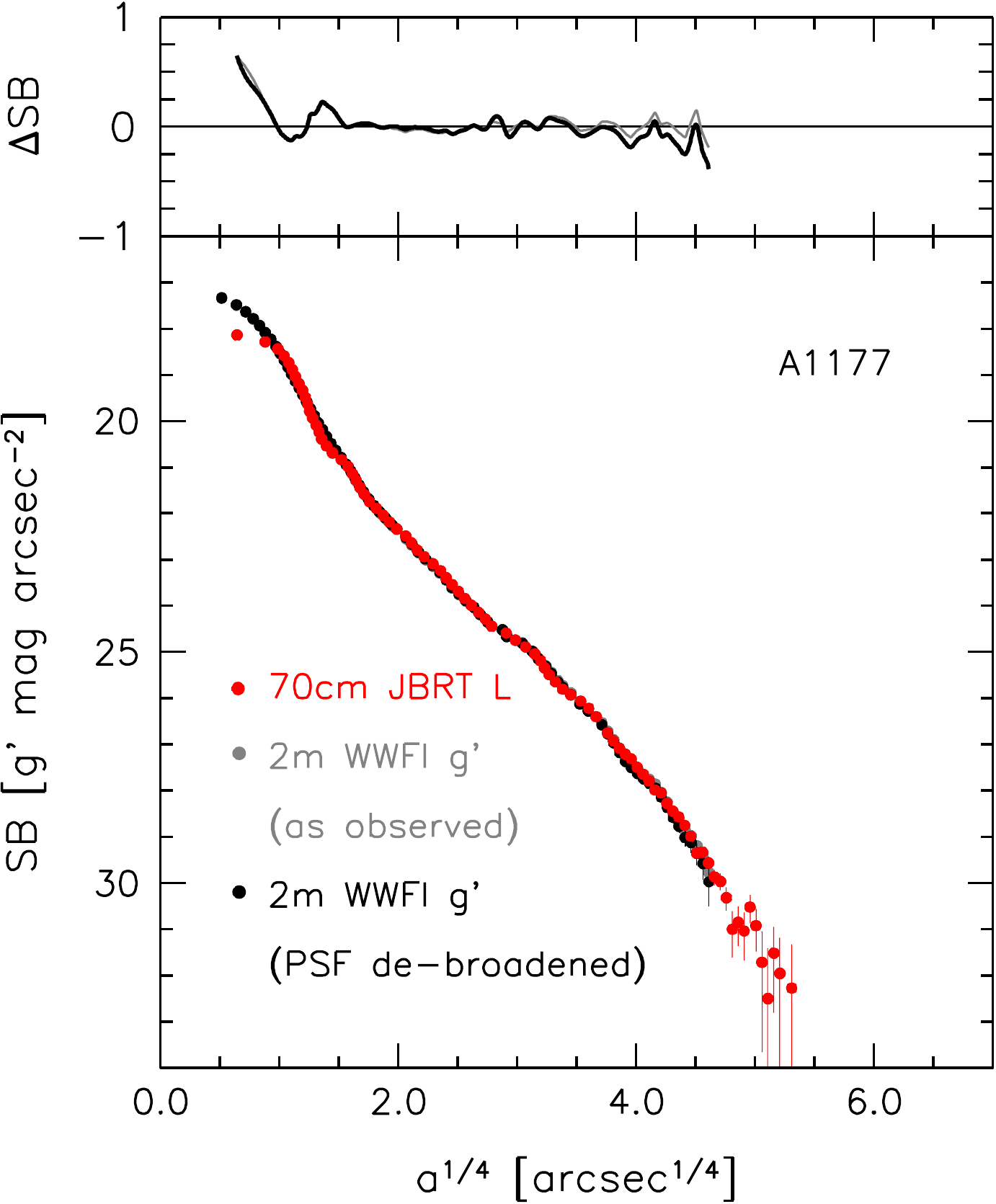}~~~~~~~~~~~~
	\includegraphics[width=0.46\linewidth]{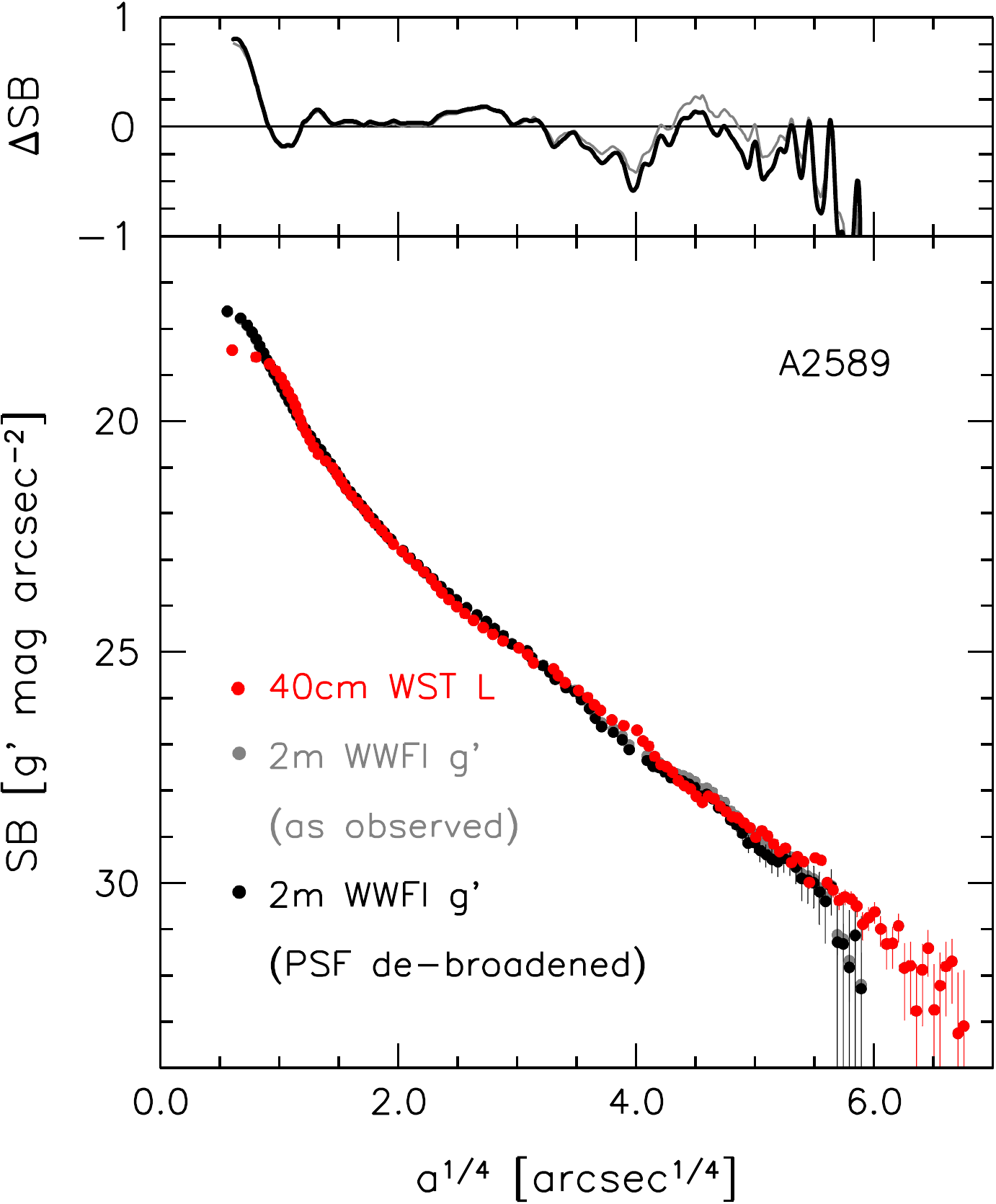}
	\caption{Control sample of two BCGs that were observed independently with different telescopes (red data points). Left panel: 70 cm Jay Baum Rich Telescope (JBRT) at the \textit{WISE} observatory, owned and operated by Tel Aviv University. Right panel: 40 cm telescope at the Wendelstein observatory. The photometric zero-points of the $L$-band profiles are adjusted so that the $L$-band profiles match the WWFI $g'$-band profiles for comparison. Color gradients in the two filters are assumed to be negligible. The WWFI $g'$-band profiles are shown with (black) and without (gray) PSF debroadening correction. The deviations of the spline-interpolated comparison profiles from the WWFI profiles are shown in the top panels.\label{fig:comparison}}
\end{figure*}

After fitting S\'ersic functions to the SB profiles before and after PSF broadening correction, we calculate the median corrections for the S\'ersic parameters of the SS BCGs:

\begin{align}
r_{\rm e,SS}^{\rm corrected}/r_{\rm e,SS}^{\rm uncorrected} &= 0.88 \pm 0.06,\\
SB_{\rm e,SS}^{\rm corrected} - SB_{\rm e,SS}^{\rm uncorrected} &= -0.25 \pm 0.06,\\
n_{\rm SS}^{\rm corrected}/n_{\rm SS}^{\rm uncorrected} &= 0.94 \pm 0.03,
\end{align}

and for the S\'ersic parameters of the DS BCGs:

\begin{align}
r_{\rm e,DS1}^{\rm corrected}/r_{\rm e,DS1}^{\rm uncorrected} &= 0.99 \pm 0.04,\\
SB_{\rm e,DS1}^{\rm corrected} - SB_{\rm e,DS1}^{\rm uncorrected} &= -0.03 \pm 0.08,\\
n_{\rm DS1}^{\rm corrected}/n_{\rm DS1}^{\rm uncorrected} &= 0.99 \pm 0.04,\\
r_{\rm e,DS2}^{\rm corrected}/r_{\rm e,DS2}^{\rm uncorrected} &= 0.95 \pm 0.03,\\
SB_{\rm e,DS2}^{\rm corrected} - SB_{\rm e,DS2}^{\rm uncorrected} &= -0.04 \pm 0.08,\\
n_{\rm DS2}^{\rm corrected}/n_{\rm DS2}^{\rm uncorrected} &= 0.96 \pm 0.04.
\end{align}

\subsection{Undetected ICL below the Limiting Magnitude} \label{sec:integrationcrop}

The SB limit of our survey is SB$_{\rm lim} = 30~g'$ mag arcsec$^{-2}$. Below that limit, we have no reliable information on how the SB profiles continue. An educated guess is the extrapolation of the fitted S\'ersic profiles because there is no indication for a truncation just above this limit (see Section \ref{sec:avgprofiles}). The following median corrections have to be applied when the lower SB boundaries are increased from 30 to infinity $g'$ mag arcsec$^{-2}$:

\begin{align}
r_{\rm e,\infty}/r_{\rm e,30} &= 1.20 \pm 0.15,\\
SB_{\rm e,\infty} - SB_{\rm e,30} &= 0.31 \pm 0.22,\\
M_{\rm tot,\infty} - M_{\rm tot,30} &= -0.09 \pm 0.06.
\end{align}

The indices "30" and "$\infty$" indicate the SB of the outermost considered isophote. \textit{The averages of both values are listed in Section} \ref{sec:strucparams}\textit{, and the uncertainties derived from both integration limits are taken as the error.} We stress again that all median correction factors in Sections \ref{sec:psfeffects} and \ref{sec:integrationcrop} are only given for illustrative purposes. Each SB profile was corrected individually.

\subsection{Comparison to Data Obtained with Other Telescopes}\label{sec:comparison}

The key obstacle for deep imaging is the task of background subtraction. In addition to the mock-BCG test described in Section \ref{sec:mockbcg}, we perform another test to make sure that the ICL is not oversubtracted in the WWFI data. For a control sample, we have obtained independent imaging data for A1177 with the 70cm JBRT at the \textit{WISE} observatory (2 hr target integration time) and for A2589 with the 40cm telescope at the Wendelstein observatory (12 hr integration time). Both imagers span an even wider field of view than the WWFI and are made of one single CCD chip. That makes them less susceptible to systematic errors during background subtraction because the BCG+ICLs cover a smaller fraction of the field of view and less masking is required.

The control sample data were observed, dithered, and reduced in a similar way to the WWFI data. The only difference is the method of background subtraction. The background in the 70cm JBRT data was modeled by fourth-order 2D polynomials in each exposure. Nonphotometric observing conditions degraded the stability of the background pattern so that the NSF method failed. However, the polynomial approach works sufficiently well because of the large field of view.

The background in the 40cm WST data was modeled by scaling and averaging the two bracketing source-masked exposures that were taken before and after each exposure. No offset sky exposures were taken. The sky background is modeled from the science exposures themselves. The large dither pattern ensures that empty sky regions in the bracketing exposures always fall around the BCG's position so that the sky is modeled accurately across the whole field of view. Regions that happen to be masked in both bracketing exposures are replaced by fourth-order 2D polynomials that were fitted to the whole average sky images.

Figure \ref{fig:comparison} shows a comparison of the SB profiles measured in both datasets. The WWFI-obtained profiles are plotted before (gray) and after (black) PSF broadening correction (see Section \ref{sec:psfeffects}). The two comparison profiles are not PSF broadening corrected. The overall agreement especially in the outermost data points provides further confidence in the accuracy of the background subtraction method. Moreover, it confirms that $<$1m class telescopes can be sufficient to perform deep imaging projects by reaching the 30 $g'$ mag arcsec$^{-2}$ SB limit.

\section{Results}\label{sec:results}

\subsection{Accretion and Merging Signatures} \label{sec:accretion}

The advent of low-surface-brightness photometry has unveiled a myriad of fine structure in the outskirts of galaxies. These relics of violent accretion have been predicted by numerical simulations as the most direct evidence for hierarchical clustering (\citealt{Bullock2005,Johnston2008,Rudick2009,Cooper2010,Puchwein2010,Cooper2013,Rodriguez2016,Harris2017,Mancillas2019}). They have been discovered around local late-type galaxies (e.g., \citealt{Martinez2010,Chonis2011,Foster2014,Ibata2014,Amorisco2015,Merritt2016}) and local early-type galaxies (\citealt{Tal2009,Duc2015,Bilek2016,Crnojevic2016,Duc2017}), as well as galaxy groups \citep{DaRocha2005,DaRocha2008,Watkins2014,Watkins2015,Okamoto2015,Spavone2018} and galaxy clusters (\citealt{Feldmeier2002,Arnaboldi2012,Iodice2016,Iodice2017,Iodice2018,Mihos2017}). A review on the topic can be found in \cite{Carlin2016}. For a comparison between literature data on the frequency of disturbed morphologies, see \cite{Atkinson2013}.

We visually inspect the clusters for accretion and merging signatures and classify them into four categories: (1) two BCGs, (2) shells, (3) tidal streams, and (4) multiple nuclei. There was no a priori knowledge of the galaxies' S\'ersic type during the identification procedure. To maximize our likelihood of finding structures on various surface brightnesses, we have visually scanned linearly and logarithmically scaled, minimum filtered \citep{Bilek2016}, isophote-model and parametric-model subtracted images. The 2D models for the latter two methods are created from the isophotal shape profiles. For the parametric models, we replaced the SB values in the data tables by their corresponding SS or DS fit values.

\begin{figure}
	\centering
	\includegraphics[width=\linewidth]{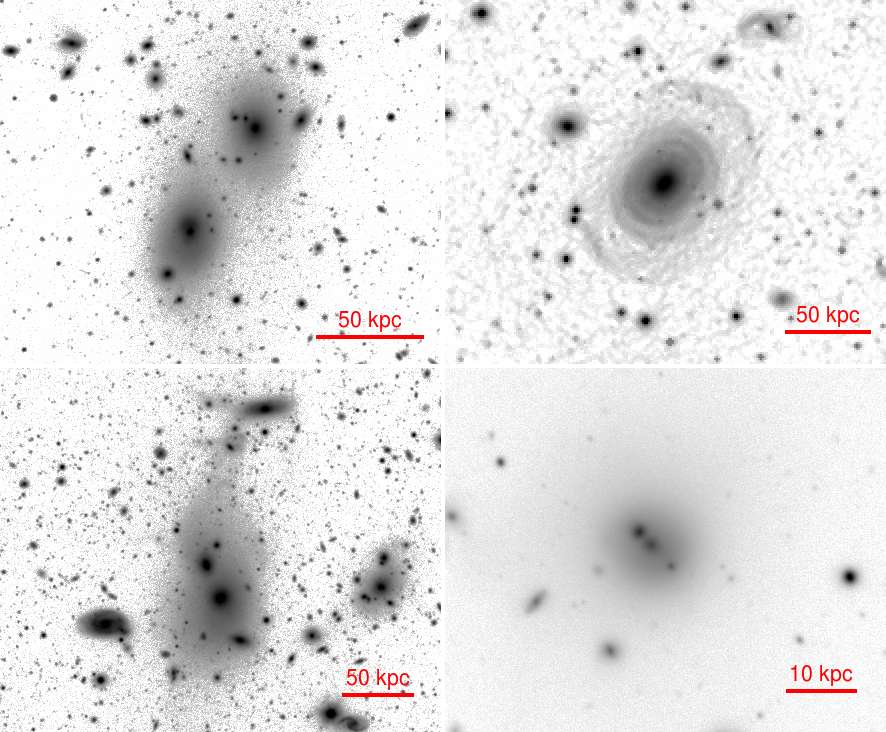}\\
	\includegraphics[width=\linewidth]{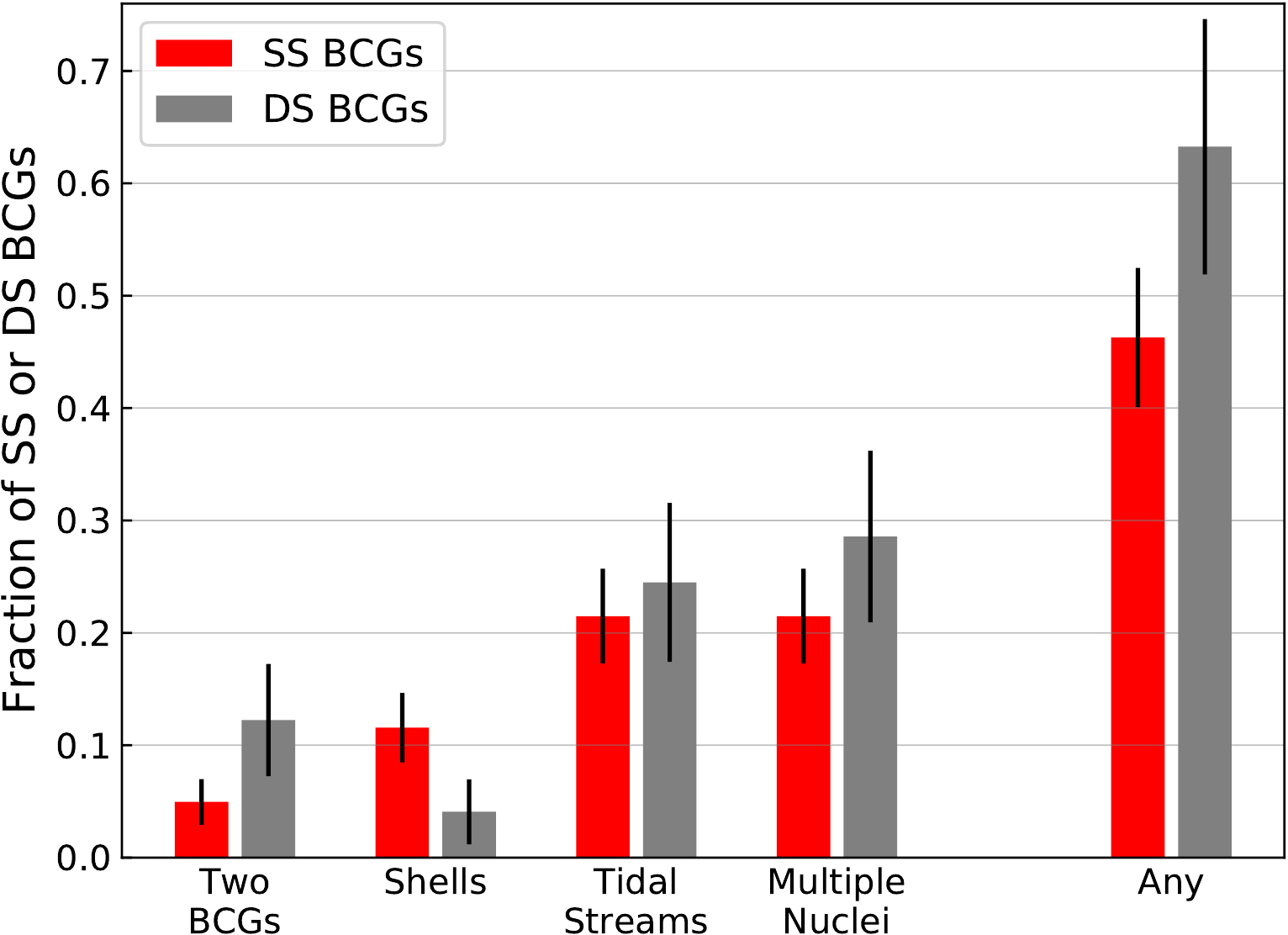}
	\caption{Accretion and merging signatures. Top panel: representative examples for each category. Top left: two BCGs in A1190. Top right: shells in A2197. Bottom left: tidal streams in A1257. Bottom right: multiple nuclei in A1185. Bottom panel: relative abundance of each of the four features and of any of the four features.\label{fig:histo_features}}
\end{figure}

One prototypical example for each category is shown in Figure \ref{fig:histo_features}, top panel. An explanation of the characteristics of each category follows.

The "two BCGs" category is not a direct sign of interactions between galaxies, but a likely indicator for merging clusters, as it is the case for the Coma cluster. Since the appearance of at least two similar-sized BCGs is a hint of a nonrelaxed state of the cluster, we include this category in our analysis. We find that 7.0\% of the clusters have two BCGs (4.1\% for SS BCG clusters and 12.2\% for DS BCG clusters).

Shells are accumulations of stars that align in circle segments around the BCG center (e.g., \citealt{Malin1980,Malin1983,Hernquist1988,Hernquist1989} and many more). These segments can be more or less concentric, depending on the type of shell system. They form when a satellite galaxy falls onto the BCG on a nearly radial trajectory with pericentric distance $<15$ kpc \citep{Karademir2018} and is disrupted (\citealt{Bilek2016,Pop2018}). The shells mark the turnaround lines in the orbits of the progenitor's stars. Shells have been reproduced in simulations with mass ratios of the merging galaxies ranging from 1/100 \citep{Quinn1984,Karademir2018} to $>1/3$ (\citealt{Karademir2018,Pop2018}). See also \cite{Bilek2015} for a review. Shells are found in between $\sim10\%$ \citep{Schweizer1988} and $\sim22\%$ of elliptical galaxies \citep{Tal2009}. The frequency in the Illustris simulation is $18\pm3\%$, which increases with increasing mass cut \citep{Pop2018}. We find that 9.4\% of our analyzed BCGs show shells (11.6\% of SS BCGs and 4.1\% of DS BCGs). A lower frequency could be explained with the degrading angular resolution because the BCGs in our sample are a factor of $\sim$10 more distant than the local ellipticals in the \cite{Tal2009} sample, which decreases the detectability of existing shells. Our result should therefore be understood as a lower boundary for the abundance of shells in BCGs.

Tidal streams are made of stars that were liberated from a satellite galaxy by a collision \citep{Moore1996} or due to the mean tidal field of the cluster \citep{Merritt1984}. These unbound stars then virialize in the cluster and add up to the ICL budget. For instance, unprecedentedly deep photometric surveys of the Virgo \citep{Mihos2017} and Fornax clusters \citep{Iodice2016,Iodice2017,Iodice2018} have unveiled multitudes of tidal streams. Other examples have been discovered in the Coma and Centaurus clusters (\citealt{Gregg1998,Trentham1998,Calcaneo2000}). We do not make a strict differentiation between tidal tails and tidal streams as proposed by \cite{Duc2015} because we lack color information. Finally, we find that 22\% of our observed BCGs show some form of stream-like features (21\% of SS BCGs and 24\% of DS BCGs). The features are usually dynamically hotter than the ones reported for field galaxies (e.g., \citealt{Martinez2010}) and thus dissolve quicker. The observed abundance therefore implies ongoing ICL accretion.

Multiple nuclei are in $\sim47\%$ of the cases simply chance superpositions as concluded from their undisturbed morphology \citep{Lauer1988}. The remaining half split into high-velocity unbound interactions (24\%) with radial velocity differences $\Delta V \gtrsim 400$ km s$^{-1}$ that lead to tidal stripping of the secondaries' envelopes and possible low-velocity, strong merger interactions (29\%) that lead to cannibalism of the secondary nucleus (\citealt{Lauer1988}; see also \citealt{Tonry1985a,Tonry1985b,Beers1986}). Without differentiating between the cases of real interactions and pure projections, we identify at least one secondary nucleus in 24\% of all BCGs (21\% of SS BCGs and 24\% of DS BCGs). That is a slightly lower fraction than values reported in the literature (28\%, \citealt{Hoessel1980}; 45\%, \citealt{Schneider1983}).

The relative abundances of the four discussed types of accretion signatures are also shown in Figure \ref{fig:histo_features}. The error bars are determined using Poisson statistics. We cannot tell whether SS or DS BCGs have higher abundances of specific accretion signature types, due to small-number statistics. However, the total frequency of accretion signatures is 46\% for SS BCGs and 63\% for DS BCGs. The latter show more indications of ongoing merging processes with a 1$\sigma$ certainty. The frequency for all BCGs is 51\%.

\begin{figure*}
	\centering
	\includegraphics[width=\linewidth]{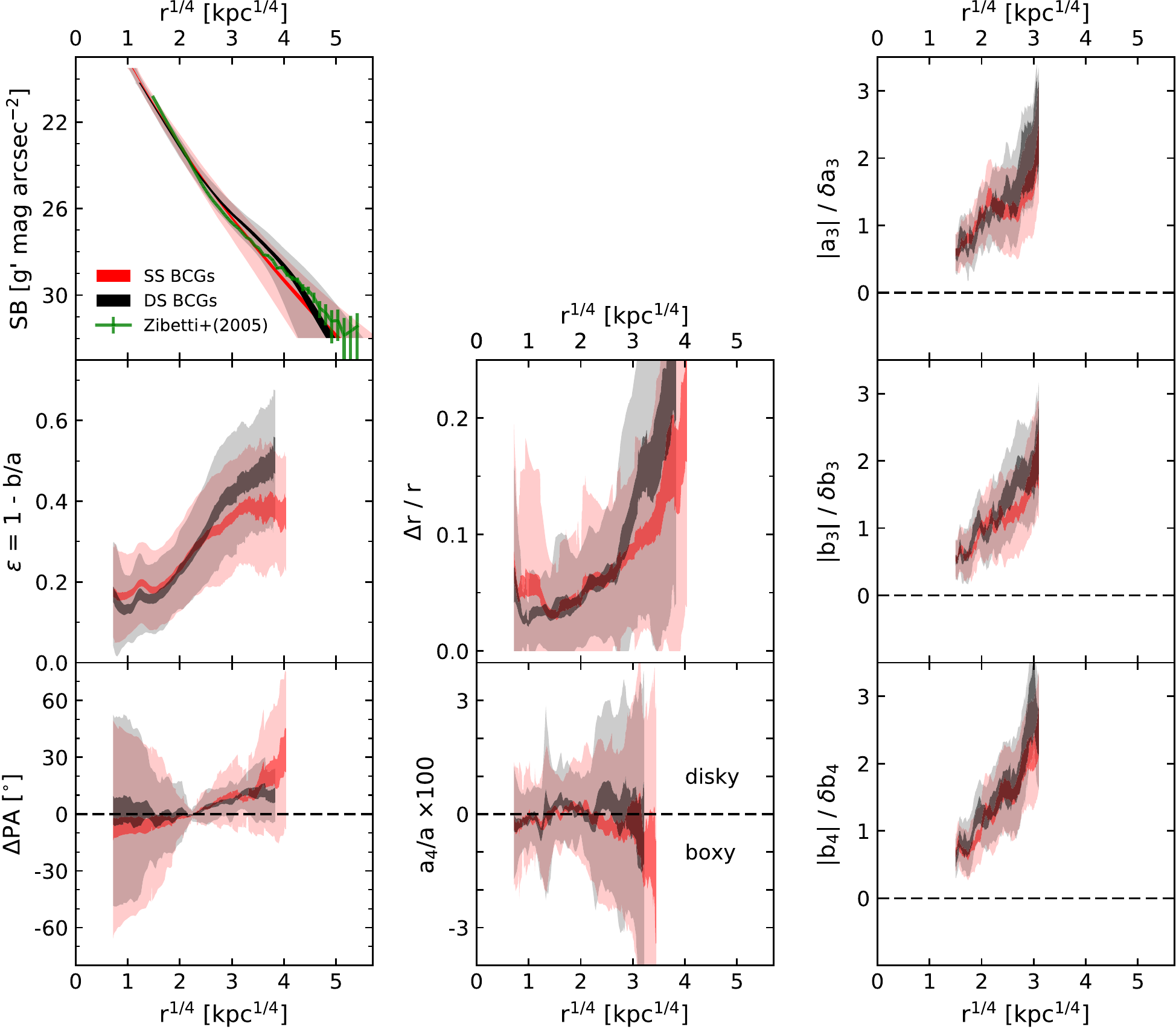}
	\caption{Average profiles of SS (DS) BCGs in red (black). The radius on the x-axis is given along the effective axis $r = \sqrt{ab}$. The more transparent shades correspond to the intrinsic 1$\sigma$ scatter, and the more opaque shades correspond to the standard error of the mean. The average SB profiles are created by averaging the measured SB data points of all BCG+ICLs inside of $a<16$ kpc and the extrapolated S\'ersic fits outside of that that semimajor-axis radius. Isophote twist $\Delta$PA profiles are normalized to the median in the range $16 < r[{\rm kpc}] < 40$. Profiles with a negative gradient are flipped. Coefficients of the Fourier expansion of the deviations from elliptical isophotes (see Equation (\ref{eq:fourier})) are plotted in the bottom middle and right panels. The disky-/boxiness indicator $a_4$ is expressed as a percentage of the semimajor-axis radius $a$. The more transparent error shades describe the 1$\sigma$ intrinsic scatter. The other three Fourier coefficients $a_3$, $b_3$ and $b_4$ are expressed differently because they quantify asymmetric distortions that are randomly oriented and therefore average to zero for large samples. Their absolute value is divided by the measurement uncertainty. This is a measure of the significance that the corresponding deviations are detected. The error shades correspond to the 1$\sigma$ measurement uncertainty. The comparison between SS and DS BCGs is only fair because the measurement scatter is almost identical .\label{fig:avgprofile}}
\end{figure*}

\subsection{Average Profiles}\label{sec:avgprofiles}

The average light profiles and isophotal shape parameter profiles are presented in Figure \ref{fig:avgprofile}. All profiles are averaged in radial intervals besides the SB profiles, which were calculated by averaging in SB intervals. The different method is necessary because the limited depth would otherwise result in an artificial upward trend in the SB profiles below ${\rm SB}\gtrsim28~g'$ mag arcsec$^{-2}$. Before averaging, outliers in every radius (or SB) interval are rejected that deviate more than 6.5 standard deviations from the mean. If data points for a minimum of 14 BCGs remain, then the average is plotted. The 1$\sigma$ standard deviations of the intrinsic scatter are shown as shaded regions for SB, $\epsilon$, $\Delta$ PA, $\Delta r/r$, and $a_4$. The shaded regions for $a_3$, $b_3$, and $b_4$ correspond to the measurement uncertainties.

The SB profiles are a composite of data and fits. Inner regions below $r<16$ kpc are taken directly from the measured light profiles, and the outer regions are replaced by the S\'ersic fits. The radius is given along the effective axis $r = \sqrt{ab}$. This is equivalent to measuring the profiles in circular apertures. It allows direct comparison to the SB profile that was measured by \cite{Zibetti2005} by combining 683 galaxy cluster images from the SDSS-DR1. We apply a K-correction of $g'{\rm [rest~frame]} = {\rm [observed~frame]} - 0.71$ mag to the $r$-band data from Zibetti et al. (\citealt{Chilingarian2010,Chilingarian2012}), a color correction of $g = r + 1.2$, derived from their multiband SB profiles and corrected for cosmic dimming. The average profile from Zibetti et al. is inconsistent with our average profiles within the standard error of the mean, that is, the thickness of the red and black lines.  We unsuccessfully tried to match our average SB profiles with the profile from Zibetti et al. by applying various total brightness cuts on our sample: after discarding all BCGs fainter than $M_{\rm tot} > -23~g'$ mag, the average SBs around ${\rm SB} \sim 30~g'$ mag arcsec$^{-2}$ match well, but the slope below $r<40$ kpc is too shallow. Instead, if we discard all BCGs brighter than $M_{\rm tot} < -23~g'$ mag then the slopes match well at these radii but the profiles are too faint, especially in the ICL outskirts. We conclude from this analysis that the deviations cannot be attributed to sample selection alone. A possible explanation is the different age of the galaxies. The mean redshift $\bar{z}_{\rm Z}=0.25$ of Zibetti et al.'s sample is higher than ours ($\bar{z}_{\rm K}=0.06$). That equals 2.16 Gyr in time evolution, after which the BCG's SB profiles might have evolved to become smoother.

The overall shape and scatter of SS and DS BCGs are fairly similar. The largest difference occurs around $r=240$ kpc, where DS BCGs are on average 0.65$\pm$0.18 $g'$ mag arcsec$^{-2}$ brighter than SS BCGs. The difference decreases again toward larger radii and becomes zero at $r=470$ kpc.

We now move on to discuss the isophotal shape parameters. As explained in Section \ref{sec:composite}, these parameters are kept fixed beyond the last plausible radius. Hence, the average profiles for these parameters do not extend as far out as the averaged surface brightnesses.

The ellipticities $\epsilon=1-b/a$ rise with radius. The slope is slightly shallower for SS BCGs.

The position angles PA are counted counterclockwise starting from the north--south axis. It is unambiguously defined in the range $0\degr \leq {\rm PA} < 180\degr$. It flips $\pm180\degr$ when it is crossing the north--south axis. These jumps are eliminated by the following procedure: if the difference between two subsequent PA data points is greater than ${\rm PA}_{i} - {\rm PA}_{i+1} > 90\degr$, an offset of 180\degr~is subtracted from all data points at greater radii. The opposite is done when ${\rm PA}_{i} - {\rm PA}_{i+1} < -90\degr$. All PA profiles are normalized to the median in the range $16 < r[{\rm kpc}] < 40$. Since PAs are randomly oriented, their profiles average to constant zero for a large sample. To avoid this, we flip PA profiles with negative gradients. The gradients are determined between the median of the range $16 \leq r[{\rm kpc}] < 40$ and the median of the range $r \geq 40$ kpc. We find average isophote twists of order $\Delta {\rm PA} / \Delta r \sim 10\degr/100$ kpc for both SS and DS BCGs. The scatter beyond $r\gtrsim 50$ kpc is about twice as high for DS BCGs than for SS BCGs. The higher scatter below $r\lesssim 20$ kpc can be explained by the lower ellipticities of DS BCGs at these radii.

The ICL offset $\Delta r(r)$ is the distance between the center of the BCG and the center of the isophotal ellipse with radius $r$ along the effective axis. The average and relative ICL offsets increase with radius. At $r=150$ kpc, they reach 10\% (i.e., 15 kpc) for SS BCGs and 20\% (i.e., 30 kpc) for DS BCGs. The spatial direction of these offsets will be discussed in a forthcoming paper.

Isophotal distortions from perfect ellipses are expanded in a Fourier series (see Section \ref{sec:ellfitn}). The most informative coefficient $a_4$ is expressed as a percentage of semimajor-axis radius $a$ (see Figure \ref{fig:avgprofile}, bottom middle panel). It quantifies the diskyness ($a_4>0$) or boxiness ($a_4<0$) of the isophotes (e.g., \citealt{Bender1987}). 

We find that the inner isophotes in 10 kpc $\lesssim r\lesssim25$ kpc are on average slightly disky. SS BCGs become boxy in the outskirts beyond $r\gtrsim40$ kpc, whereas DS BCGs are slightly disky at that radius.

We show the first three coefficients of the Fourier expansion $a_3$, $b_3$ and $b_4$ in Figure \ref{fig:avgprofile}. The two parameters $a_3$ and $b_3$ quantify the triangularity of the isophotal shapes. The last parameter $b_4$ quantifies distortions similar to the disky/boxy parameter $a_4$, but includes a $\sim 45\degr$ rotation because it is the amplitude of the sine component from the Fourier expansion (see Equation (\ref{eq:fourier})). The values of all of these parameters average to zero for a large sample because asymmetric distortions are randomly oriented. In order to gain knowledge from them, we have to look at their moduli. What makes the analysis difficult is that the measurement errors are of the order of the values themselves. A better diagnostic is the significance whether $a_3$, $b_3$ and $b_4$ type deviations are detected at all. We therefore express these parameters in the form of their modulus, divided by the measurement uncertainty. The error shades in the bottom three panels on the right are the average measurement uncertainties and, like for the other parameters, the intrinsic scatter. A comparison between SS and DS profiles is only fair because the average uncertainties are very similar, as it is expected for a large sample size of similar galaxies. We find that SS BCGs are characterized by lower values for $a_3$, $b_3$ and $b_4$ than DS BCGs. In other words, SS BCGs have less pronounced asymmetric isophotal distortions, indicating a more relaxed state of SS BCGs.

\subsection{Structural Parameters} \label{sec:strucparams}

\begin{figure*}
	\centering
	\includegraphics[width=0.75\linewidth]{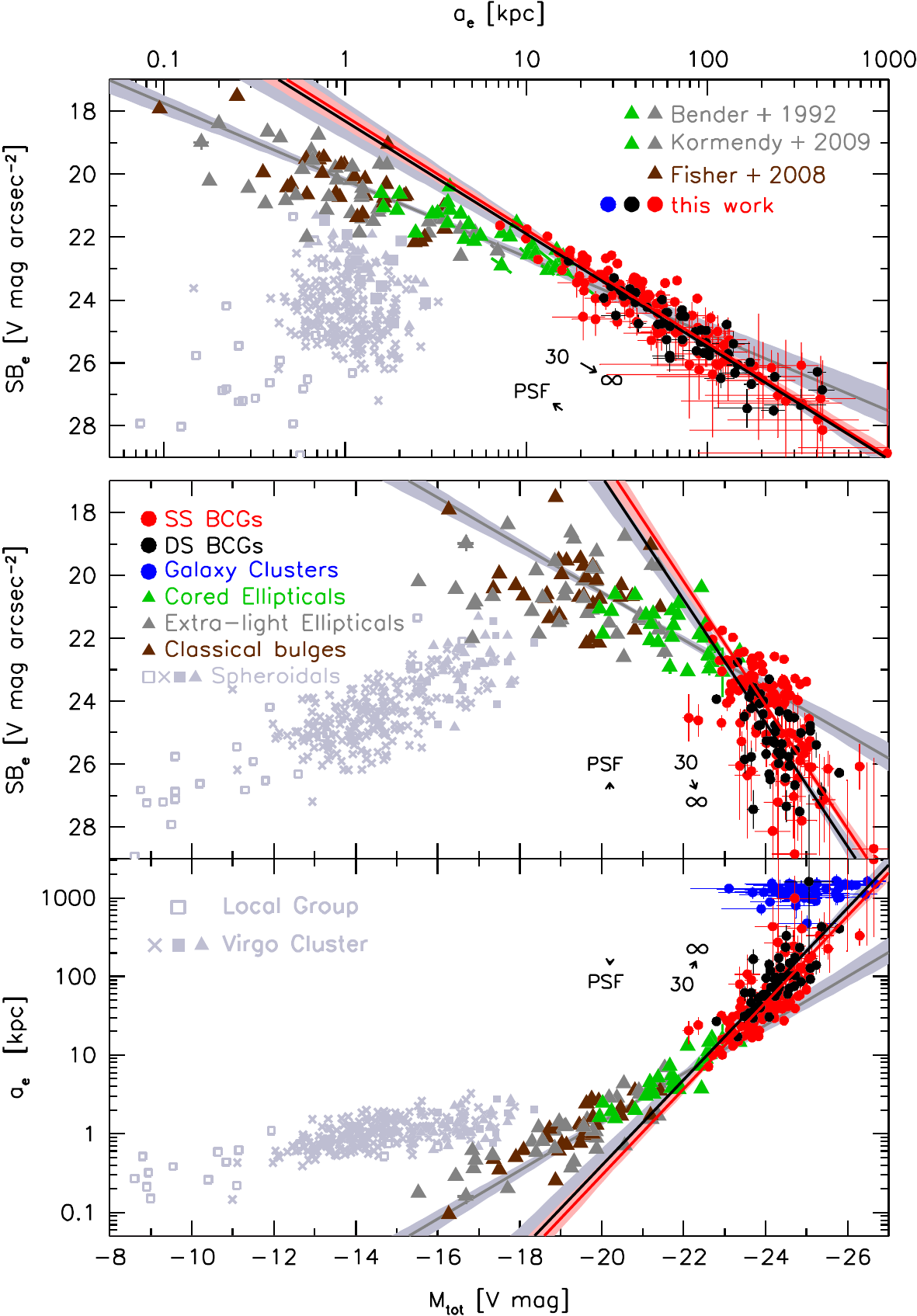}
	\caption{Comparison between integrated absolute brightnesses $M_{\rm tot}$, effective radii $a_{\rm e}$ along the major axis, and effective surface brightness SB$_{\rm e}$ of BCGs, regular ellipticals, and galaxy clusters. The basis for this plot is Figure 37 in \cite{Kormendy2009} with updates in Figure 2 in \cite{Kormendy2012} and Figure 14 in \cite{Bender2015}. The galaxy data points from the literature and from this work are calculated from 2D profile integration and evaluated along the major axes. For the clusters, the (circular) gravitational radius $r_{\rm g}$ is used, which will be described in a forthcoming paper. The arrow \textit{$30\rightarrow\infty$} shows the median shift of the BCG parameters when using no upper integration limit compared to SB$_{\rm lim} = 30~g'$ mag arcsec$^{-2}$. The arrow \mbox{$\rightarrow$ PSF} shows the median shift of the BCG parameters due to the PSF broadening correction. The $g'$-band magnitudes were converted to $V$-band magnitudes via $V = g - 0.45$ mag \citep{Jester2005} for $g-r=0.78$ \citep{Tojeiro2013}.\label{fig:profileparams}}
\end{figure*}

In this section, we examine first how BCGs populate the parameter space ($r_{\rm e}$, SB$_{\rm e}$, $M_{\rm tot}$) and secondly how the S\'ersic indices $n$ are distributed. We therefore overplot the structural parameters listed in columns (9), (10), and (11) in Appendix \ref{sec:apclusterparams} on Figure 2 from \cite{Kormendy2012}. The result is shown in Figure \ref{fig:profileparams}. The literature parameters are determined by integrating the extrapolated SB profiles down to ${\rm SB}\sim$ 29.7 $V$ mag arcsec$^{-2}$ for core ellipticals and down to an arbitrarily faint SB for coreless ellipticals and dwarf spheroidal galaxies \citep{Kormendy2009}. Hence, they include most of the stellar halos. Since no attempt at a galaxy--halo decomposition was done for the literature parameters, we avoid doing so for the BCGs, too; that is, we use the structural parameters determined for the whole light distribution including ICL for a consistent comparison.

The average corrections for PSF broadening and the average systematic error due to the finite depth of our survey are indicated by the arrows. These average corrections and also the individual errors are neglected for the fitting of the parameter correlations. Otherwise, a significant number of data points had almost zero weight, due to the high inhomogeneity of the errors. We find that BCGs extend the population of regular ellipticals in parameter space to larger integrated brightnesses, dimmer effective SBs, and larger effective radii, but their parameter correlations have different slopes (see Table \ref{tab:parrelations}). In the next paragraphs, we compare our derived parameter correlations to those derived by \cite{Donzelli2011} and \cite{Bernardi2007} from shallower datasets and offer an explanation for the discrepancies. Later, we argue that the broken slopes appear because the growth of BCGs, compared to regular ellipticals, is more dominated by accretion of stellar material in their outskirts.

The \cite{Kormendy1977} relation $SB_{\rm e} = \alpha\log(r_{\rm e}) + \beta$ that we fit to our data has a slope of $\alpha=3.61\pm0.13$ for the SS BCGs and $\alpha=3.58\pm0.26$ for the DS BCGs (see Table \ref{tab:parrelations}). Both results are in tension with the Kormendy relations found by \cite{Donzelli2011}. Using shallower imaging data, they measured a slope of $\alpha=3.29\pm0.06$ for the SS BCGs and $\alpha=2.79\pm0.08$ for the DS BCGs. By also taking the offsets $\beta$ after color corrections into consideration, we notice that the data points from Donzelli et al. are systematically shifted off our measured Kormendy relation toward smaller effective radii. A plausible explanation for this offset is the underestimation of the ICL amount in Donzelli et al.'s data because some upward curvature of the light profiles remained undetected. Their limiting SB converted to the $g'$-band is SB$_{\rm lim}^{\rm Donzelli+2011} = 25.7~g'$ mag arcsec$^{-2}$. In a forthcoming paper, we show that about half of the transitions between the two S\'ersic components (and therefore a strong upward curvature in the SB profiles) occur below Donzelli et al.'s detection limit. The authors themselves pointed out that their measured correlation coefficients depend on the applied radius or surface brightness cuts.

\begin{deluxetable}{ccccr}
	\tablewidth{700pt}
	\tabletypesize{\scriptsize}
	\tablecaption{Correlations between Structural Parameters.\label{tab:parrelations}}
	\tablehead{
		\colhead{Galaxy Type} & \colhead{$X$} & \colhead{$Y$} & \colhead{Slope $\alpha$} & \colhead{Offset $\beta$}
	}
	\colnumbers
	\startdata
	Regular Es  & $M$ & log($r_{\rm e}$) & -0.309 $\pm$ 0.020 &  -6.02 $\pm$ 0.43\\
	SS BCGs     & $M$ & log($r_{\rm e}$) & -0.550 $\pm$ 0.037 & -11.29 $\pm$ 0.87\\
	DS BCGs     & $M$ & log($r_{\rm e}$) & -0.547 $\pm$ 0.060 & -11.10 $\pm$ 1.43\\
	All BCGs    & $M$ & log($r_{\rm e}$) & -0.563 $\pm$ 0.032 & -11.57 $\pm$ 0.75\\
	\hline
	Regular Es  & $M$ & SB$_{\rm e}$ & -0.75 $\pm$ 0.13 &   5.47 $\pm$ 2.65\\
	SS BCGs     & $M$ & SB$_{\rm e}$ & -1.90 $\pm$ 0.20 & -20.32 $\pm$ 4.70\\
	DS BCGs     & $M$ & SB$_{\rm e}$ & -1.96 $\pm$ 0.35 & -20.95 $\pm$ 8.26\\
	All BCGs    & $M$ & SB$_{\rm e}$ & -2.02 $\pm$ 0.18 & -22.93 $\pm$ 4.26\\	
	\hline
	Regular Es  & log($r_{\rm e}$) & SB$_{\rm e}$ & 2.44 $\pm$ 0.22 & 20.18 $\pm$ 0.12\\
	SS BCGs     & log($r_{\rm e}$) & SB$_{\rm e}$ & 3.61 $\pm$ 0.13 & 18.61 $\pm$ 0.22\\
	DS BCGs     & log($r_{\rm e}$) & SB$_{\rm e}$ & 3.58 $\pm$ 0.26 & 18.77 $\pm$ 0.51\\
	All BCGs    & log($r_{\rm e}$) & SB$_{\rm e}$ & 3.63 $\pm$ 0.11 & 18.61 $\pm$ 0.88\\
	\enddata
	\tablecomments{The correlations are in the form of $Y=\alpha X + \beta$. Orthogonal distance regression was applied to find the best-fit parameters. All values are for the $g'$-band. The galaxies of type "regular Es" are from \cite{Bender1992}, that is, the dark gray data points in Figure \ref{fig:profileparams}.}
\end{deluxetable}

\begin{figure}
	\includegraphics[width=\linewidth]{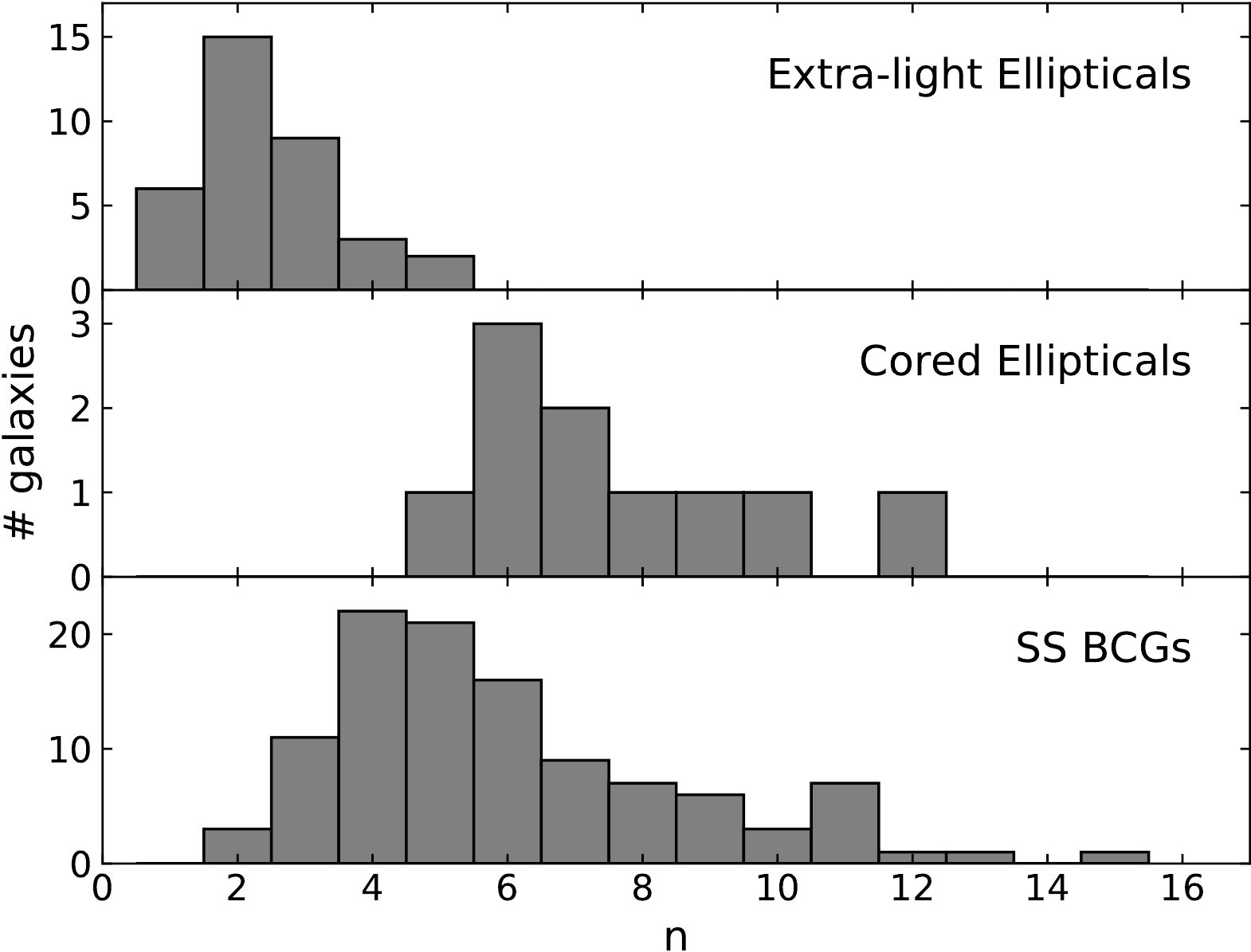}
	\caption{Distribution of S\'ersic indices $n$ for extra-light ellipticals (top panel), cored ellipticals (middle panel), and SS BCGs (bottom panel). The data for the extra-light and cored ellipticals are taken from \cite{Kormendy2009}. \label{fig:sersicn}}
\end{figure}

Our measured size-luminosity relation $\log(r_{\rm e}) = \alpha M + \beta$ has a slope $\alpha=-0.550 \pm 0.037$ for SS BCGs and $\alpha=-0.547 \pm 0.060$ for DS BCGs. That is significantly steeper than $\alpha=-0.354$ as measured by \cite{Bernardi2007}. They fit SS functions to the semimajor-axis SB profiles of 215 BCGs ($z<0.12$), measured on SDSS DR2 $r$-band data. The average total brightness of the BCGs in the Bernardi et al. sample is 1 - 2 $g'$ mag fainter than the BCGs in our sample. A different sample selection, preferentially toward lower mass clusters, could explain that discrepancy. Unfortunately, direct comparison of individual BCGs was not possible. Furthermore, the authors concluded from mock observations that undetected ICL shifts the data points along the size-luminosity relation but does not change its slope. However, our results disagree with their conclusion. For the brightest BCGs, we find larger effective radii than predicted by the size-luminosity relation as measured by \cite{Bernardi2007}. In a forthcoming paper, we show that the fraction of light that is encompassed in the low-SB outskirts increases with total BCG+ICL brightness. That can lead to a shallower size-luminosity relation when that trend is not included in the models of Bernardi et al.

The trend that brighter BCGs have larger luminosity fractions in their low-SB outskirts offers an explanation for the broken slope in the size-luminosity relation and consequently also in the other two relations shown in Figure \ref{fig:profileparams} and listed in Table \ref{tab:parrelations}. If BCGs are unique in growing predominantly by accreting stellar material into their outskirts, then their effective radii increase for a fixed brightness.

The same argument can also be made for galaxy clusters as a whole that grow purely by accretion. The slope of the size-luminosity relation is broken here once more. Compared to the BCGs, they are located at much larger radii but only slightly larger brightnesses (blue data points in Figure \ref{fig:profileparams}). The measurements of the cluster parameters will be discussed in a forthcoming paper.

We notice from further inspecting Figure \ref{fig:profileparams} that SS BCGs have a smaller median effective radius and a brighter median effective surface brightness ($45\pm24$ kpc; $22.4\pm0.9~g'$ mag arcsec$^{-2}$) compared to their DS counterparts ($72\pm31$ kpc; $25.5\pm0.8~g'$ mag arcsec$^{-2}$). The slightly more compact shape of SS BCGs is also seen in the average SB profiles in Figure \ref{fig:avgprofile}. In integrated brightnesses, SS BCGs are 14.8\% fainter ($-23.68\pm0.53~g'$ mag) than DS BCGs ($-23.83\pm0.41~g'$ mag).

The distribution of S\'ersic indices $n$ for all SS BCGs is shown in Figure \ref{fig:sersicn}. Most SS BCGs have $n\geq4$, but 20/121 (17\%) have significantly lower $n$. That value approximately separates the two classes of disky--coreless--rotating ellipticals (denoted extra-light ellipticals in the histogram) and boxy-core-nonrotating ellipticals \citep{Kormendy2009}. We further elaborate on that dichotomy in the next subsection. The high S\'ersic indices can be explained by accretion that is predominantly happening in the outskirts, which subsequently increases the upward curvature of the SB profiles and consequently the S\'ersic indices.

\section{Discussion}\label{sec:discussion}

\subsection{Do BCGs Form a Unique Class of Elliptical Galaxies?}

We compare the structural parameters of BCGs to those of regular ellipticals in Figure \ref{fig:profileparams}. Spheroidals and ellipticals including classical bulges populate different areas in parameter space, which indicates that they are formed by different formation scenarios \citep{Kormendy2009}. BCGs also do not follow the correlations for regular ellipticals. The slopes are steeper. The downward trend of the Kormendy and $M_{\rm tot} \propto {\rm SB}_{\rm e}$ relations in Figure \ref{fig:profileparams} illustrates the growing importance of the low-surface-brightness stellar halo and ICL contribution of ellipticals with increasing luminosity because half of the light is below the effective surface brightness. The broken slopes of these relations underline that the stellar halos and ICL are even more important for BCGs, that is, their growth is even more dominated by accretion in their low-SB outskirts (e.g., \citealt{Oser2010}). We emphasize that we do not try to dissect the ICL or stellar halos from the BCGs in this work and instead consider their combined stellar light.

The effective radii are also larger than what would be expected from extrapolating the size-luminosity relation $M_{\rm tot} \propto \log(r_{\rm e})$ for regular ellipticals. All these findings confirm the picture that regular ellipticals and BCGs differ from each other in the importance of accretion in their formation history. BCGs reside near the center of their host cluster. Contrary to regular ellipticals, that enables them to accumulate enormous amounts of stellar material instead of being tidally stripped by the cluster potential.

\subsection{Is the Inner Component of DS BCGs ``Extra Light"?}

An alternative explanation for the origin of DS profile shapes could be due to a central poststarburst stellar population as it is often seen in extra-light ellipticals (e.g., \citealt{Faber1997,Kormendy1999,Kormendy2009,Kormendy2013}). There are two families of ellipticals: boxy--core--nonrotating and disky--extralight--rotating (\citealt{Bender1988a,Bender1988,Bender1989,Bender1991,Kormendy1996,Kormendy2009}). Most BCGs are categorized as boxy--core--nonrotating galaxies which is further confirmed by the distribution of S\'ersic indices (see Section \ref{sec:strucparams}). Those ellipticals are believed to have formed via dissipationless merging and subsequent violent relaxation. However, judging from the SB profiles in Appendix \ref{sec:sbprofiles}, there are some BCGs that could potentially be categorized as unusually massive extra-light ellipticals. Non-BCG extra-light ellipticals have small transition radii of $r_{\times} \lesssim 1$ kpc \citep{Hopkins2009,Kormendy2009} or $r_{\times}<0.04 r_{\rm e}$ \citep{Mihos1994}. A light excess above the inward extrapolation of the outer S\'ersic profile is interpreted to arise from a poststarburst stellar population that was formed after a wet merger. The origin of the DS shape would then be unrelated to the ICL phenomenon. Those BCGs can bias the structural parameter relations and correlations with cluster properties.

By conservatively selecting only DS BCGs that have transition radii $r_{\times}\gtrsim0.1 r_{\rm e}$ and transition surface brightness SB$_{\times}>23~g'$ mag arcsec$^{-2}$, we discard 49/98 BCGs from the DS sample that are potentially extra-light ellipticals and classify them as SS BCGs. The structural parameter relations between $r_{\rm e}$, SB$_{\rm e}$, and $M_{\rm tot}$ for both split samples do not differ significantly from each other.

Furthermore, we inspect the isophotal distortions $a_4$ of the potential extra-light ellipticals. Extra light is frequently observed to have disky isophotes when viewed edge-on (e.g., Section 9.3 in \citealt{Kormendy2009}). At least some of the 49 potential extra-light BCGs in our sample should have high inclinations. Therefore, we expect the average $a_4$ to be higher in the inner regions for the potential extra-light BCG subsample than for the rest of the DS BCGs. We do not find that. The isophotes of the potential extra-light ellipticals are not diskier near the transition radii than those of the BCGs that have no potential extra light. 

The abundance of multiple cores for potential extra-light ellipticals would increase if some of them are the remnants of the wet mergers. Contrary to that expectation, it is even less. Also, malicious handling of overlapping galaxies is thereby excluded as an artificial origin of small $r_{\times}$ DS profiles.

All 18 BCGs that overlap with the \cite{Lauer2007} sample (A76, A193, A260, A347, A376, A397, A634, A999, A1016, A1020, A1142, A1177, A1656, A1831, A2052, A2147, A2199, and A2589) are classified by the authors as cored ellipticals. The decisions were made based on high-resolution \textit{HST} images. Six out of those BCGs are classified by our criteria as potential extra-light ellipticals: A193, A260, A397, A1020, A2147, and A2589; that is, the SB profiles have a core inside a potential extra-light region. This will break the known dichotomy if the extra light will be confirmed to form in the same poststarburst scenario as it is the case for lower mass ellipticals.

\subsection{Do DS BCGs Differ from SS BCGs in Their Evolutionary State?}

The members of both S\'ersic types are, in general, very similar in their appearance. Both families follow the same structural parameter correlations (see Figure \ref{fig:profileparams} and Table \ref{tab:parrelations}). Any characteristic that qualifies each S\'ersic type as distinct might be subtle. Nevertheless, there are differences beyond the simple number of analytic functions that are needed to fit their light profiles well.

First of all, we take a closer look at the average profiles presented in Figure \ref{fig:avgprofile}. The clearest discrepancy is found in the ellipticity $\epsilon$ profiles. DS BCGs are, on average, rounder at small radii $r = \sqrt{ab} < 16$ kpc and become more elliptical at larger radii. This is qualitatively consistent with the discovery by \cite{Donzelli2011}. DS BCGs furthermore have lower scatter in position angle drifts $\Delta {\rm PA}(r)$, that is, smaller isophote twists for individual galaxies.\footnote{The larger scatter at small radii can be attributed to the smaller ellipticities that increase the uncertainty of the PA measurement.} And finally, DS BCGs have on average less boxy isophotes ($a_4^{\rm DS} > a_4^{\rm SS}$). We must note here that boxy isophotes also result from shells \citep{Gonzalez2005b}, which are actually marginally more common for SS BCGs (see Figure \ref{fig:histo_features}). Nevertheless, all of these tendencies are identical with those of more rotationally supported and thus less evolved systems. The spatial offsets between the ICL and the BCG are on average larger for DS BCGs. This is related to the higher abundance of multiple nuclei (see Figure \ref{fig:histo_features}) that drag the main nucleus along by their gravitational attraction. The analog to ICL offsets in velocity space is systemic velocity offsets, that is, the line-of-sight velocity difference between the BCG and the average cluster line-of-sight velocity. We have examined the distributions of systemic velocity offsets for SS and DS BCGs separately using published data from \cite{Lauer2014}. A Kolmogorov--Smirnov test showed that no conclusion can be drawn from those data.

The isophotal parameters that describe asymmetric distortions $a_3$, $b_3$ and $b_4$ are higher for DS BCGs (see Figure \ref{fig:avgprofile}). Such shapes are not stable and are therefore evidence for ongoing accretion. The larger abundance of signatures from these accretion processes is also documented in Figure \ref{fig:histo_features}. These features are relatively short-lived because they are dynamically hot. They originate from collisions and stripping events with high relative velocities of order $\sim$1000 km s$^{-1}$. Because these remnants are visible today, DS BCGs must have undergone more of these events recently.

We mentioned before that any dichotomy between SS and DS BCGs is subtle. Most of our distinctions are not very significant, but they all point to the same conclusion. SS BCGs are currently in a more relaxed state because they have experienced fewer accretion events in the recent past. Either the earlier accreted stellar mass has already virialized by violent relaxation, or the events that would create a distinctly visible envelope have not yet taken place.

\section{Summary and Conclusion}

We have obtained optical $g'$-band observations of 170 galaxy clusters with the Wendelstein Wide Field Imager. The data reduction pipeline was developed and assembled specifically for that instrument and optimized for low-surface-brightness photometry. 

We have measured semimajor axis surface brightness profiles of the BCGs down to a limiting surface brightness of SB$_{\rm lim} = 30~g'$ mag arcsec$^{-2}$, which is an unprecedented depth for a large sample size. 

Our results are summarized as follows:

(1) BCGs have larger effective radii, dimmer effective surface brightnesses, and brighter absolute magnitudes than expected for an extrapolation of the parameter correlations for regular ellipticals. The Kormendy, the size--luminosity and the $M_{\rm tot}$--SB$_{\rm e}$ relations have broken slopes at least in part because of the presence of ICL around the BCGs.

(2) By fitting S\'ersic functions to the semimajor-axis SB profiles, we find that 71\% of the observed BCGs are well described by a single S\'ersic function (SS BCGs). The remaining 29\% of BCGs have variations in the slope of their SB profiles that require using two S\'ersic functions to obtain a good fit (DS BCGs). DS BCGs with transition radii $r_{\times}<0.1 r_{\rm e}$ and transition surface brightnesses SB$_{\times}<23~g'$ mag arcsec$^{-2}$ were fitted with a single S\'ersic function excluding the inner excess light. The DS profile shape is more likely to arise in those cases because of a poststarburst stellar population following a wet merger than because of the ICL phenomenon.

(3) SS and DS BCGs do not deviate significantly from each other in their parameter correlations between effective radii $a_{\rm e}$ along the major axis, effective surface brightnesses SB$_{\rm e}$, and integrated absolute brightnesses $M_{\rm tot}$.

(4) SS BCGs are slightly more compact ($r_{\rm e}^{\rm SS} = 45\pm24$ kpc) than DS BCGs ($r_{\rm e}^{\rm DS} = 72\pm31$ kpc). In integrated brightnesses, SS BCGs are 14.8\% fainter ($-23.68\pm0.53~g'$ mag) than DS BCGs ($-23.83\pm0.41~g'$ mag).

(5) The S\'ersic indices of SS BCGs are significantly larger than $n\geq4$ in 83\% of the cases. That value approximately separates the two classes of disky--coreless--rotating ellipticals and boxy--core--nonrotating ellipticals.

(6) The radial profiles of their structural parameters show that SS BCGs have on average

\begin{enumerate}
	\item shallower ellipticity profiles,
	\item stronger individual isophote twists,
	\item smaller ICL offsets,
	\item boxier isophotes,
	\item less pronounced asymmetric isophotal distortions, and
	\item fewer accretion signatures
\end{enumerate}

than DS BCGs. We deduce from these results that SS BCGs are on average marginally more relaxed because they have encountered fewer accretion events in the recent past. The tendencies are identical to those of more triaxial and dispersion supported, that is, more evolved systems.

(7) The average isophote twists are $\Delta {\rm PA} / \Delta r \sim 10\degr / 100$ kpc.

(8) The average ellipticity increases with radius and reaches $\epsilon = 0.4 - 0.5$ at a circular radius of $r \approx 200$ kpc.

(9) The isophotal offset with respect to the nucleus increases with radius. At 200 kpc circular galactocentric radius from the nucleus, the average offset is 37 kpc with 34 kpc intrinsic scatter. 

We conclude from our study that BCG+ICLs have scaling relations with steeper slopes than those for normal non-BCG ellipticals. That is likely because the faint ICL outskirts around BCGs have a significant influence on the structural parameters. Our deep SB profiles enable us furthermore to decide more consistently whether an SB profile is well described by an SS or DS profile. The former case is more common (71\%) at redshift $\bar{z}=0.06$, that is, most of the BCG+ICLs have relatively smooth SB profiles. Whether the photometrically distinct stellar envelopes around the rarer DS BCGs trace the ICL is debated. We have shown that the isophotal shapes of DS BCGs are more disturbed and accretion signatures are more common inside them than in SS BCGs. Hence, it is possible that the envelopes are simply the result of unrelaxed, recently accreted material and not necessarily the signature of pure ICL. On the other hand, it could also be that SS BCGs have not yet accumulated sufficient stellar material to build up a distinct ICL envelope. We will further address that question in a forthcoming paper where we compare different photometric methods to dissect the ICL from the BCGs. Lastly, we have shown in Figure \ref{fig:profileparams} that the size--brightness relation curves upward toward host cluster data points. In other words, the ICL transitions smoothly into the galaxy clusters. We will explore further correlations between BCG/ICL and host cluster properties in the same forthcoming paper.

\acknowledgments

We are grateful to Rhea-Silvia Remus, Klaus Dolag, Stella Seitz, John Kormendy, Tod Lauer, and Walter Dehnen for helpful conversations and constructive feedback on the manuscript. We also wish to thank the anonymous referee for his or her comments and suggestions that allowed us to significantly improve the paper.

The 2m telescope project is funded by the Bavarian government and by the German Federal government through a common funding process. Part of the 2m instrumentation including some of the upgrades for the infrastructure and the 40cm telescope housing were funded by the Cluster of Excellence "Origin of the Universe" of the German Science foundation DFG. The 40cm telescope was funded by the Ludwig-Maximilians-University, Munich.

This work made use of data products based on observations made with the NASA/ESA \textit{Hubble Space Telescope}, and obtained from the Hubble Legacy Archive, which is a collaboration between the Space Telescope Science Institute (STScI/NASA), the Space Telescope European Coordinating Facility (ST-ECF/ESA), and the Canadian Astronomy Data Centre (CADC/NRC/CSA).

This work would not have been practical without extensive use of NASA's Astrophysics Data System Bibliographic Services and the SIMBAD database, operated at CDS, Strasbourg, France.

We also used the image display tool SAOImage DS9 developed by Smithsonian Astrophysical Observatory and the image display tool Fitsedit, developed by Johannes Koppenhoefer.

This research made use of Astropy, a community-developed core Python package for Astronomy \citep{Astropy2013}.

\facilities{WO:2m (Wide-field camera), \textit{HST} (WFPC2, ACS), Planck (HFI)}

\clearpage
\appendix

\section{Surface Brightness Profiles and S\'ersic Fits} \label{sec:sbprofiles}

The SB profiles, ellipticity profiles, and PA profiles of all BCGs in our sample are shown in Figure \ref{fig:sersicfits01} as data points. S\'ersic profiles with the best-fit parameters (see Appendix \ref{sec:apclusterparams}) are overplotted as lines.

\begin{figure*}[b]
	\centering
	\includegraphics[width=0.915\linewidth]{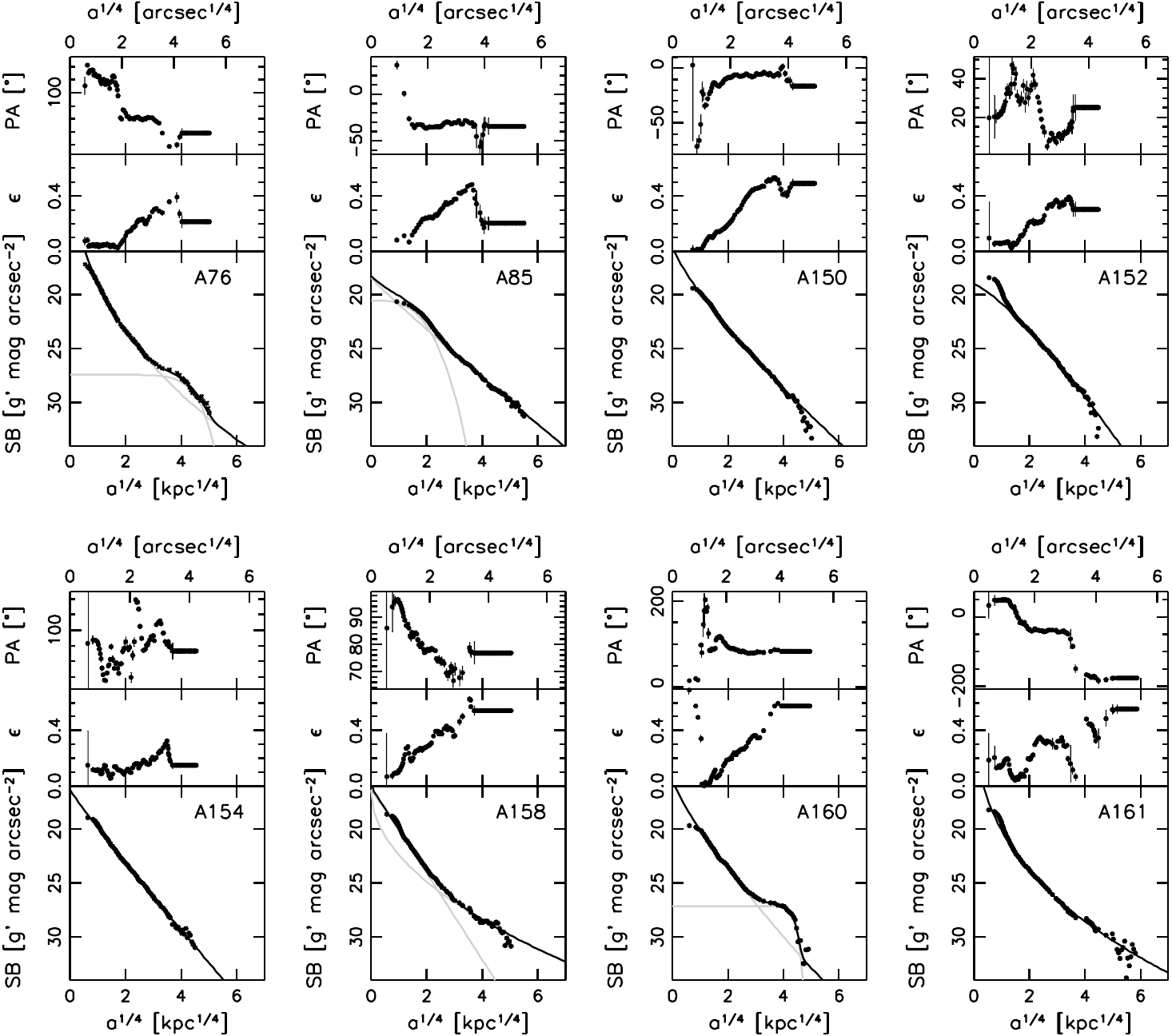}
	\caption{The SB profiles are corrected for PSF broadening. No K-correction and no corrections for dust extinction and cosmic dimming are applied. If the SB profiles were fitted by a double S\'ersic function, then the light gray lines show the contributions of each component. Ellipticity and position angle profiles are presented in the middle and top panels, respectively. \label{fig:sersicfits01}}
\end{figure*}

\clearpage

\begin{figure*}
	\centering
	\includegraphics[width=0.915\linewidth]{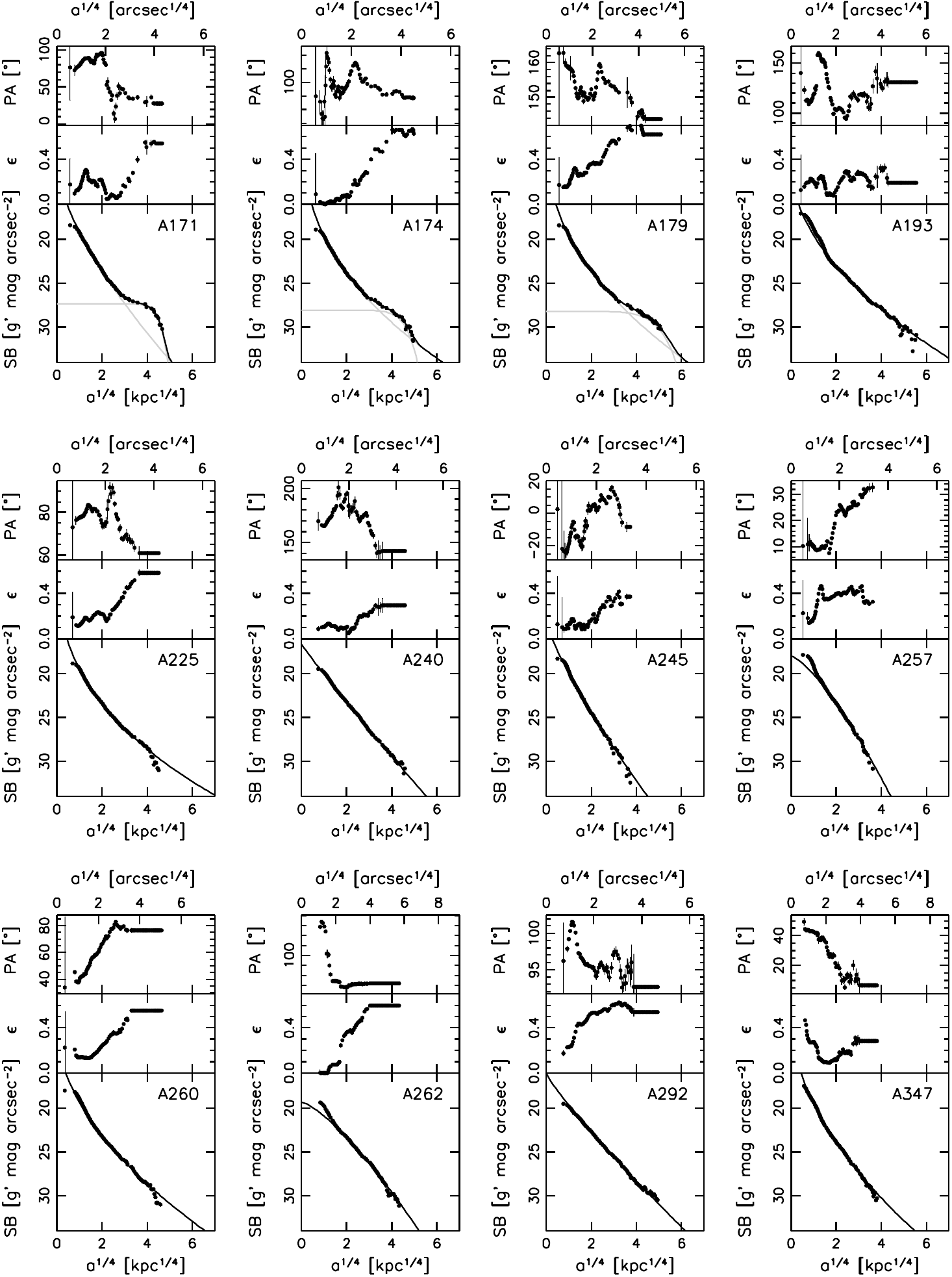}\\~\\
	\textbf{Figure \ref*{fig:sersicfits01}} \textit{(continued)}
\end{figure*}

\clearpage

\begin{figure*}
	\centering
	\includegraphics[width=0.915\linewidth]{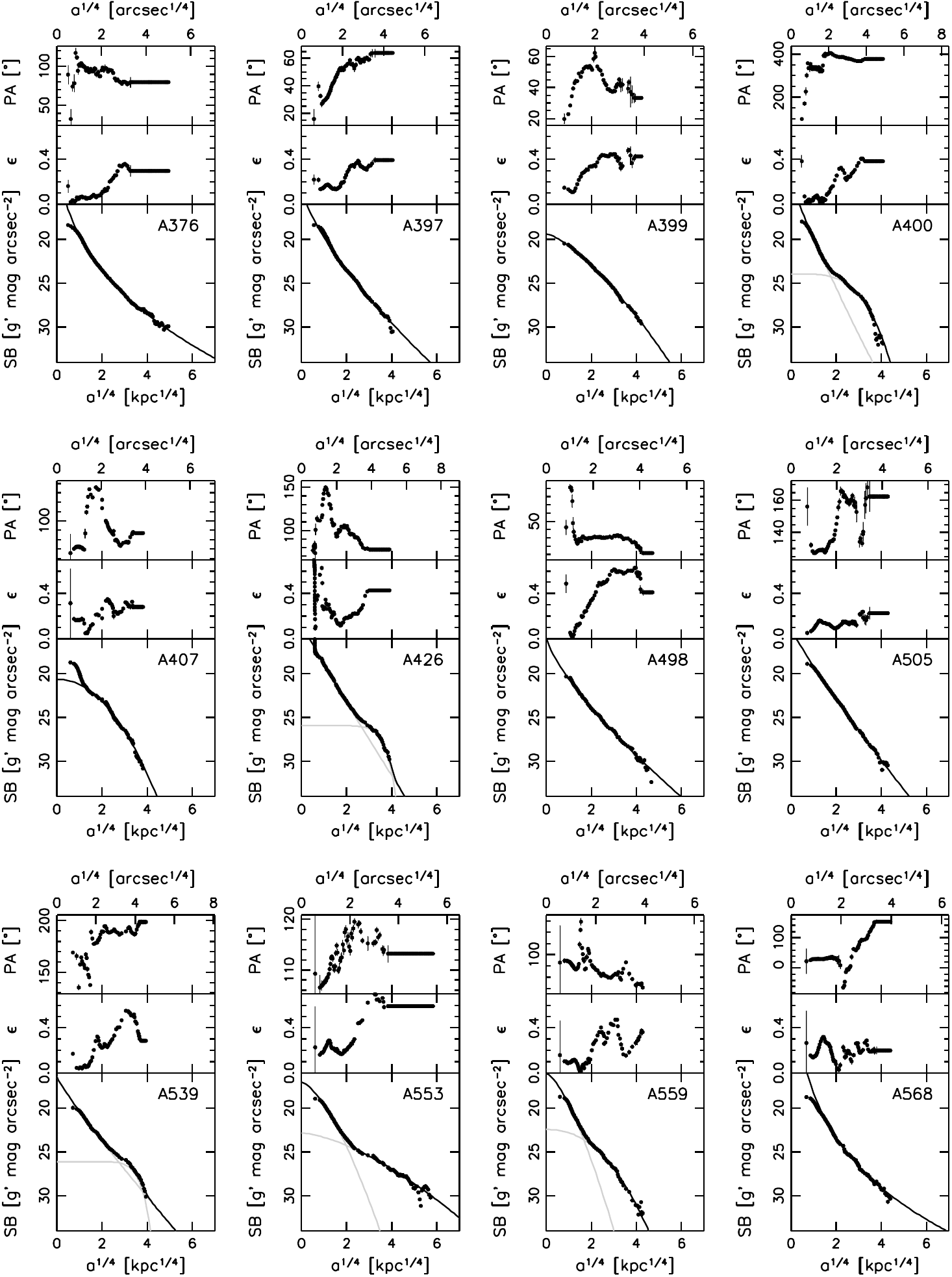}\\~\\
	\textbf{Figure \ref*{fig:sersicfits01}} \textit{(continued)}
\end{figure*}

\clearpage

\begin{figure*}
	\centering
	\includegraphics[width=0.915\linewidth]{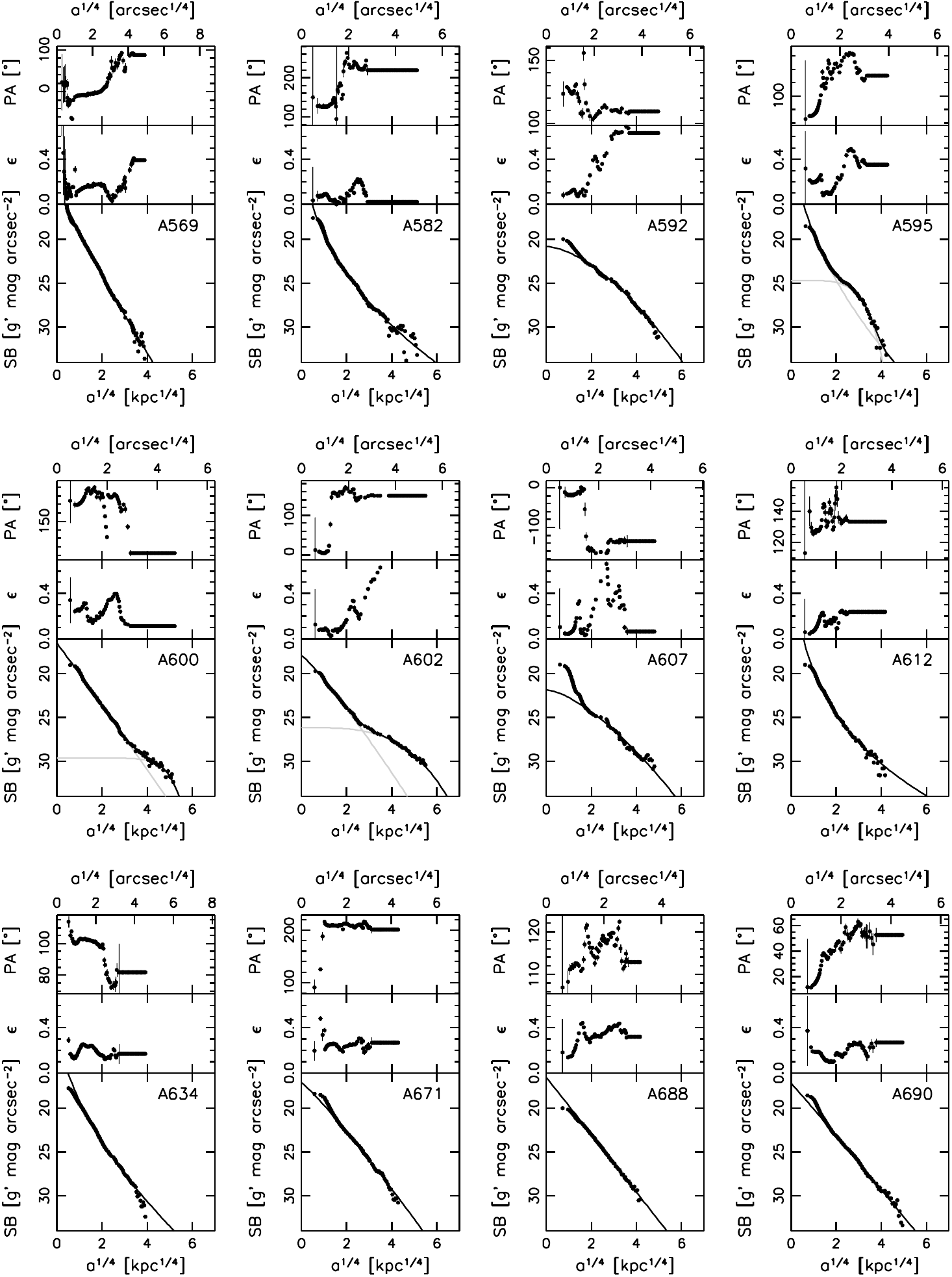}\\~\\
	\textbf{Figure \ref*{fig:sersicfits01}} \textit{(continued)}
\end{figure*}

\clearpage

\begin{figure*}
	\centering
	\includegraphics[width=0.915\linewidth]{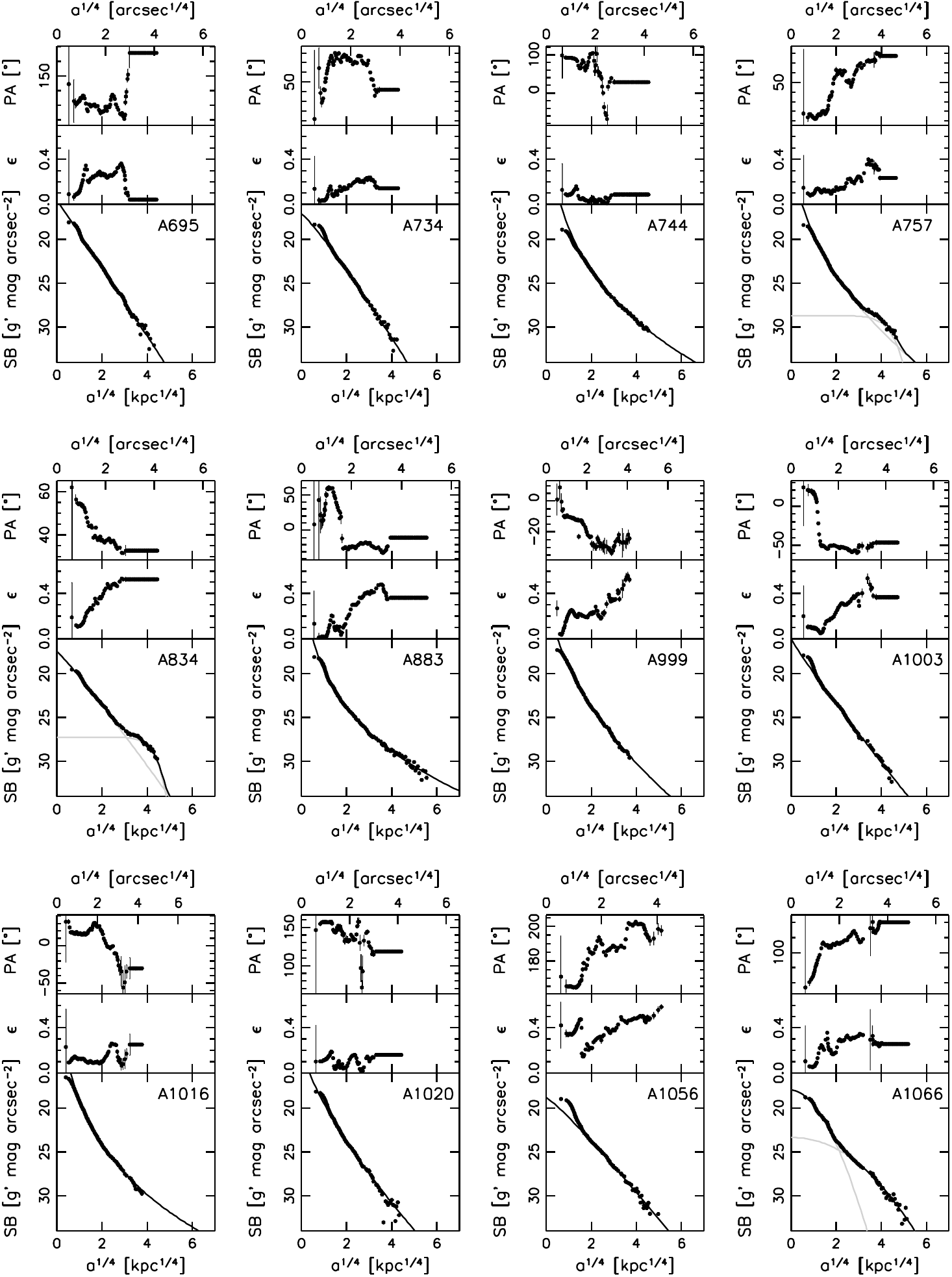}\\~\\
	\textbf{Figure \ref*{fig:sersicfits01}} \textit{(continued)}
\end{figure*}

\clearpage

\begin{figure*}
	\centering
	\includegraphics[width=0.915\linewidth]{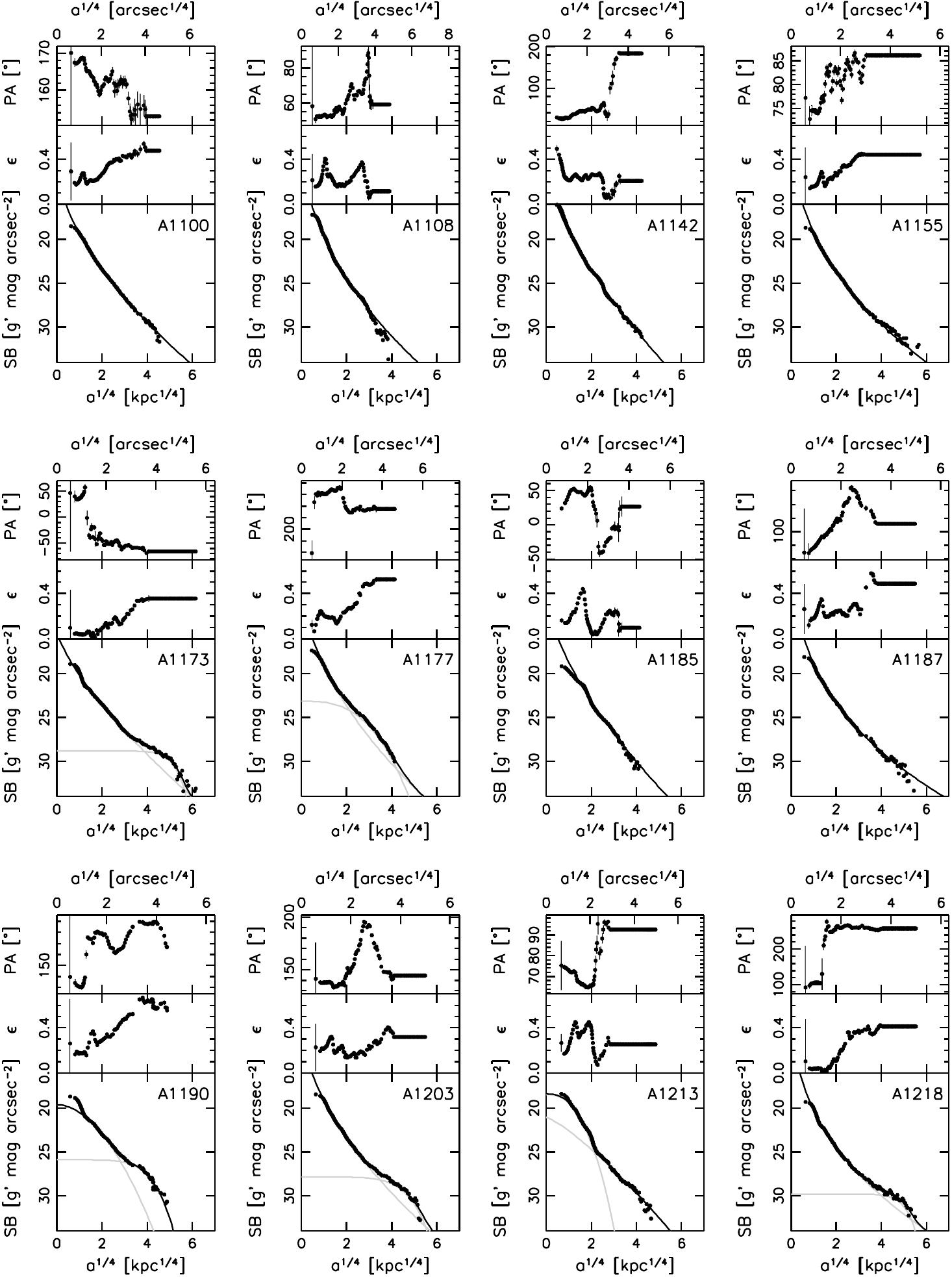}\\~\\
	\textbf{Figure \ref*{fig:sersicfits01}} \textit{(continued)}
\end{figure*}

\clearpage

\begin{figure*}
	\centering
	\includegraphics[width=0.915\linewidth]{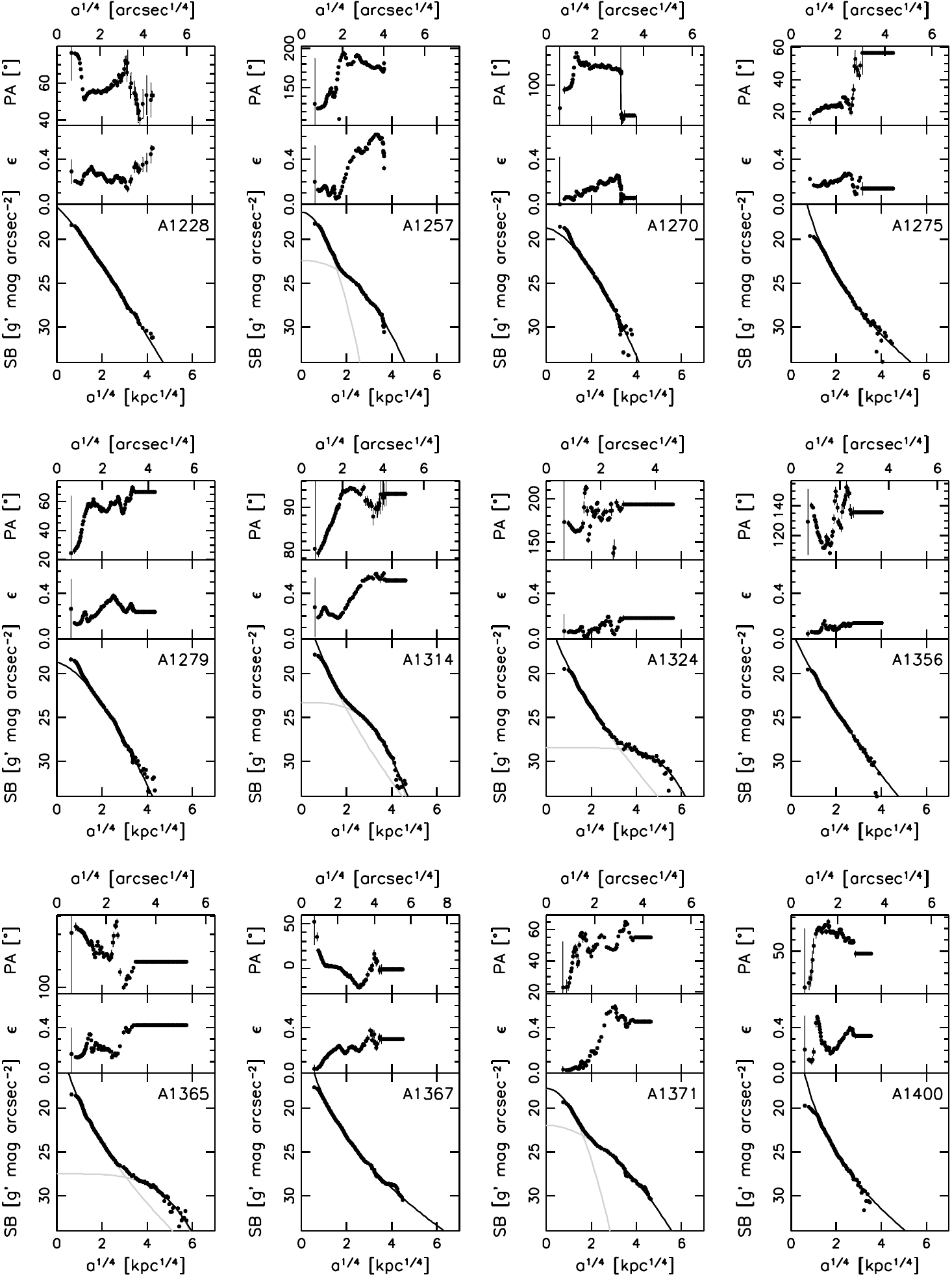}\\~\\
	\textbf{Figure \ref*{fig:sersicfits01}} \textit{(continued)}
\end{figure*}

\clearpage

\begin{figure*}
	\centering
	\includegraphics[width=0.915\linewidth]{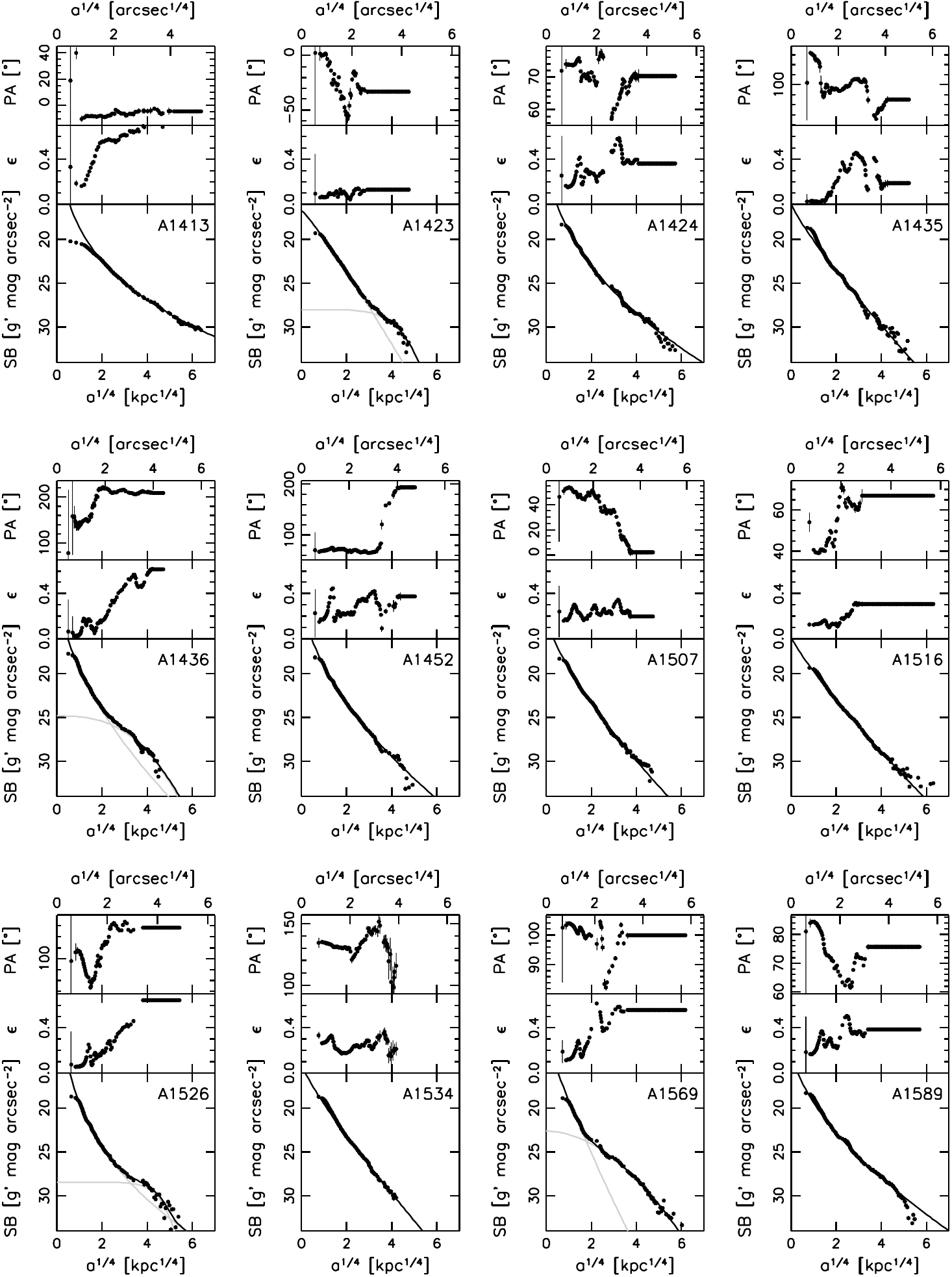}\\~\\
	\textbf{Figure \ref*{fig:sersicfits01}} \textit{(continued)}
\end{figure*}

\clearpage

\begin{figure*}
	\centering
	\includegraphics[width=0.915\linewidth]{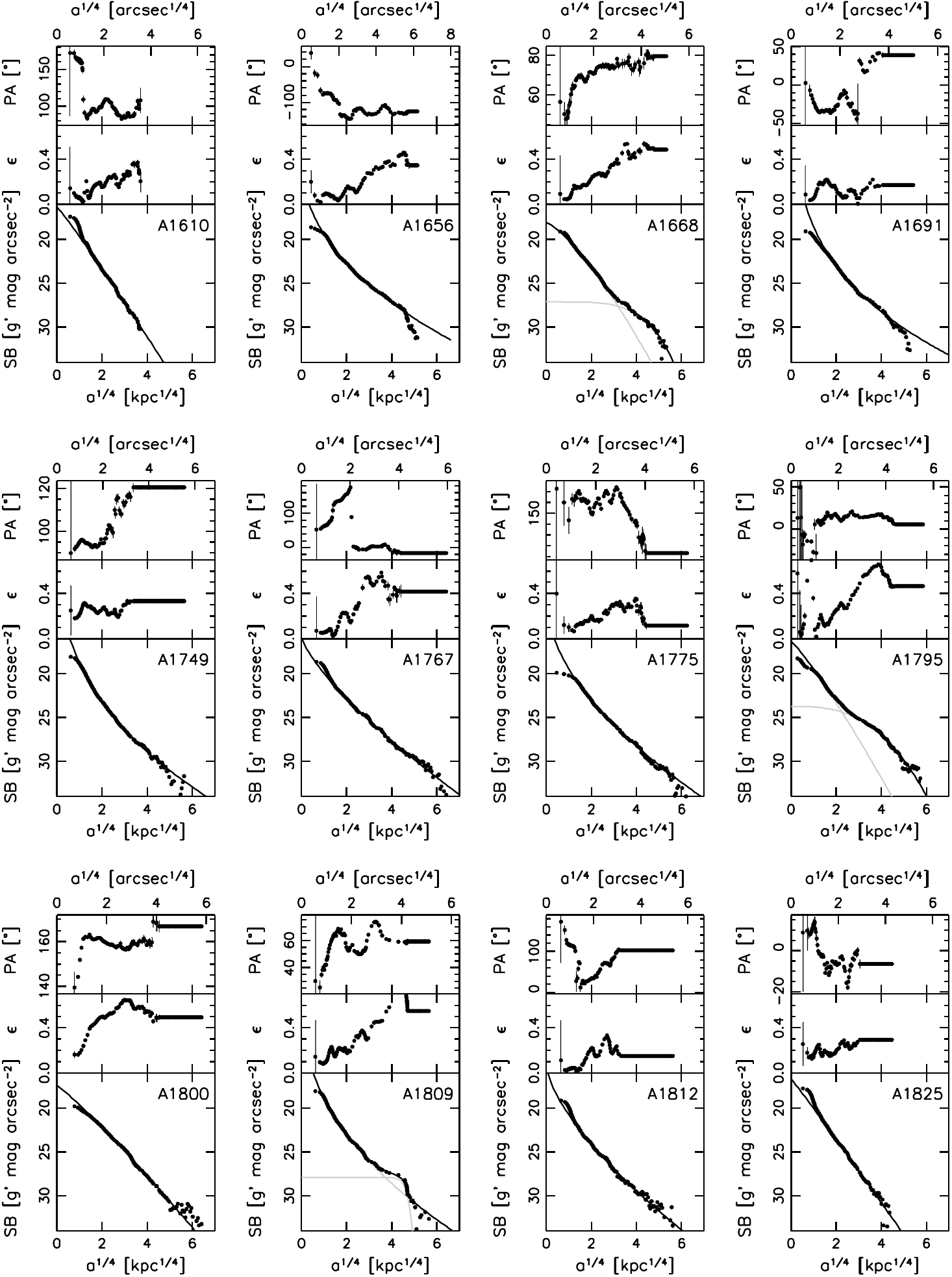}\\~\\
	\textbf{Figure \ref*{fig:sersicfits01}} \textit{(continued)}
\end{figure*}

\clearpage

\begin{figure*}
	\centering
	\includegraphics[width=0.915\linewidth]{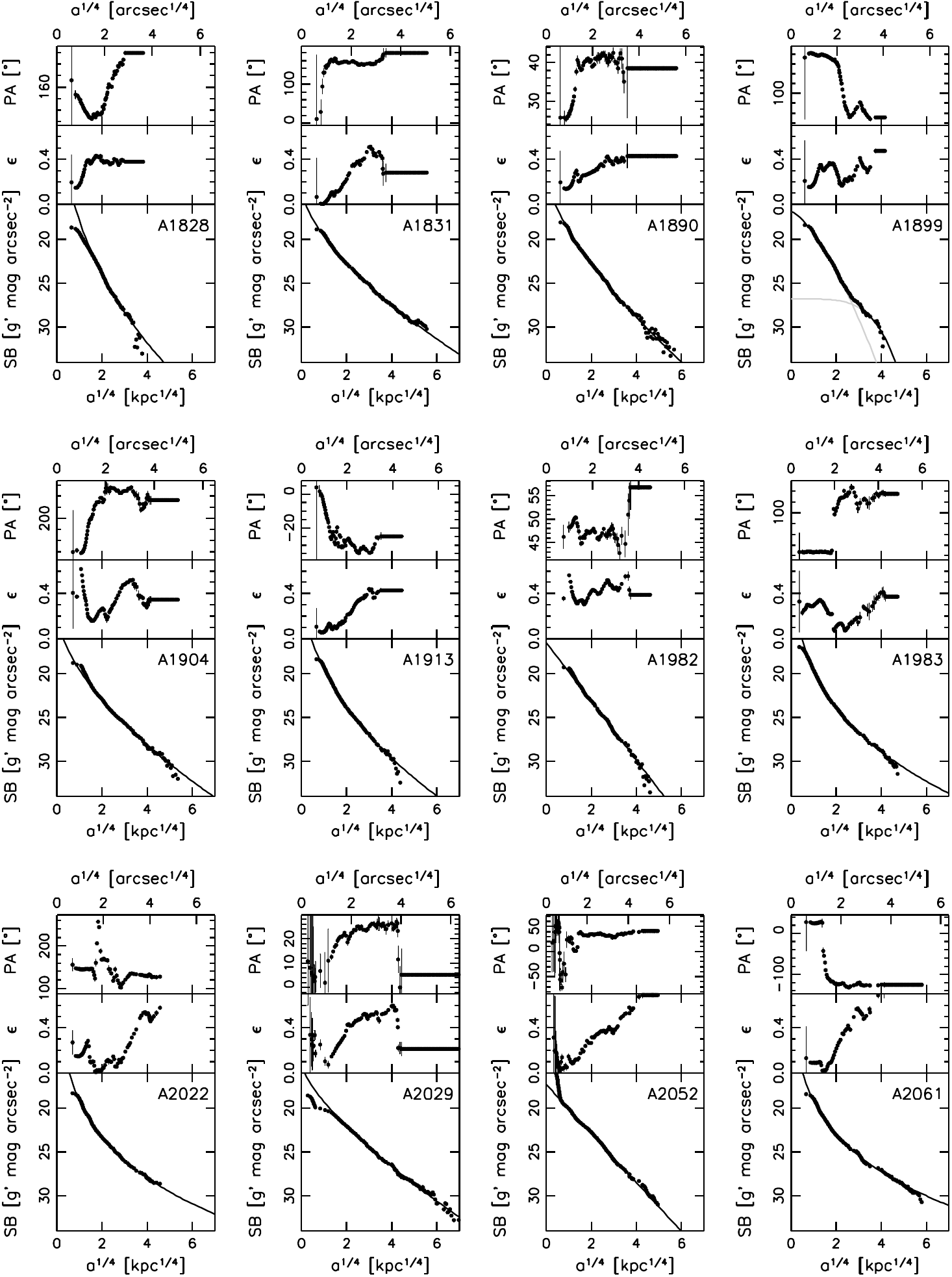}\\~\\
	\textbf{Figure \ref*{fig:sersicfits01}} \textit{(continued)}
\end{figure*}

\clearpage

\begin{figure*}
	\centering
	\includegraphics[width=0.915\linewidth]{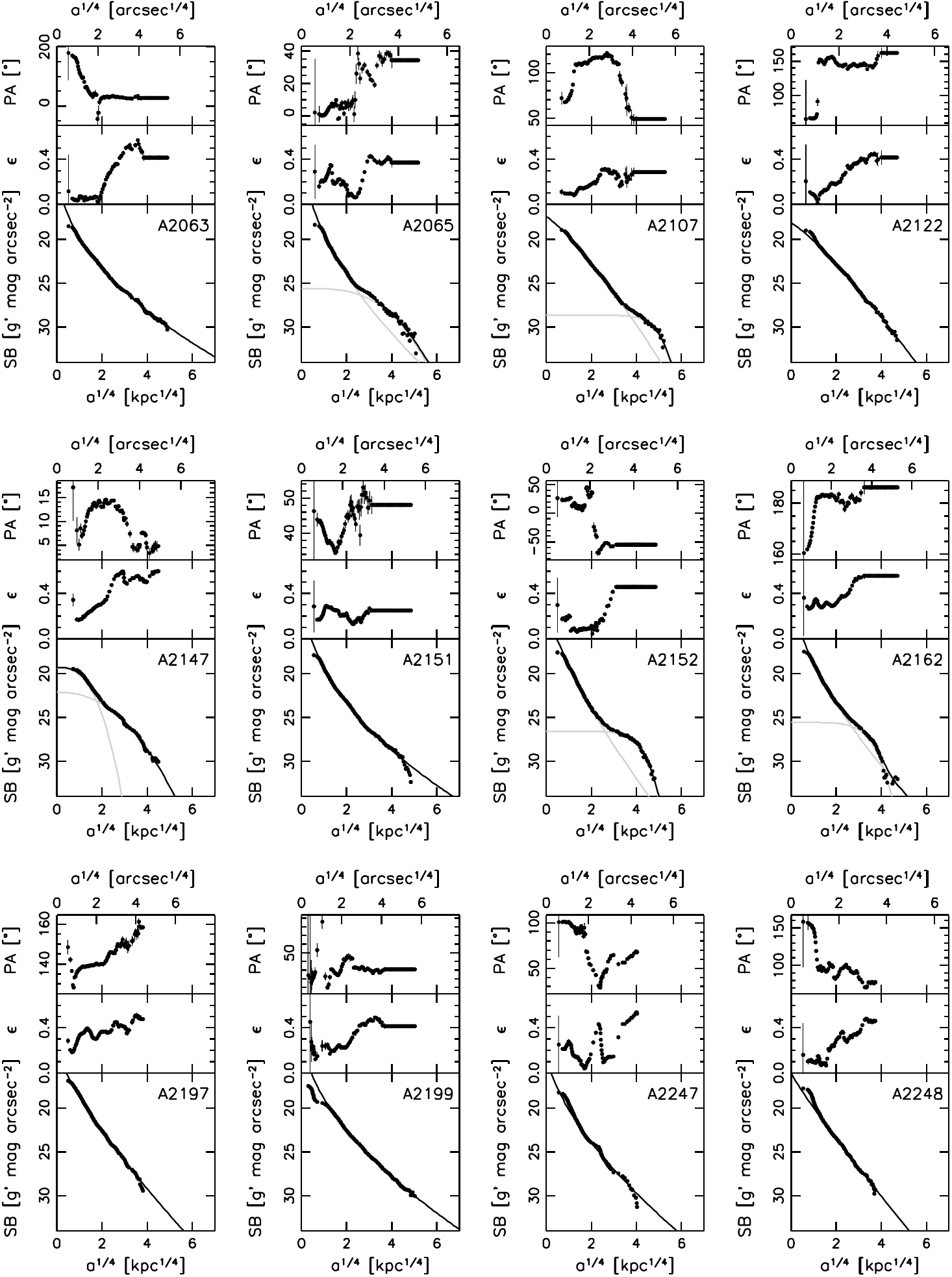}\\~\\
	\textbf{Figure \ref*{fig:sersicfits01}} \textit{(continued)}
\end{figure*}

\clearpage

\begin{figure*}
	\centering
	\includegraphics[width=0.915\linewidth]{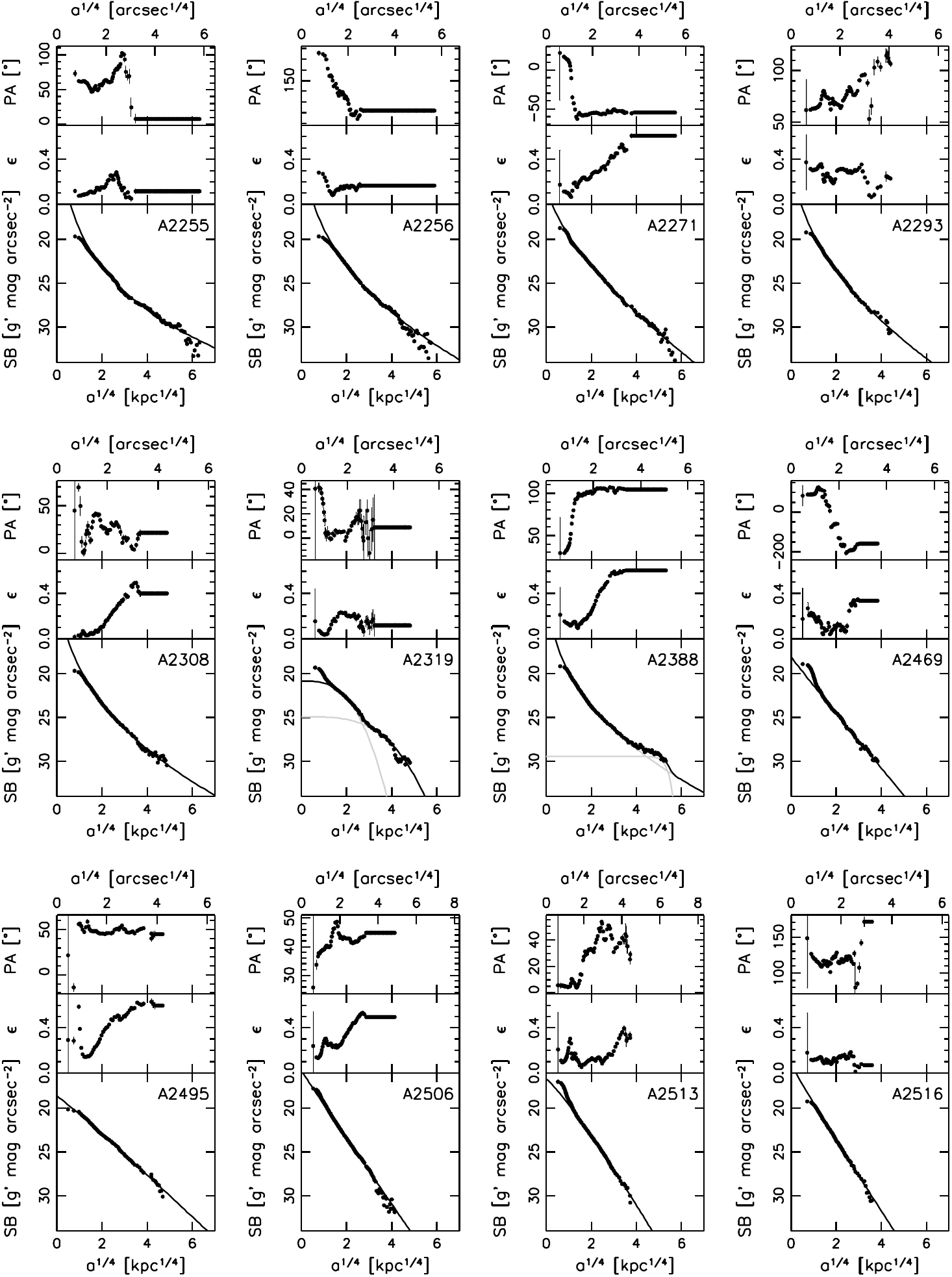}\\~\\
	\textbf{Figure \ref*{fig:sersicfits01}} \textit{(continued)}
\end{figure*}

\clearpage

\begin{figure*}
	\centering
	\includegraphics[width=0.915\linewidth]{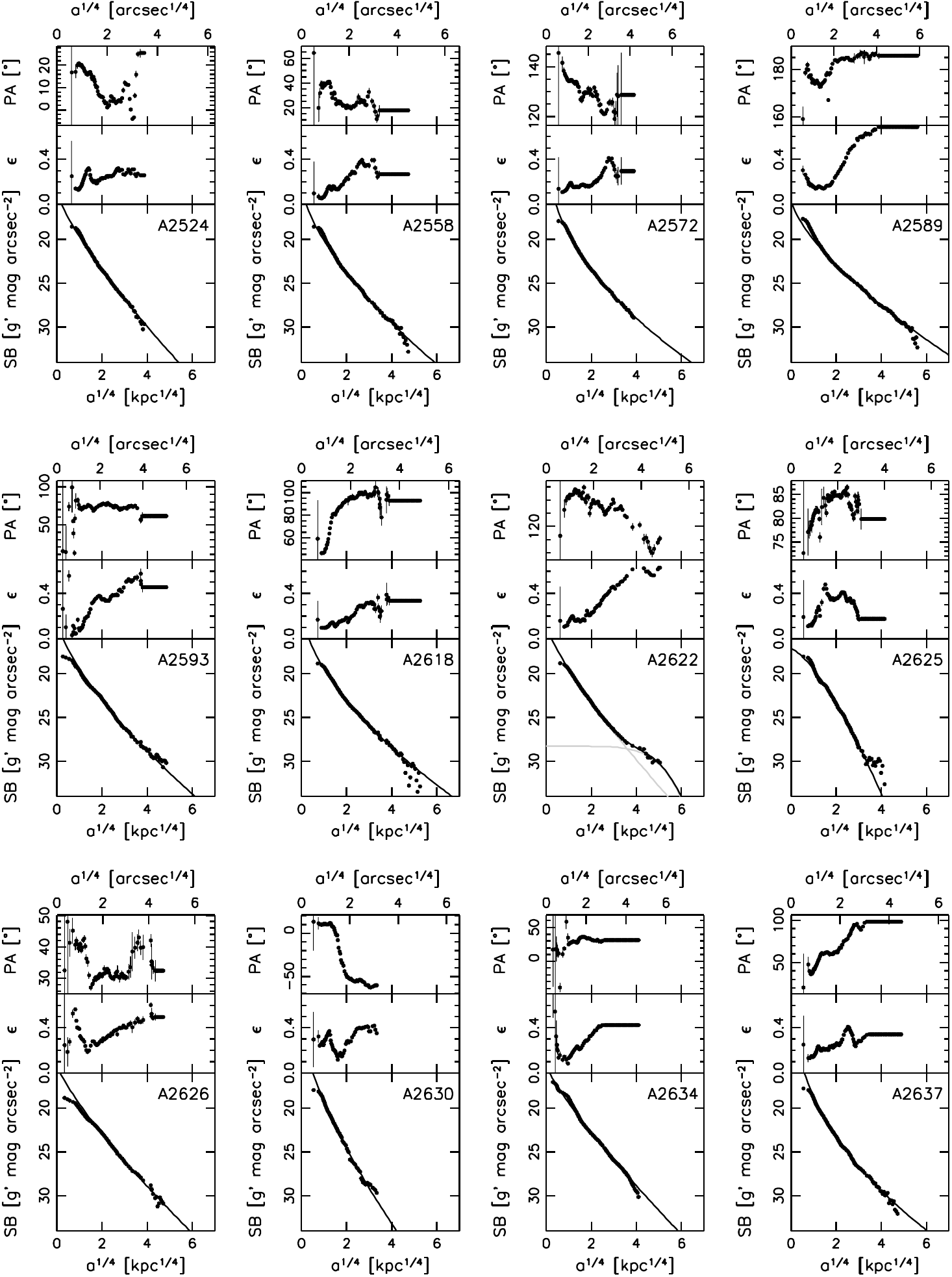}\\~\\
	\textbf{Figure \ref*{fig:sersicfits01}} \textit{(continued)}
\end{figure*}

\clearpage

\begin{figure*}
	\centering
	\includegraphics[width=0.915\linewidth]{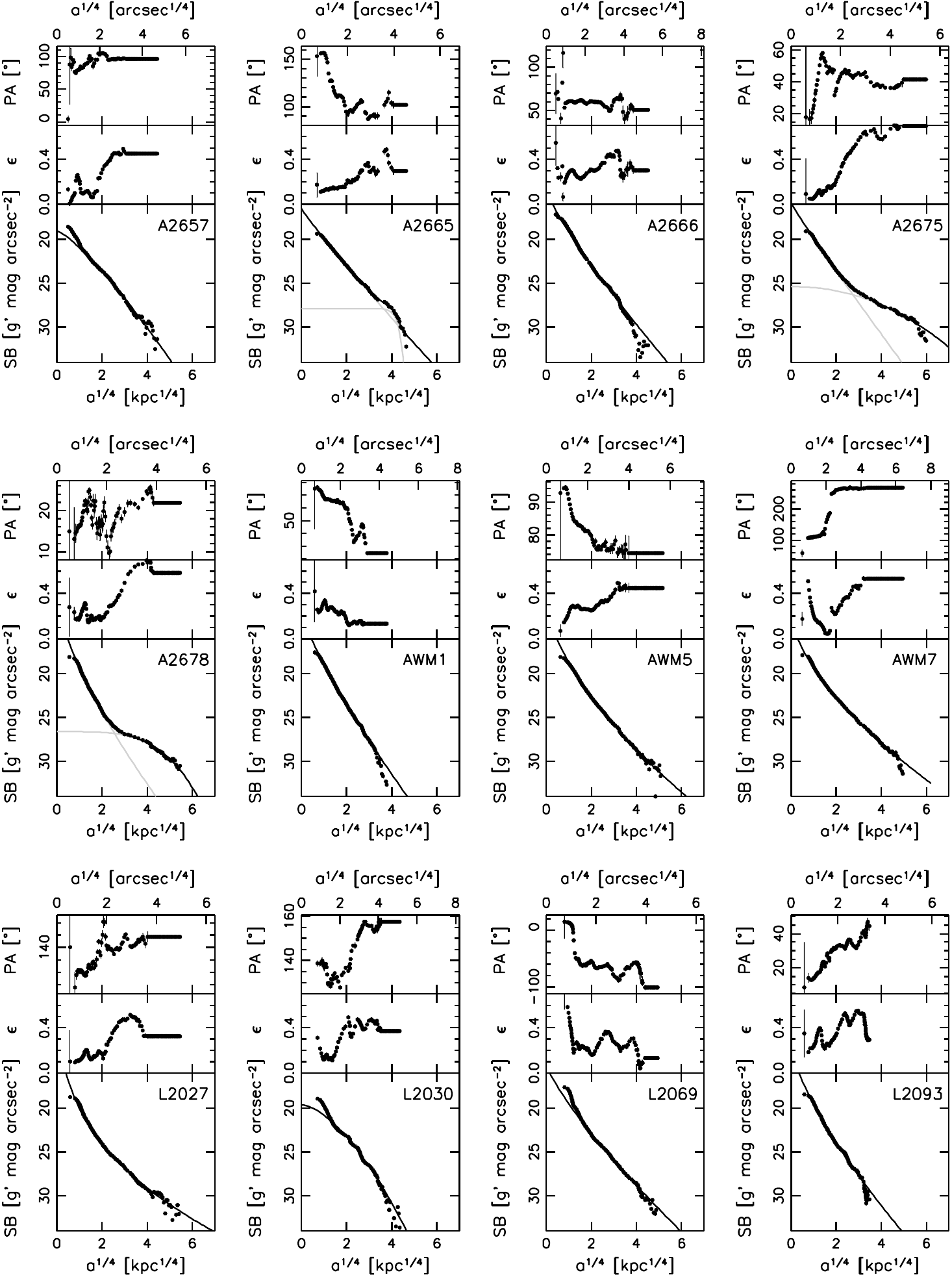}\\~\\
	\textbf{Figure \ref*{fig:sersicfits01}} \textit{(continued)}
\end{figure*}

\clearpage

\begin{figure*}
	\centering
	\includegraphics[width=0.915\linewidth]{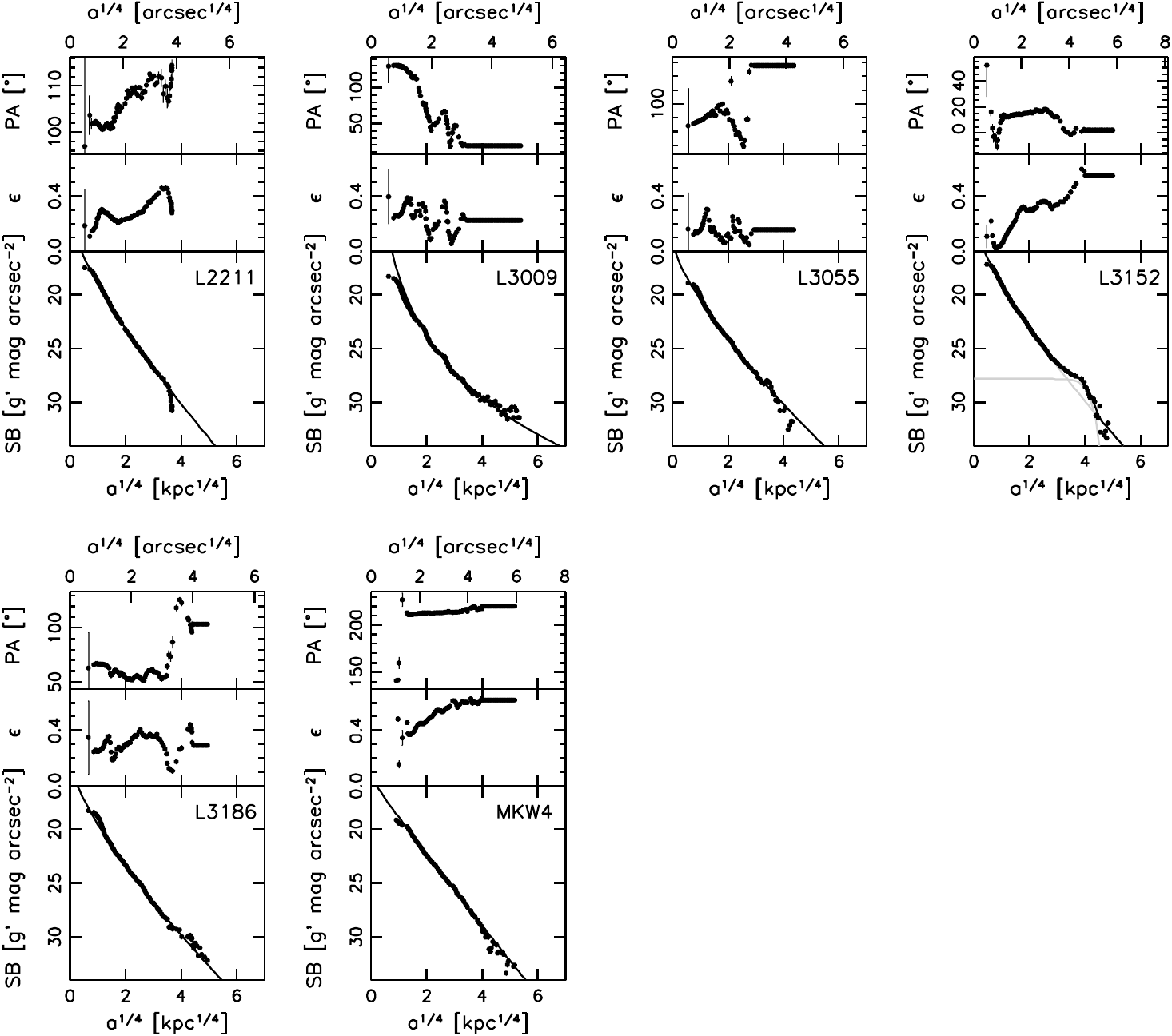}\\~\\
	\textbf{Figure \ref*{fig:sersicfits01}} \textit{(continued)}
\end{figure*}

\section{Image Cutouts}\label{sec:screenshots}

\begin{figure*}
	\centering
	\includegraphics[width=\linewidth]{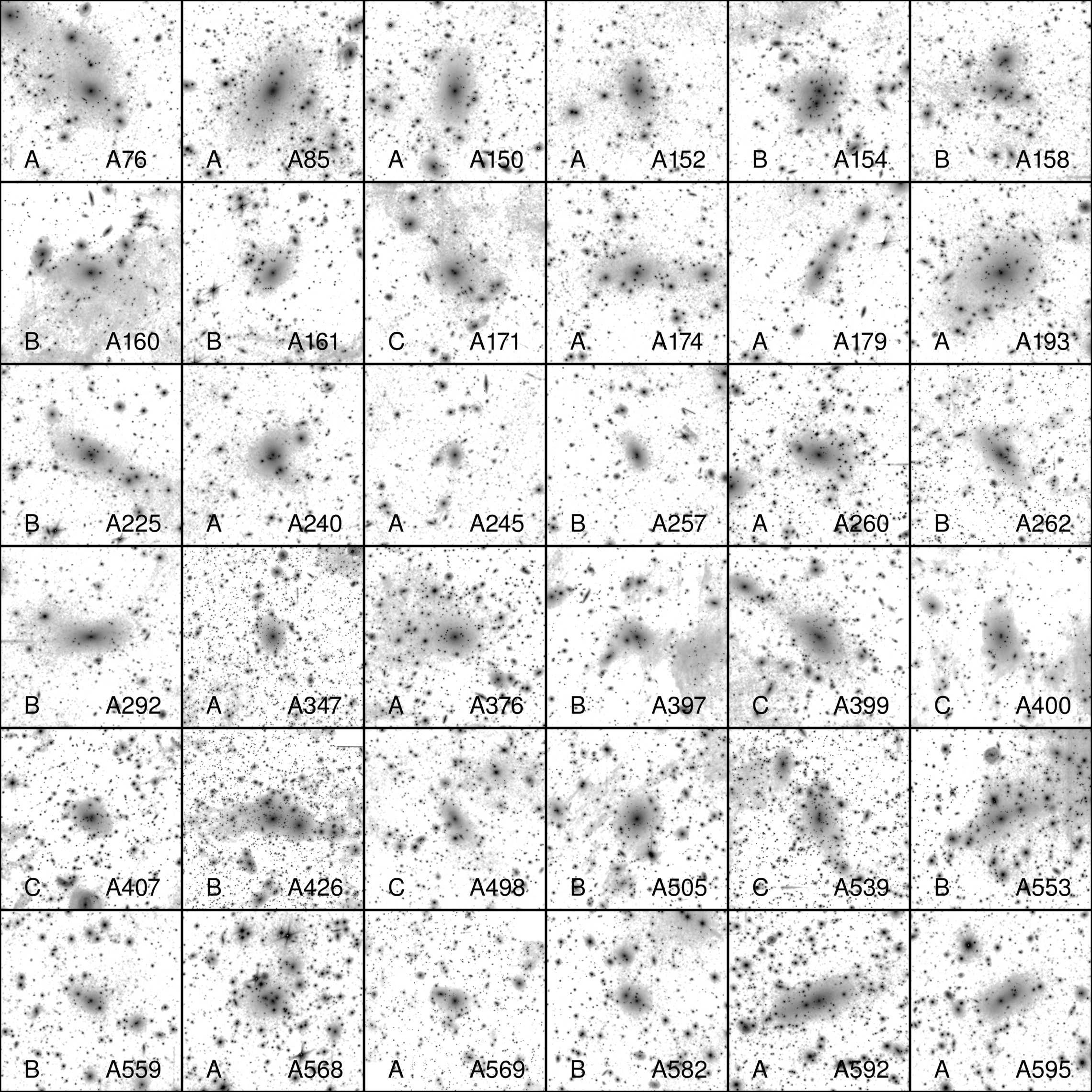}\\~\\~\\
	\includegraphics[width=\linewidth]{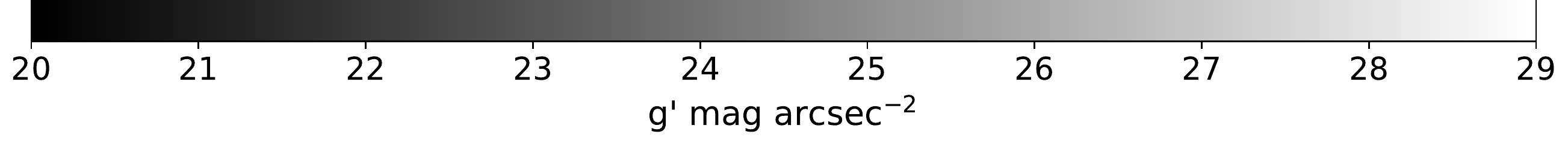}
	\caption{Image cutouts, centered on all BCGs that are analyzed in this study. The side length of each box is 750 kpc. North is up and east is left.	\label{fig:screenshots1}}
\end{figure*}

Figure \ref{fig:screenshots1} shows our observations, centered on the BCGs. They are all rescaled to the same physical size.

\clearpage

\clearpage

\begin{figure*}
	\centering
	\includegraphics[width=\linewidth]{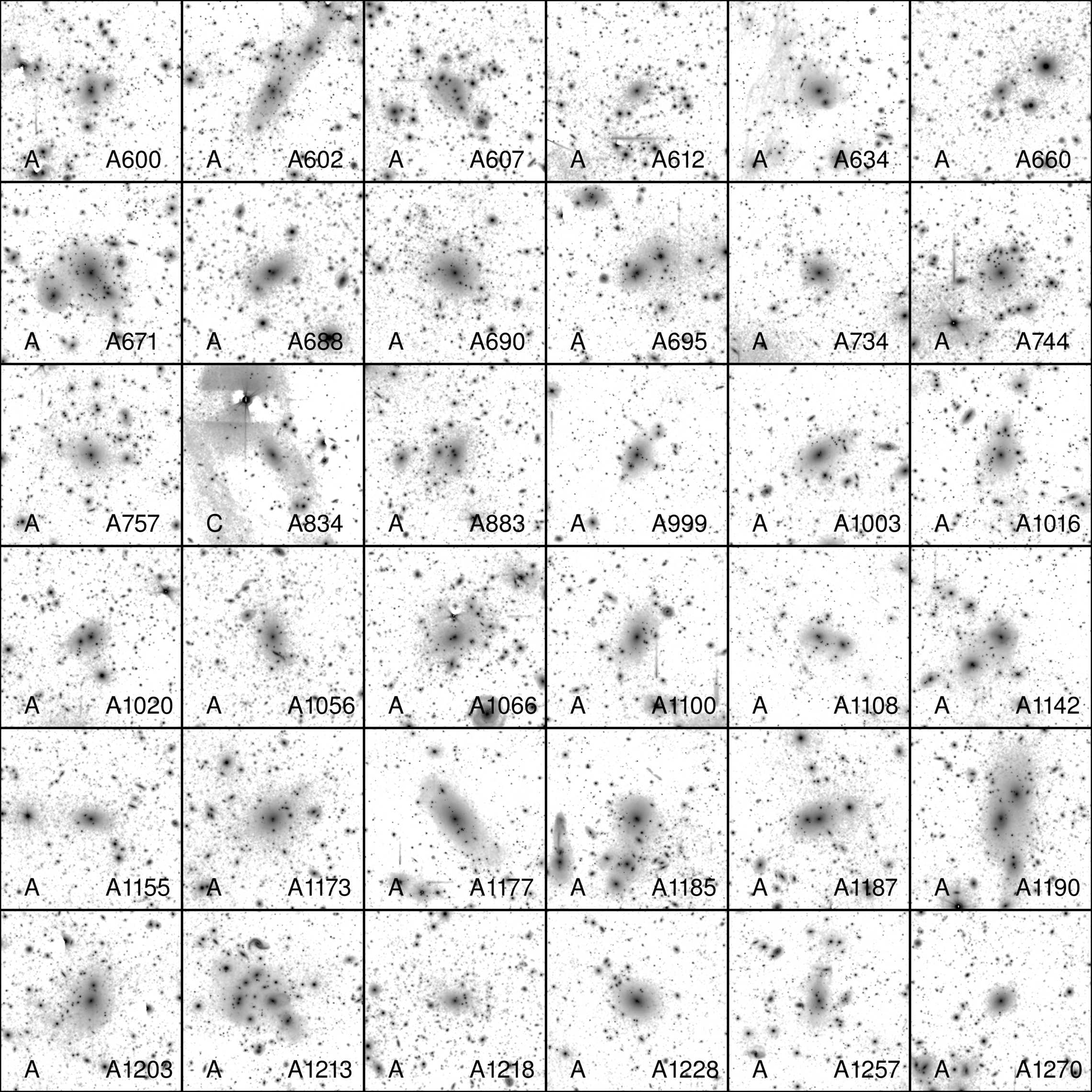}\\~\\~\\
	\includegraphics[width=\linewidth]{fig19f.pdf}\\
	\textbf{Figure \ref*{fig:screenshots1}} \textit{(continued)}
\end{figure*}

\clearpage

\begin{figure*}
	\centering
	\includegraphics[width=\linewidth]{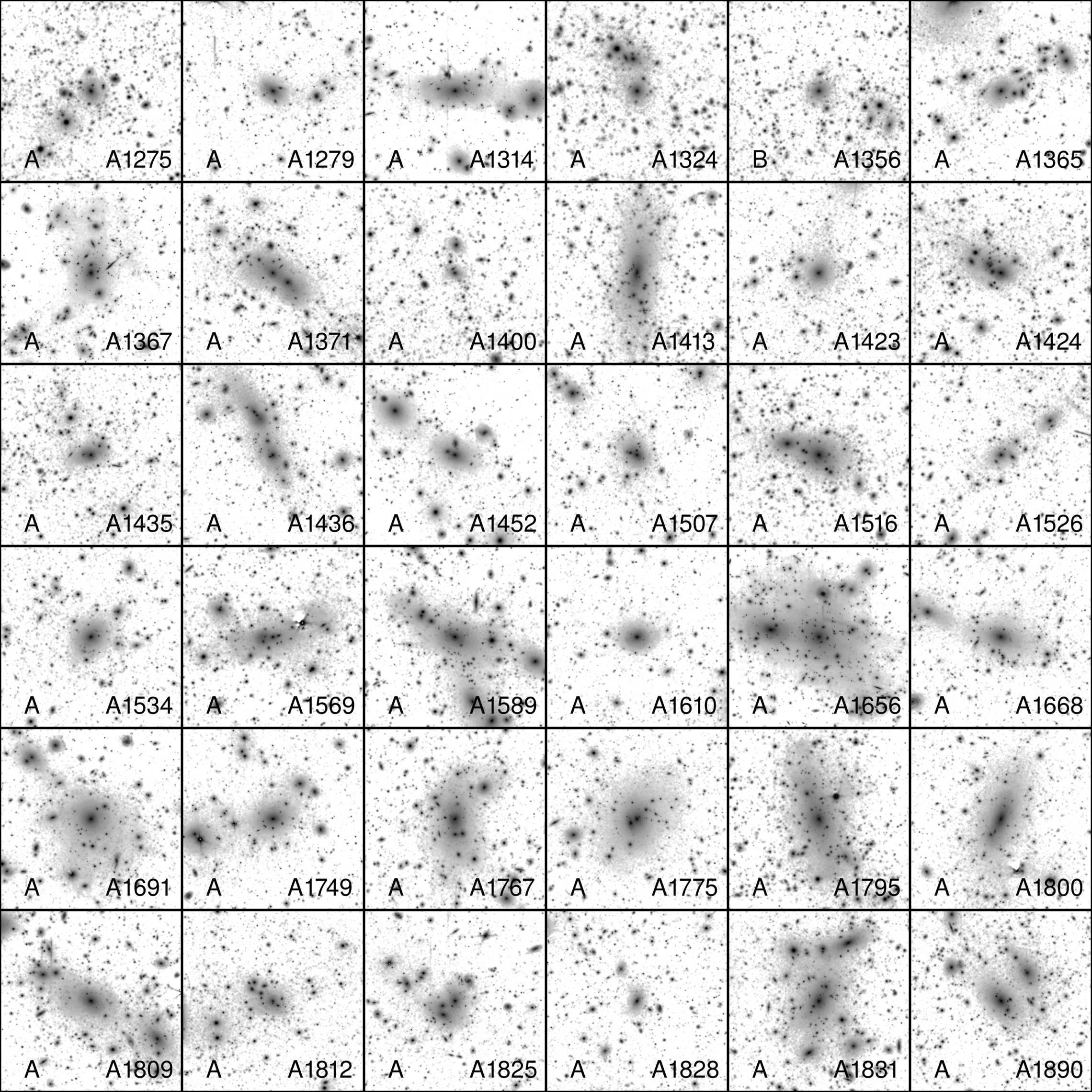}\\~\\~\\
	\includegraphics[width=\linewidth]{fig19f.pdf}\\
	\textbf{Figure \ref*{fig:screenshots1}} \textit{(continued)}
\end{figure*}

\clearpage

\begin{figure*}
	\centering
	\includegraphics[width=\linewidth]{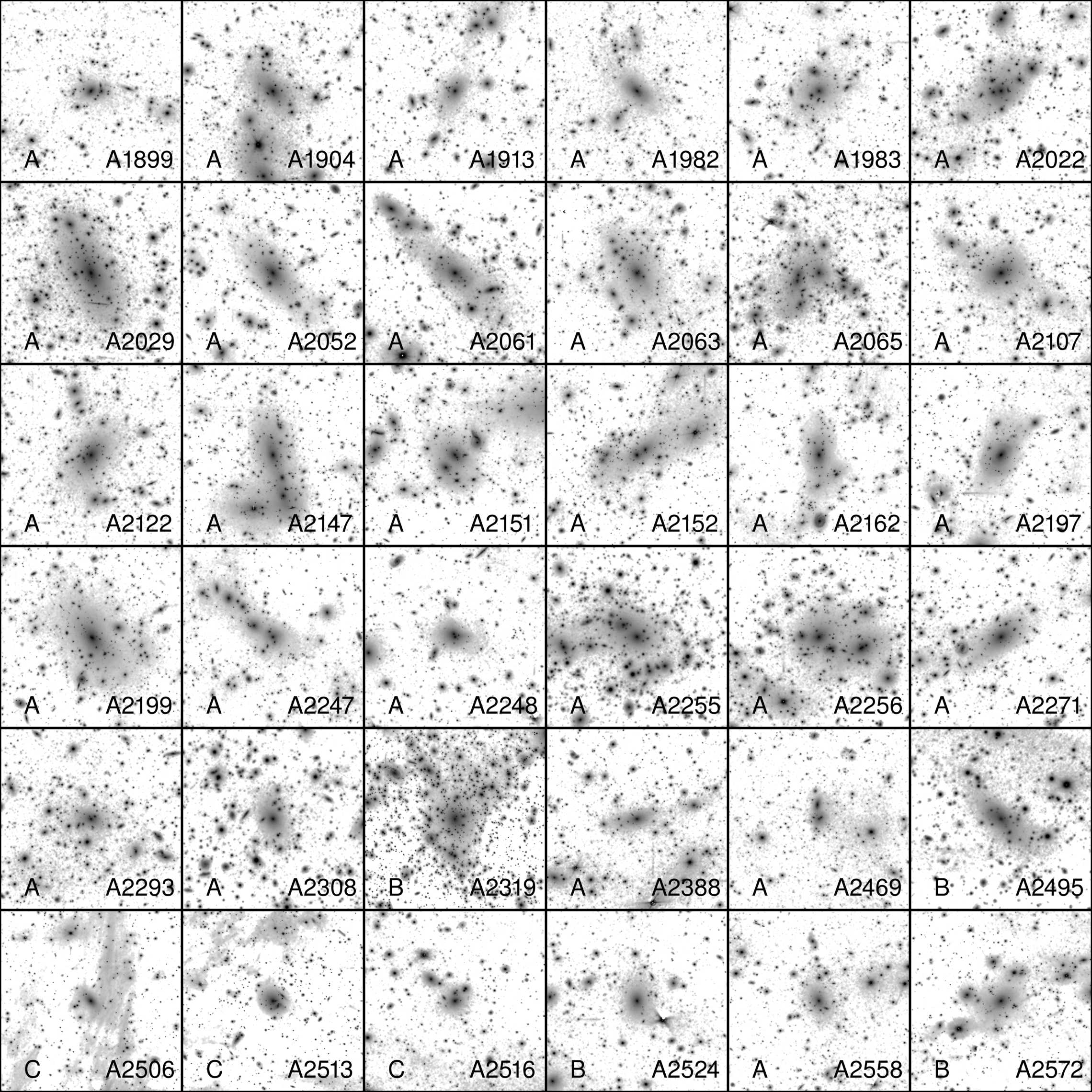}\\~\\~\\
	\includegraphics[width=\linewidth]{fig19f.pdf}\\
	\textbf{Figure \ref*{fig:screenshots1}} \textit{(continued)}
\end{figure*}

\clearpage

\begin{figure*}
	\centering
	\includegraphics[width=\linewidth]{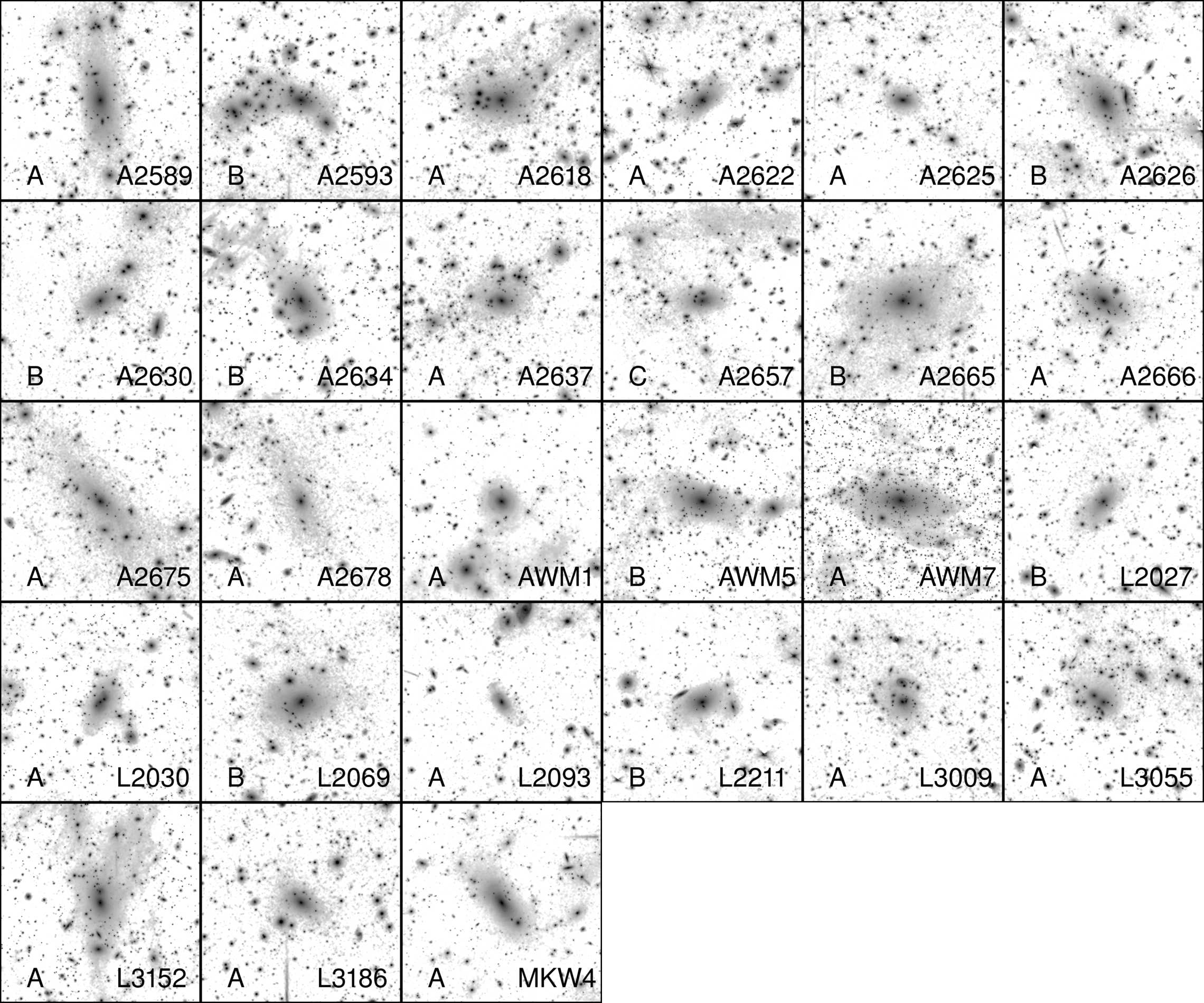}\\~\\~\\
	\includegraphics[width=\linewidth]{fig19f.pdf}\\
	\textbf{Figure \ref*{fig:screenshots1}} \textit{(continued)}
\end{figure*}

\section{Structural Parameters of the BCG+ICLs} \label{sec:apclusterparams}

Table \ref{tab:profileparams} lists the structural parameters of the BCG+ICLs, that is, the best-fit S\'ersic parameters, and parameters, which are derived by direct integration of the light profiles.

\clearpage

\begin{longrotatetable}
	\begin{deluxetable}{lccccccccccl}
		\tablewidth{0pt}
		\tabletypesize{\scriptsize}
		
		\tablecaption{Structural parameters of the BCG+ICLs. \label{tab:profileparams}}
		\tablehead{
			\colhead{} & \multicolumn{3}{c}{1st Semimajor Axis S\'ersic Component} & \multicolumn{3}{c}{2nd Semimajor Axis S\'ersic Component} & \colhead{} & \multicolumn{3}{c}{Parameters from 2D Profile Integration} & \\
			\cline{2-4} \cline{5-7} \cline{9-11} & \\
			\colhead{Cluster} & \colhead{$n_1$} & \colhead{$r_{\rm e,1}$} & \colhead{SB$_{\rm e,1}$} & \colhead{$n_2$} & \colhead{$r_{\rm e,2}$} & \colhead{SB$_{\rm e,2}$} & \colhead{$S_2 / (S_1+S_2)$} & \colhead{$r_{\rm e}$} & \colhead{SB$_{\rm e}$} & \colhead{$M_{\rm tot}$} & Feature \\
			\colhead{} & \colhead{} & \colhead{(kpc)} & \colhead{($g'$ mag \arcsec$^{-2}$)} & \colhead{} & \colhead{(kpc)} & \colhead{($g'$ mag $\arcsec^{-2}$)} & \colhead{} & \colhead{(kpc)} & \colhead{($g'$ mag \arcsec$^{-2}$)} & \colhead{($g'$ mag)} & \\
		}
		\colnumbers
		\startdata
		A76   &  10.24 $\pm$   1.14   &    $   79.43^{+   29.66}_{  -23.16}$  &   26.15 $\pm$  0.63    &      0.46 $\pm$  0.08 &  $  249.7^ {+   6.95}_{  -6.81}$  &  27.66 $\pm$   0.10 & 0.38 $\pm$ 0.06 &    147.0  $\pm$    20.37 & 26.51 $\pm$  0.06 & -24.14 $\pm$   0.07 & d \\ 
		A85   &   1.26 $\pm$   0.29   &    $   15.81^{+    0.87}_{   -0.84}$  &   22.43 $\pm$  0.18    &      4.21 $\pm$  0.70 &  $  208.1^ {+  10.48}_{ -10.10}$  &  26.74 $\pm$   0.10 & 0.83 $\pm$ 0.04 &    139.1  $\pm$    32.81 & 25.84 $\pm$  0.40 & -24.79 $\pm$   0.10 & \ldots \\ 
		A150  &   5.06 $\pm$   0.25   &    $   84.98^{+    4.48}_{   -4.31}$  &   25.61 $\pm$  0.11    &             \ldots        &           \ldots             &      \ldots             &  \ldots     &     52.19 $\pm$     5.66 & 24.57 $\pm$  0.18 & -23.81 $\pm$   0.06 & cd \\ 
		A152  &   3.13 $\pm$   0.09   &    $   56.51^{+    1.14}_{   -1.13}$  &   24.97 $\pm$  0.05    &             \ldots        &           \ldots             &      \ldots             &  \ldots     &     44.02 $\pm$     2.15 & 24.50 $\pm$  0.06 & -23.55 $\pm$   0.03 & \ldots \\ 
		A154  &   4.28 $\pm$   0.18   &    $   51.08^{+    1.38}_{   -1.35}$  &   24.59 $\pm$  0.06    &             \ldots        &           \ldots             &      \ldots             &  \ldots     &     43.09 $\pm$     2.93 & 24.24 $\pm$  0.10 & -24.01 $\pm$   0.04 & ad \\ 
		A158  &   3.90 $\pm$   2.39   &    $   19.24^{+   44.43}_{  -15.78}$  &   23.78 $\pm$  1.10    &      7.95 $\pm$ 34.21 &  $ 3980^  {+7.0E5}_{-2.9E4}$  &  32.65 $\pm$  19.64 & 0.84 $\pm$ 0.26 &   1601     $\pm$  1409     & 29.74 $\pm$  2.32 & -24.60 $\pm$   0.60 & \ldots \\ 
		A160  &   4.96 $\pm$   0.31   &    $   46.13^{+    5.73}_{   -5.24}$  &   25.08 $\pm$  0.22    &      0.17 $\pm$  0.02 &  $  167.4^ {+  91.10}_{ -64.60}$  &  26.58 $\pm$   0.05 & 0.54 $\pm$ 0.08 &    130.4  $\pm$     2.31 & 26.16 $\pm$  0.01 & -23.91 $\pm$   0.01 & d \\ 
		A161  &  11.37 $\pm$   1.55   &    $  863.5^ {+  436.9 }_{ -316.3 }$  &   30.28 $\pm$  0.83    &             \ldots        &           \ldots             &      \ldots             &  \ldots     &    406.5  $\pm$   277.3  & 28.25 $\pm$  1.53 & -24.43 $\pm$   0.36 & c \\ 
		A171  &   5.98 $\pm$   0.46   &    $   21.39^{+    1.64}_{   -1.55}$  &   23.36 $\pm$  0.15    &      0.27 $\pm$  0.02 &  $  236.3^ {+   3.06}_{  -3.03}$  &  26.76 $\pm$   0.04 & 0.52 $\pm$ 0.02 &    103.2  $\pm$     1.33 & 26.01 $\pm$  0.03 & -24.27 $\pm$   0.01 & \ldots \\ 
		A174  &   9.89 $\pm$   1.11   &    $  113.9^ {+   45.75}_{  -35.14}$  &   26.97 $\pm$  0.66    &      0.38 $\pm$  0.08 &  $  264.6^ {+   8.22}_{  -8.03}$  &  27.85 $\pm$   0.11 & 0.30 $\pm$ 0.05 &     86.95 $\pm$    13.94 & 26.06 $\pm$  0.10 & -23.85 $\pm$   0.06 & \ldots \\ 
		A179  &   8.66 $\pm$   0.88   &    $  100.9^ {+   34.47}_{  -27.43}$  &   26.63 $\pm$  0.57    &      0.53 $\pm$  0.17 &  $  382.3^ {+  28.75}_{ -27.21}$  &  28.46 $\pm$   0.13 & 0.37 $\pm$ 0.06 &    141.8  $\pm$    26.63 & 26.77 $\pm$  0.16 & -23.64 $\pm$   0.07 & cd \\ 
		A193  &   7.79 $\pm$   0.98   &    $  243.1^ {+   41.03}_{  -36.42}$  &   27.38 $\pm$  0.33    &             \ldots        &           \ldots             &      \ldots             &  \ldots     &    181.1  $\pm$    67.92 & 26.55 $\pm$  0.64 & -24.67 $\pm$   0.18 & d \\ 
		A225  &  10.54 $\pm$   1.37   &    $  284.1^ {+   91.64}_{  -73.76}$  &   28.14 $\pm$  0.58    &             \ldots        &           \ldots             &      \ldots             &  \ldots     &     98.82 $\pm$    42.02 & 25.79 $\pm$  0.79 & -24.04 $\pm$   0.18 & \ldots \\ 
		A240  &   4.22 $\pm$   0.16   &    $   53.47^{+    1.47}_{   -1.44}$  &   24.89 $\pm$  0.06    &             \ldots        &           \ldots             &      \ldots             &  \ldots     &     40.36 $\pm$     2.35 & 24.21 $\pm$  0.10 & -23.74 $\pm$   0.04 & \ldots \\ 
		A245  &   5.05 $\pm$   0.53   &    $   16.09^{+    0.81}_{   -0.78}$  &   23.78 $\pm$  0.10    &             \ldots        &           \ldots             &      \ldots             &  \ldots     &     11.61 $\pm$     0.50 & 23.16 $\pm$  0.09 & -22.43 $\pm$   0.02 & \ldots \\ 
		A257  &   2.92 $\pm$   0.16   &    $   21.81^{+    0.51}_{   -0.50}$  &   23.33 $\pm$  0.05    &             \ldots        &           \ldots             &      \ldots             &  \ldots     &     17.46 $\pm$     0.33 & 22.88 $\pm$  0.03 & -23.02 $\pm$   0.01 & \ldots \\ 
		A260  &   7.58 $\pm$   0.53   &    $  115.0^ {+   13.55}_{  -12.45}$  &   26.47 $\pm$  0.23    &             \ldots        &           \ldots             &      \ldots             &  \ldots     &     53.07 $\pm$    10.64 & 24.87 $\pm$  0.42 & -23.60 $\pm$   0.10 & \ldots \\ 
		A262  &   2.96 $\pm$   0.11   &    $   54.26^{+    1.32}_{   -1.29}$  &   25.02 $\pm$  0.06    &             \ldots        &           \ldots             &      \ldots             &  \ldots     &     32.70 $\pm$     1.18 & 24.11 $\pm$  0.06 & -22.89 $\pm$   0.02 & d \\ 
		A292  &   4.57 $\pm$   0.19   &    $   81.07^{+    2.85}_{   -2.78}$  &   24.92 $\pm$  0.08    &             \ldots        &           \ldots             &      \ldots             &  \ldots     &     66.02 $\pm$     6.35 & 24.51 $\pm$  0.13 & -24.01 $\pm$   0.05 & \ldots \\ 
		A347  &   7.54 $\pm$   0.54   &    $   36.97^{+    3.37}_{   -3.15}$  &   25.26 $\pm$  0.19    &             \ldots        &           \ldots             &      \ldots             &  \ldots     &     27.57 $\pm$     3.94 & 24.51 $\pm$  0.26 & -22.60 $\pm$   0.08 & d \\ 
		A376  &  11.17 $\pm$   1.00   &    $  446.3^ {+  116.5 }_{  -97.41}$  &   29.10 $\pm$  0.47    &             \ldots        &           \ldots             &      \ldots             &  \ldots     &    244.3  $\pm$   147.5  & 27.48 $\pm$  1.34 & -24.24 $\pm$   0.29 & \ldots \\ 
		A397  &   5.66 $\pm$   0.35   &    $   55.34^{+    3.93}_{   -3.74}$  &   25.14 $\pm$  0.15    &             \ldots        &           \ldots             &      \ldots             &  \ldots     &     38.04 $\pm$     4.09 & 24.23 $\pm$  0.24 & -23.42 $\pm$   0.06 & c \\ 
		A399  &   2.85 $\pm$   0.11   &    $   65.11^{+    1.72}_{   -1.68}$  &   24.20 $\pm$  0.06    &             \ldots        &           \ldots             &      \ldots             &  \ldots     &     58.36 $\pm$     1.51 & 23.91 $\pm$  0.05 & -24.39 $\pm$   0.02 & c \\ 
		A400  &   4.52 $\pm$   0.90   &    $    4.92^{+    1.46}_{   -1.20}$  &   21.94 $\pm$  0.51    &      1.05 $\pm$  0.09 &  $   65.23^{+   1.67}_{  -1.64}$  &  25.10 $\pm$   0.07 & 0.79 $\pm$ 0.02 &     46.56 $\pm$     0.39 & 24.68 $\pm$  0.01 & -23.20 $\pm$   0.01 & d \\ 
		A407  &   1.90 $\pm$   0.14   &    $   34.91^{+    1.28}_{   -1.25}$  &   23.45 $\pm$  0.09    &             \ldots        &           \ldots             &      \ldots             &  \ldots     &     29.50 $\pm$     0.22 & 23.03 $\pm$  0.02 & -23.77 $\pm$   0.01 & d \\ 
		A426  &   4.25 $\pm$   0.28   &    $   14.55^{+    1.32}_{   -1.23}$  &   22.42 $\pm$  0.17    &      0.44 $\pm$  0.04 &  $   97.04^{+   1.42}_{  -1.41}$  &  25.86 $\pm$   0.07 & 0.37 $\pm$ 0.03 &     30.46 $\pm$     0.30 & 23.75 $\pm$  0.02 & -23.64 $\pm$   0.00 & d \\ 
		A498  &   5.38 $\pm$   0.39   &    $   98.41^{+    9.35}_{   -8.73}$  &   25.39 $\pm$  0.19    &             \ldots        &           \ldots             &      \ldots             &  \ldots     &     55.21 $\pm$     6.92 & 24.29 $\pm$  0.21 & -24.12 $\pm$   0.06 & b \\ 
		A505  &   4.45 $\pm$   0.17   &    $   30.15^{+    0.66}_{   -0.65}$  &   23.43 $\pm$  0.05    &             \ldots        &           \ldots             &      \ldots             &  \ldots     &     27.45 $\pm$     0.90 & 23.12 $\pm$  0.06 & -24.00 $\pm$   0.02 & \ldots \\ 
		A539  &   4.44 $\pm$   0.72   &    $   45.65^{+   21.11}_{  -15.66}$  &   24.99 $\pm$  0.66    &      0.42 $\pm$  0.10 &  $   97.99^{+   2.44}_{  -2.39}$  &  25.93 $\pm$   0.16 & 0.37 $\pm$ 0.11 &     59.06 $\pm$     3.11 & 24.85 $\pm$  0.09 & -23.49 $\pm$   0.03 & d \\ 
		A553  &   2.66 $\pm$   0.38   &    $    7.14^{+    1.19}_{   -1.06}$  &   21.55 $\pm$  0.27    &      2.62 $\pm$  0.33 &  $  505.6^ {+  64.47}_{ -58.84}$  &  27.22 $\pm$   0.23 & 0.93 $\pm$ 0.01 &    405.4  $\pm$    47.01 & 26.73 $\pm$  0.06 & -25.35 $\pm$   0.09 & \ldots \\ 
		A559  &   2.73 $\pm$   0.59   &    $    3.34^{+    0.57}_{   -0.50}$  &   20.81 $\pm$  0.28    &      1.75 $\pm$  0.18 &  $   51.18^{+   2.53}_{  -2.44}$  &  25.08 $\pm$   0.11 & 0.75 $\pm$ 0.04 &     29.36 $\pm$     0.80 & 24.02 $\pm$  0.03 & -23.25 $\pm$   0.02 & \ldots \\ 
		A568  &  25.83 $\pm$  11.10   &    $  228.5^ {+  284.6 }_{ -145.7 }$  &   28.10 $\pm$  1.88    &             \ldots        &           \ldots             &      \ldots             &  \ldots     &    191.4  $\pm$   139.2  & 26.89 $\pm$  2.01 & -24.35 $\pm$   0.32 & \ldots \\ 
		A569  &   4.60 $\pm$   0.21   &    $    9.76^{+    0.29}_{   -0.29}$  &   22.59 $\pm$  0.06    &             \ldots        &           \ldots             &      \ldots             &  \ldots     &      9.95 $\pm$     0.20 & 22.49 $\pm$  0.04 & -22.26 $\pm$   0.01 & d \\ 
		A582  &   9.41 $\pm$   1.38   &    $   69.87^{+   12.36}_{  -10.91}$  &   26.14 $\pm$  0.35    &             \ldots        &           \ldots             &      \ldots             &  \ldots     &     52.96 $\pm$    18.66 & 25.51 $\pm$  1.02 & -23.56 $\pm$   0.14 & \ldots \\ 
		A592  &   2.44 $\pm$   0.13   &    $  122.2^ {+    5.81}_{   -5.61}$  &   25.21 $\pm$  0.10    &             \ldots        &           \ldots             &      \ldots             &  \ldots     &     87.88 $\pm$     2.98 & 24.40 $\pm$  0.06 & -24.31 $\pm$   0.02 & cd \\ 
		A595  &   8.26 $\pm$   1.46   &    $    8.71^{+    2.46}_{   -2.03}$  &   22.68 $\pm$  0.50    &      0.77 $\pm$  0.07 &  $   61.35^{+   1.08}_{  -1.07}$  &  25.34 $\pm$   0.06 & 0.55 $\pm$ 0.04 &     34.36 $\pm$     0.83 & 24.32 $\pm$  0.02 & -23.15 $\pm$   0.01 & a \\ 
		A600  &   3.93 $\pm$   0.20   &    $   27.38^{+    1.64}_{   -1.57}$  &   23.97 $\pm$  0.12    &      0.42 $\pm$  0.18 &  $  355.8^ {+  29.61}_{ -27.87}$  &  29.48 $\pm$   0.19 & 0.29 $\pm$ 0.04 &     59.45 $\pm$     8.92 & 25.71 $\pm$  0.53 & -23.57 $\pm$   0.06 & d \\ 
		A602  &   3.53 $\pm$   0.31   &    $   30.25^{+    6.15}_{   -5.34}$  &   24.66 $\pm$  0.31    &      1.14 $\pm$  0.16 &  $  391.2^ {+  20.67}_{ -19.88}$  &  27.70 $\pm$   0.11 & 0.71 $\pm$ 0.03 &    172.0  $\pm$    13.04 & 26.43 $\pm$  0.04 & -23.87 $\pm$   0.04 & c \\ 
		A607  &   2.67 $\pm$   0.28   &    $  126.0^ {+   14.45}_{  -13.30}$  &   26.43 $\pm$  0.22    &             \ldots        &           \ldots             &      \ldots             &  \ldots     &    121.7  $\pm$    16.37 & 26.41 $\pm$  0.45 & -23.86 $\pm$   0.08 & \ldots \\ 
		A612  &  20.24 $\pm$  14.22   &    $  199.4^ {+  389.8 }_{ -154.5 }$  &   28.98 $\pm$  2.60    &             \ldots        &           \ldots             &      \ldots             &  \ldots     &    106.7  $\pm$    78.96 & 26.81 $\pm$  2.01 & -23.11 $\pm$   0.31 & \ldots \\ 
		A634  &   7.14 $\pm$   0.57   &    $   22.81^{+    1.24}_{   -1.19}$  &   24.17 $\pm$  0.12    &             \ldots        &           \ldots             &      \ldots             &  \ldots     &     24.07 $\pm$     3.09 & 24.40 $\pm$  0.24 & -22.69 $\pm$   0.06 & b \\ 
		A671  &   3.57 $\pm$   0.16   &    $   44.18^{+    1.31}_{   -1.28}$  &   24.05 $\pm$  0.07    &             \ldots        &           \ldots             &      \ldots             &  \ldots     &     40.45 $\pm$     1.44 & 23.76 $\pm$  0.05 & -23.99 $\pm$   0.02 & \ldots \\ 
		A688  &   3.98 $\pm$   0.19   &    $   41.54^{+    0.83}_{   -0.82}$  &   23.87 $\pm$  0.04    &             \ldots        &           \ldots             &      \ldots             &  \ldots     &     39.34 $\pm$     1.61 & 23.73 $\pm$  0.09 & -24.27 $\pm$   0.03 & \ldots \\ 
		A690  &   3.88 $\pm$   0.19   &    $   53.45^{+    1.73}_{   -1.69}$  &   24.59 $\pm$  0.07    &             \ldots        &           \ldots             &      \ldots             &  \ldots     &     41.56 $\pm$     2.17 & 24.04 $\pm$  0.08 & -24.09 $\pm$   0.03 & \ldots \\ 
		A695  &   3.92 $\pm$   0.25   &    $   20.80^{+    0.75}_{   -0.73}$  &   23.11 $\pm$  0.08    &             \ldots        &           \ldots             &      \ldots             &  \ldots     &     20.48 $\pm$     0.69 & 23.02 $\pm$  0.08 & -23.40 $\pm$   0.02 & d \\ 
		A734  &   3.52 $\pm$   0.18   &    $   24.94^{+    0.66}_{   -0.65}$  &   23.80 $\pm$  0.06    &             \ldots        &           \ldots             &      \ldots             &  \ldots     &     20.49 $\pm$     0.65 & 23.36 $\pm$  0.07 & -23.23 $\pm$   0.02 & \ldots \\ 
		A744  &  17.44 $\pm$   2.28   &    $  142.8^ {+   32.98}_{  -28.10}$  &   27.39 $\pm$  0.45    &             \ldots        &           \ldots             &      \ldots             &  \ldots     &    114.8  $\pm$    66.44 & 26.53 $\pm$  1.34 & -24.03 $\pm$   0.25 & \ldots \\ 
		A757  &   9.20 $\pm$   1.09   &    $   38.77^{+   10.41}_{   -8.66}$  &   25.53 $\pm$  0.48    &      0.46 $\pm$  0.16 &  $  227.8^ {+  14.11}_{ -13.49}$  &  28.93 $\pm$   0.15 & 0.26 $\pm$ 0.05 &     62.10 $\pm$    12.01 & 26.30 $\pm$  0.44 & -23.04 $\pm$   0.07 & \ldots \\ 
		A834  &   3.70 $\pm$   0.32   &    $   36.13^{+    5.80}_{   -5.17}$  &   24.14 $\pm$  0.27    &      0.38 $\pm$  0.08 &  $  199.8^ {+   5.89}_{  -5.76}$  &  26.75 $\pm$   0.09 & 0.48 $\pm$ 0.04 &     98.28 $\pm$     1.96 & 25.81 $\pm$  0.07 & -23.94 $\pm$   0.01 & \ldots \\ 
		A883  &  18.08 $\pm$   3.28   &    $ 2.1E3^{+ 2.0E3}_{-1.1E3}$  &   32.40 $\pm$  1.45    &             \ldots        &           \ldots             &      \ldots             &  \ldots     &   9.8E2 $\pm$  8.9E2 & 29.31 $\pm$  2.90 & -24.26 $\pm$   0.54 & \ldots \\ 
		A999  &   8.02 $\pm$   0.66   &    $   32.61^{+    2.37}_{   -2.24}$  &   24.86 $\pm$  0.15    &             \ldots        &           \ldots             &      \ldots             &  \ldots     &     20.98 $\pm$     2.34 & 23.90 $\pm$  0.16 & -22.71 $\pm$   0.05 & c \\ 
		A1003 &   4.31 $\pm$   0.33   &    $   36.78^{+    1.82}_{   -1.76}$  &   24.54 $\pm$  0.11    &             \ldots        &           \ldots             &      \ldots             &  \ldots     &     25.45 $\pm$     1.36 & 23.63 $\pm$  0.11 & -23.18 $\pm$   0.03 & \ldots \\ 
		A1016 &  21.23 $\pm$   4.11   &    $  163.0^ {+   90.82}_{  -63.95}$  &   28.56 $\pm$  0.97    &             \ldots        &           \ldots             &      \ldots             &  \ldots     &     79.53 $\pm$    53.95 & 26.50 $\pm$  1.72 & -22.93 $\pm$   0.29 & c \\ 
		A1020 &   5.62 $\pm$   0.53   &    $   24.36^{+    1.17}_{   -1.13}$  &   24.11 $\pm$  0.10    &             \ldots        &           \ldots             &      \ldots             &  \ldots     &     19.58 $\pm$     1.57 & 23.69 $\pm$  0.13 & -23.14 $\pm$   0.04 & \ldots \\ 
		A1056 &   3.55 $\pm$   0.20   &    $   65.15^{+    2.44}_{   -2.38}$  &   25.38 $\pm$  0.08    &             \ldots        &           \ldots             &      \ldots             &  \ldots     &     43.42 $\pm$     2.53 & 24.50 $\pm$  0.08 & -23.64 $\pm$   0.04 & \ldots \\ 
		A1066 &   2.20 $\pm$   0.22   &    $    7.67^{+    0.90}_{   -0.82}$  &   21.82 $\pm$  0.18    &      1.91 $\pm$  0.30 &  $  125.9^ {+   6.42}_{  -6.19}$  &  26.62 $\pm$   0.11 & 0.73 $\pm$ 0.03 &     72.69 $\pm$     6.01 & 25.71 $\pm$  0.15 & -23.68 $\pm$   0.04 & d \\ 
		A1100 &   6.87 $\pm$   0.27   &    $   53.62^{+    2.03}_{   -1.98}$  &   25.34 $\pm$  0.08    &             \ldots        &           \ldots             &      \ldots             &  \ldots     &     33.82 $\pm$     4.71 & 24.32 $\pm$  0.26 & -23.18 $\pm$   0.07 & c \\ 
		A1108 &   8.45 $\pm$   1.01   &    $   28.66^{+    3.57}_{   -3.27}$  &   25.50 $\pm$  0.26    &             \ldots        &           \ldots             &      \ldots             &  \ldots     &     24.12 $\pm$     6.22 & 25.06 $\pm$  0.52 & -21.91 $\pm$   0.11 & cd \\ 
		A1142 &   6.10 $\pm$   0.56   &    $   23.57^{+    1.38}_{   -1.32}$  &   23.96 $\pm$  0.13    &             \ldots        &           \ldots             &      \ldots             &  \ldots     &     23.30 $\pm$     1.83 & 23.76 $\pm$  0.10 & -22.95 $\pm$   0.04 & bcd \\ 
		A1155 &  10.04 $\pm$   1.02   &    $   61.36^{+    5.66}_{   -5.29}$  &   25.80 $\pm$  0.19    &             \ldots        &           \ldots             &      \ldots             &  \ldots     &     34.94 $\pm$     8.40 & 24.52 $\pm$  0.54 & -23.27 $\pm$   0.11 & \ldots \\ 
		A1173 &   5.52 $\pm$   0.49   &    $   83.91^{+   19.15}_{  -16.35}$  &   26.04 $\pm$  0.39    &      0.35 $\pm$  0.12 &  $  443.2^ {+  27.33}_{ -26.12}$  &  28.68 $\pm$   0.16 & 0.39 $\pm$ 0.05 &    174.7  $\pm$    16.46 & 27.12 $\pm$  0.09 & -24.26 $\pm$   0.04 & d \\ 
		A1177 &   7.46 $\pm$   1.61   &    $   28.62^{+   29.08}_{  -16.40}$  &   24.61 $\pm$  1.31    &      1.38 $\pm$  0.24 &  $   72.57^{+   3.32}_{  -3.21}$  &  25.51 $\pm$   0.10 & 0.53 $\pm$ 0.14 &     40.15 $\pm$     1.45 & 24.21 $\pm$  0.03 & -23.41 $\pm$   0.02 & bc \\ 
		A1185 &   6.36 $\pm$   1.01   &    $   24.46^{+    2.18}_{   -2.04}$  &   23.69 $\pm$  0.19    &             \ldots        &           \ldots             &      \ldots             &  \ldots     &     29.76 $\pm$     2.89 & 24.31 $\pm$  0.17 & -23.25 $\pm$   0.05 & d \\ 
		A1187 &  12.79 $\pm$   2.03   &    $  149.0^ {+   41.06}_{  -34.02}$  &   27.12 $\pm$  0.52    &             \ldots        &           \ldots             &      \ldots             &  \ldots     &     66.69 $\pm$    25.34 & 25.48 $\pm$  0.90 & -23.95 $\pm$   0.16 & a \\ 
		A1190 &   2.09 $\pm$   0.24   &    $   25.66^{+    3.29}_{   -3.00}$  &   23.26 $\pm$  0.21    &      0.66 $\pm$  0.09 &  $  188.7^ {+   7.98}_{  -7.74}$  &  26.43 $\pm$   0.12 & 0.54 $\pm$ 0.04 &     71.25 $\pm$     1.08 & 24.96 $\pm$  0.02 & -24.13 $\pm$   0.01 & a \\ 
		A1203 &   7.09 $\pm$   0.51   &    $   41.24^{+    6.46}_{   -5.78}$  &   24.71 $\pm$  0.29    &      0.63 $\pm$  0.11 &  $  309.7^ {+  11.10}_{ -10.81}$  &  28.31 $\pm$   0.09 & 0.37 $\pm$ 0.04 &    102.5  $\pm$     9.87 & 26.22 $\pm$  0.13 & -24.11 $\pm$   0.04 & \ldots \\ 
		A1213 &   1.60 $\pm$   0.17   &    $    6.38^{+    0.42}_{   -0.40}$  &   21.23 $\pm$  0.08    &      3.20 $\pm$  1.36 &  $  106.8^ {+  14.77}_{ -13.39}$  &  27.30 $\pm$   0.28 & 0.60 $\pm$ 0.08 &     31.21 $\pm$     5.41 & 24.94 $\pm$  0.24 & -23.03 $\pm$   0.06 & c \\ 
		A1218 &  10.98 $\pm$   1.69   &    $  149.7^ {+   86.54}_{  -60.26}$  &   28.26 $\pm$  0.92    &      0.35 $\pm$  0.17 &  $  407.0^ {+  37.39}_{ -34.98}$  &  29.66 $\pm$   0.22 & 0.28 $\pm$ 0.09 &    165.7  $\pm$    58.25 & 27.89 $\pm$  0.61 & -23.25 $\pm$   0.15 & \ldots \\ 
		A1228 &   3.49 $\pm$   0.11   &    $   21.41^{+    0.37}_{   -0.37}$  &   23.18 $\pm$  0.04    &             \ldots        &           \ldots             &      \ldots             &  \ldots     &     21.59 $\pm$     0.40 & 23.15 $\pm$  0.04 & -23.23 $\pm$   0.01 & cd \\ 
		A1257 &   2.09 $\pm$   0.32   &    $    2.39^{+    0.22}_{   -0.20}$  &   20.77 $\pm$  0.16    &      1.76 $\pm$  0.16 &  $   52.39^{+   1.63}_{  -1.60}$  &  25.56 $\pm$   0.07 & 0.77 $\pm$ 0.02 &     26.73 $\pm$     0.73 & 24.39 $\pm$  0.03 & -22.35 $\pm$   0.02 & c \\ 
		A1270 &   2.51 $\pm$   0.10   &    $   18.01^{+    0.42}_{   -0.41}$  &   23.28 $\pm$  0.05    &             \ldots        &           \ldots             &      \ldots             &  \ldots     &     15.03 $\pm$     0.18 & 22.94 $\pm$  0.02 & -22.89 $\pm$   0.01 & \ldots \\ 
		A1275 &  14.93 $\pm$   2.60   &    $    9.17^{+    1.60}_{   -1.41}$  &   22.32 $\pm$  0.36    &             \ldots        &           \ldots             &      \ldots             &  \ldots     &     18.98 $\pm$     4.11 & 23.89 $\pm$  0.49 & -22.76 $\pm$   0.10 & \ldots \\ 
		A1279 &   2.57 $\pm$   0.10   &    $   20.53^{+    0.42}_{   -0.42}$  &   23.55 $\pm$  0.04    &             \ldots        &           \ldots             &      \ldots             &  \ldots     &     16.44 $\pm$     0.27 & 23.16 $\pm$  0.03 & -22.70 $\pm$   0.01 & \ldots \\ 
		A1314 &   6.56 $\pm$   1.51   &    $    7.29^{+    2.44}_{   -1.95}$  &   22.05 $\pm$  0.58    &      1.18 $\pm$  0.07 &  $   71.91^{+   0.99}_{  -0.98}$  &  25.23 $\pm$   0.07 & 0.63 $\pm$ 0.05 &     37.27 $\pm$     0.52 & 24.10 $\pm$  0.02 & -23.29 $\pm$   0.01 & c \\ 
		A1324 &   6.12 $\pm$   0.72   &    $   19.46^{+    2.15}_{   -1.98}$  &   23.54 $\pm$  0.22    &      0.69 $\pm$  0.19 &  $  478.3^ {+  50.14}_{ -46.48}$  &  28.83 $\pm$   0.20 & 0.59 $\pm$ 0.04 &    232.0  $\pm$    33.94 & 27.96 $\pm$  0.18 & -24.38 $\pm$   0.07 & \ldots \\ 
		A1356 &   4.97 $\pm$   0.38   &    $   22.59^{+    0.81}_{   -0.79}$  &   24.30 $\pm$  0.08    &             \ldots        &           \ldots             &      \ldots             &  \ldots     &     20.84 $\pm$     1.48 & 24.15 $\pm$  0.14 & -22.68 $\pm$   0.04 & \ldots \\ 
		A1365 &   7.67 $\pm$   1.11   &    $   20.60^{+    4.61}_{   -3.95}$  &   23.93 $\pm$  0.42    &      0.96 $\pm$  0.18 &  $  347.5^ {+  23.21}_{ -22.10}$  &  28.67 $\pm$   0.13 & 0.50 $\pm$ 0.05 &    118.3  $\pm$    22.60 & 26.94 $\pm$  0.29 & -23.67 $\pm$   0.07 & d \\ 
		A1367 &   9.80 $\pm$   0.77   &    $   74.47^{+    9.17}_{   -8.40}$  &   26.12 $\pm$  0.25    &             \ldots        &           \ldots             &      \ldots             &  \ldots     &     52.23 $\pm$    15.99 & 25.32 $\pm$  0.66 & -23.32 $\pm$   0.13 & \ldots \\ 
		A1371 &   2.03 $\pm$   0.63   &    $    3.87^{+    0.73}_{   -0.64}$  &   21.24 $\pm$  0.30    &      2.12 $\pm$  0.21 &  $  104.7^ {+   4.49}_{  -4.35}$  &  25.62 $\pm$   0.09 & 0.89 $\pm$ 0.02 &     72.45 $\pm$     3.19 & 24.77 $\pm$  0.05 & -23.89 $\pm$   0.03 & a \\ 
		A1400 &  16.49 $\pm$   5.48   &    $   15.57^{+    2.30}_{   -2.07}$  &   24.50 $\pm$  0.30    &             \ldots        &           \ldots             &      \ldots             &  \ldots     &     20.56 $\pm$     6.68 & 24.98 $\pm$  0.75 & -21.67 $\pm$   0.15 & \ldots \\ 
		A1413 &  16.34 $\pm$   1.91   &    $ 7256^  {+ 4667    }_{-3141    }$  &   32.50 $\pm$  1.01    &             \ldots        &           \ldots             &      \ldots             &  \ldots     &   3083     $\pm$  2766     & 29.14 $\pm$  2.90 & -26.19 $\pm$   0.53 & \ldots \\ 
		A1423 &   3.58 $\pm$   0.13   &    $   19.46^{+    0.93}_{   -0.90}$  &   23.54 $\pm$  0.09    &      0.76 $\pm$  0.15 &  $  225.1^ {+  10.64}_{ -10.27}$  &  28.73 $\pm$   0.10 & 0.33 $\pm$ 0.03 &     41.62 $\pm$     3.99 & 25.20 $\pm$  0.29 & -23.43 $\pm$   0.04 & d \\ 
		A1424 &   8.93 $\pm$   1.58   &    $  151.7^ {+   40.16}_{  -33.50}$  &   26.67 $\pm$  0.51    &             \ldots        &           \ldots             &      \ldots             &  \ldots     &     97.78 $\pm$    32.81 & 25.63 $\pm$  0.95 & -24.35 $\pm$   0.15 & c \\ 
		A1435 &   4.66 $\pm$   0.37   &    $   49.23^{+    3.08}_{   -2.94}$  &   25.17 $\pm$  0.13    &             \ldots        &           \ldots             &      \ldots             &  \ldots     &     31.17 $\pm$     3.19 & 24.12 $\pm$  0.32 & -23.26 $\pm$   0.05 & c \\ 
		A1436 &   7.29 $\pm$   2.50   &    $   18.40^{+   20.06}_{  -10.95}$  &   23.98 $\pm$  1.50    &      1.44 $\pm$  0.37 &  $  147.5^ {+   8.09}_{  -7.77}$  &  27.10 $\pm$   0.21 & 0.54 $\pm$ 0.12 &     53.41 $\pm$     2.82 & 25.22 $\pm$  0.08 & -23.38 $\pm$   0.03 & ac \\ 
		A1452 &   7.27 $\pm$   0.72   &    $   51.62^{+    4.06}_{   -3.83}$  &   25.15 $\pm$  0.16    &             \ldots        &           \ldots             &      \ldots             &  \ldots     &     38.79 $\pm$     6.49 & 24.48 $\pm$  0.34 & -23.50 $\pm$   0.08 & a \\ 
		A1507 &   5.49 $\pm$   0.41   &    $   35.45^{+    1.64}_{   -1.59}$  &   24.48 $\pm$  0.10    &             \ldots        &           \ldots             &      \ldots             &  \ldots     &     29.02 $\pm$     2.57 & 24.00 $\pm$  0.20 & -23.30 $\pm$   0.05 & \ldots \\ 
		A1516 &   4.63 $\pm$   0.22   &    $   63.37^{+    2.29}_{   -2.23}$  &   24.89 $\pm$  0.08    &             \ldots        &           \ldots             &      \ldots             &  \ldots     &     47.19 $\pm$     3.66 & 24.27 $\pm$  0.13 & -24.16 $\pm$   0.04 & \ldots \\ 
		A1526 &  15.44 $\pm$   4.15   &    $   56.23^{+   47.20}_{  -28.84}$  &   26.32 $\pm$  1.33    &      0.50 $\pm$  0.14 &  $  265.9^ {+  15.58}_{ -14.92}$  &  28.53 $\pm$   0.14 & 0.29 $\pm$ 0.11 &     61.94 $\pm$    12.16 & 26.22 $\pm$  0.29 & -23.20 $\pm$   0.07 & \ldots \\ 
		A1534 &   4.69 $\pm$   0.30   &    $   38.13^{+    1.38}_{   -1.35}$  &   24.33 $\pm$  0.08    &             \ldots        &           \ldots             &      \ldots             &  \ldots     &     32.85 $\pm$     2.33 & 23.95 $\pm$  0.13 & -23.62 $\pm$   0.04 & \ldots \\ 
		A1569 &   4.58 $\pm$   2.62   &    $    3.76^{+    1.31}_{   -1.04}$  &   21.05 $\pm$  0.61    &      2.11 $\pm$  0.25 &  $  146.9^ {+   8.02}_{  -7.70}$  &  26.31 $\pm$   0.11 & 0.83 $\pm$ 0.04 &     98.19 $\pm$     5.54 & 25.42 $\pm$  0.08 & -23.79 $\pm$   0.04 & c \\ 
		A1589 &   6.67 $\pm$   0.66   &    $  160.4^ {+   23.56}_{  -21.22}$  &   26.39 $\pm$  0.29    &             \ldots        &           \ldots             &      \ldots             &  \ldots     &    123.5  $\pm$    30.95 & 25.91 $\pm$  0.49 & -24.59 $\pm$   0.11 & bc \\ 
		A1610 &   3.74 $\pm$   0.19   &    $   22.92^{+    0.57}_{   -0.56}$  &   23.61 $\pm$  0.05    &             \ldots        &           \ldots             &      \ldots             &  \ldots     &     16.78 $\pm$     0.48 & 23.09 $\pm$  0.06 & -23.22 $\pm$   0.02 & c \\ 
		A1656 &   9.00 $\pm$   0.90   &    $  784.0^ {+  291.1 }_{ -227.6 }$  &   29.30 $\pm$  0.62    &             \ldots        &           \ldots             &      \ldots             &  \ldots     &    421.2  $\pm$   234.7  & 27.58 $\pm$  1.35 & -24.97 $\pm$   0.29 & ac \\ 
		A1668 &   2.97 $\pm$   0.17   &    $   27.35^{+    2.36}_{   -2.22}$  &   23.68 $\pm$  0.15    &      0.86 $\pm$  0.17 &  $  274.4^ {+  13.58}_{ -13.09}$  &  28.12 $\pm$   0.11 & 0.40 $\pm$ 0.04 &     56.30 $\pm$     3.68 & 25.20 $\pm$  0.12 & -23.78 $\pm$   0.03 & c \\ 
		A1691 &  16.87 $\pm$   3.63   &    $  552.0^ {+  404.7 }_{ -260.3 }$  &   29.33 $\pm$  1.19    &             \ldots        &           \ldots             &      \ldots             &  \ldots     &    333.8  $\pm$   220.2  & 27.73 $\pm$  1.61 & -24.87 $\pm$   0.33 & c \\ 
		A1749 &  11.20 $\pm$   0.95   &    $  104.6^ {+   12.95}_{  -11.85}$  &   26.48 $\pm$  0.25    &             \ldots        &           \ldots             &      \ldots             &  \ldots     &     68.99 $\pm$    24.29 & 25.46 $\pm$  0.82 & -23.84 $\pm$   0.15 & \ldots \\ 
		A1767 &   5.38 $\pm$   0.33   &    $  195.9^ {+   17.03}_{  -15.99}$  &   26.57 $\pm$  0.17    &             \ldots        &           \ldots             &      \ldots             &  \ldots     &    131.7  $\pm$    31.46 & 25.71 $\pm$  0.37 & -24.64 $\pm$   0.11 & b \\ 
		A1775 &   7.31 $\pm$   0.44   &    $  153.9^ {+   13.16}_{  -12.36}$  &   26.64 $\pm$  0.17    &             \ldots        &           \ldots             &      \ldots             &  \ldots     &    126.1  $\pm$    38.52 & 26.11 $\pm$  0.69 & -24.49 $\pm$   0.14 & \ldots \\ 
		A1795 &   3.47 $\pm$   1.27   &    $   17.68^{+    9.39}_{   -6.70}$  &   23.05 $\pm$  0.79    &      1.42 $\pm$  0.14 &  $  199.2^ {+  10.21}_{  -9.83}$  &  26.00 $\pm$   0.14 & 0.78 $\pm$ 0.06 &    127.7  $\pm$     4.72 & 25.23 $\pm$  0.05 & -24.64 $\pm$   0.02 & c \\ 
		A1800 &   3.57 $\pm$   0.15   &    $   79.54^{+    2.45}_{   -2.40}$  &   24.29 $\pm$  0.07    &             \ldots        &           \ldots             &      \ldots             &  \ldots     &     67.94 $\pm$     3.40 & 23.83 $\pm$  0.08 & -24.55 $\pm$   0.03 & bc \\ 
		A1809 &   9.12 $\pm$   0.90   &    $  111.7^ {+   33.43}_{  -27.30}$  &   26.16 $\pm$  0.52    &      0.16 $\pm$  0.01 &  $   71.76^{+3.2E8}_{-2.7E8}$  &  27.31 $\pm$   0.15 & 0.24 $\pm$ 0.01 &     85.80 $\pm$    13.65 & 25.46 $\pm$  0.17 & -24.41 $\pm$   0.06 & b \\ 
		A1812 &   6.03 $\pm$   0.55   &    $  121.9^ {+   17.05}_{  -15.43}$  &   27.31 $\pm$  0.27    &             \ldots        &           \ldots             &      \ldots             &  \ldots     &     90.05 $\pm$    30.34 & 26.67 $\pm$  0.95 & -23.20 $\pm$   0.16 & \ldots \\ 
		A1825 &   3.87 $\pm$   0.22   &    $   29.28^{+    0.85}_{   -0.84}$  &   24.21 $\pm$  0.07    &             \ldots        &           \ldots             &      \ldots             &  \ldots     &     23.63 $\pm$     1.06 & 23.75 $\pm$  0.07 & -23.08 $\pm$   0.03 & acd \\ 
		A1828 &  10.99 $\pm$   2.26   &    $    3.03^{+    1.36}_{   -1.01}$  &   19.58 $\pm$  0.90    &             \ldots        &           \ldots             &      \ldots             &  \ldots     &     10.07 $\pm$     0.57 & 22.20 $\pm$  0.14 & -22.49 $\pm$   0.03 & \ldots \\ 
		A1831 &   6.44 $\pm$   0.35   &    $  270.2^ {+   25.54}_{  -23.85}$  &   27.04 $\pm$  0.18    &             \ldots        &           \ldots             &      \ldots             &  \ldots     &    226.1  $\pm$    63.27 & 26.60 $\pm$  0.56 & -25.07 $\pm$   0.16 & \ldots \\ 
		A1890 &   5.95 $\pm$   0.37   &    $   53.23^{+    2.46}_{   -2.37}$  &   24.57 $\pm$  0.10    &             \ldots        &           \ldots             &      \ldots             &  \ldots     &     38.85 $\pm$     3.46 & 23.83 $\pm$  0.23 & -23.95 $\pm$   0.05 & \ldots \\ 
		A1899 &   2.76 $\pm$   0.16   &    $    9.60^{+    0.53}_{   -0.51}$  &   22.02 $\pm$  0.11    &      0.87 $\pm$  0.27 &  $  119.3^ {+   8.62}_{  -8.18}$  &  27.87 $\pm$   0.15 & 0.27 $\pm$ 0.03 &     17.13 $\pm$     0.41 & 23.21 $\pm$  0.07 & -22.87 $\pm$   0.01 & d \\ 
		A1904 &   6.83 $\pm$   0.43   &    $  164.3^ {+   15.21}_{  -14.22}$  &   26.61 $\pm$  0.18    &             \ldots        &           \ldots             &      \ldots             &  \ldots     &    123.6  $\pm$    37.51 & 25.86 $\pm$  0.68 & -24.43 $\pm$   0.13 & \ldots \\ 
		A1913 &   8.87 $\pm$   0.80   &    $   77.53^{+    9.44}_{   -8.65}$  &   26.47 $\pm$  0.24    &             \ldots        &           \ldots             &      \ldots             &  \ldots     &     37.30 $\pm$     9.84 & 24.86 $\pm$  0.49 & -23.08 $\pm$   0.12 & \ldots \\ 
		A1982 &   3.97 $\pm$   0.22   &    $   37.70^{+    1.30}_{   -1.27}$  &   24.30 $\pm$  0.08    &             \ldots        &           \ldots             &      \ldots             &  \ldots     &     31.80 $\pm$     1.78 & 24.01 $\pm$  0.10 & -23.18 $\pm$   0.03 & b \\ 
		A1983 &  16.03 $\pm$   2.07   &    $ 1047^  {+  613.8 }_{ -425.5 }$  &   31.42 $\pm$  0.95    &             \ldots        &           \ldots             &      \ldots             &  \ldots     &    431.5  $\pm$   352.1  & 28.58 $\pm$  2.22 & -23.72 $\pm$   0.43 & \ldots \\ 
		A2022 &  24.43 $\pm$   4.06   &    $ 3.3E4^{+ 5.7E4}_{-2.4E4}$  &   36.96 $\pm$  2.17    &             \ldots        &           \ldots             &      \ldots             &  \ldots     &   1.1E4 $\pm$  1.1E4 & 31.16 $\pm$  5.02 & -25.12 $\pm$   0.80 & \ldots \\ 
		A2029 &   5.55 $\pm$   0.26   &    $  261.2^ {+   17.47}_{  -16.64}$  &   26.08 $\pm$  0.14    &             \ldots        &           \ldots             &      \ldots             &  \ldots     &    329.3  $\pm$    82.12 & 26.53 $\pm$  0.73 & -25.85 $\pm$   0.12 & \ldots \\ 
		A2052 &   4.02 $\pm$   0.23   &    $   79.16^{+    4.74}_{   -4.54}$  &   25.32 $\pm$  0.12    &             \ldots        &           \ldots             &      \ldots             &  \ldots     &     47.87 $\pm$     2.12 & 24.33 $\pm$  0.11 & -23.64 $\pm$   0.03 & d \\ 
		A2061 &  23.24 $\pm$   6.48   &    $158347^  {+777761    }_{-152375    }$  &   38.80 $\pm$  4.12    &             \ldots        &           \ldots             &      \ldots             &  \ldots     &   64592     $\pm$  64343     & 32.39 $\pm$  5.97 & -26.02 $\pm$   1.10 & b \\ 
		A2063 &   8.46 $\pm$   0.64   &    $  345.6^ {+   68.57}_{  -59.69}$  &   28.41 $\pm$  0.36    &             \ldots        &           \ldots             &      \ldots             &  \ldots     &    202.1  $\pm$   103.1  & 27.00 $\pm$  1.21 & -24.02 $\pm$   0.23 & \ldots \\ 
		A2065 &   7.58 $\pm$   2.29   &    $   25.55^{+   23.75}_{  -13.92}$  &   24.42 $\pm$  1.34    &      1.24 $\pm$  0.30 &  $  189.5^ {+  10.79}_{ -10.35}$  &  27.35 $\pm$   0.18 & 0.58 $\pm$ 0.13 &     91.60 $\pm$     6.60 & 25.79 $\pm$  0.11 & -23.78 $\pm$   0.04 & \ldots \\ 
		A2107 &   3.31 $\pm$   0.11   &    $   35.71^{+    1.28}_{   -1.24}$  &   23.78 $\pm$  0.07    &      0.43 $\pm$  0.08 &  $  352.5^ {+  11.00}_{ -10.75}$  &  28.74 $\pm$   0.08 & 0.27 $\pm$ 0.02 &     60.10 $\pm$     3.01 & 24.85 $\pm$  0.15 & -24.09 $\pm$   0.02 & c \\ 
		A2122 &   3.35 $\pm$   0.13   &    $   58.19^{+    1.52}_{   -1.49}$  &   24.53 $\pm$  0.06    &             \ldots        &           \ldots             &      \ldots             &  \ldots     &     43.82 $\pm$     1.52 & 23.86 $\pm$  0.09 & -24.03 $\pm$   0.02 & \ldots \\ 
		A2147 &   1.43 $\pm$   0.39   &    $    6.32^{+    0.79}_{   -0.73}$  &   21.78 $\pm$  0.17    &      1.77 $\pm$  0.16 &  $   85.49^{+   2.93}_{  -2.86}$  &  25.26 $\pm$   0.07 & 0.84 $\pm$ 0.03 &     56.15 $\pm$     1.32 & 24.44 $\pm$  0.04 & -23.50 $\pm$   0.01 & a \\ 
		A2151 &   8.91 $\pm$   0.70   &    $  152.4^ {+   23.52}_{  -21.08}$  &   27.21 $\pm$  0.30    &             \ldots        &           \ldots             &      \ldots             &  \ldots     &    111.9  $\pm$    42.43 & 26.48 $\pm$  0.69 & -23.80 $\pm$   0.17 & \ldots \\ 
		A2152 &   5.51 $\pm$   0.30   &    $   11.78^{+    0.86}_{   -0.81}$  &   22.89 $\pm$  0.14    &      0.52 $\pm$  0.03 &  $  188.8^ {+   3.04}_{  -3.00}$  &  26.92 $\pm$   0.04 & 0.60 $\pm$ 0.01 &     95.39 $\pm$     2.25 & 26.13 $\pm$  0.01 & -23.56 $\pm$   0.01 & a \\ 
		A2162 &   6.36 $\pm$   0.51   &    $   20.11^{+    3.50}_{   -3.09}$  &   23.52 $\pm$  0.30    &      0.70 $\pm$  0.09 &  $   99.66^{+   2.09}_{  -2.06}$  &  26.38 $\pm$   0.09 & 0.34 $\pm$ 0.03 &     33.13 $\pm$     1.06 & 24.15 $\pm$  0.06 & -23.22 $\pm$   0.02 & c \\ 
		A2197 &   5.19 $\pm$   0.29   &    $   35.00^{+    1.42}_{   -1.37}$  &   23.89 $\pm$  0.09    &             \ldots        &           \ldots             &      \ldots             &  \ldots     &     28.29 $\pm$     1.63 & 23.51 $\pm$  0.13 & -23.48 $\pm$   0.03 & b \\ 
		A2199 &   7.23 $\pm$   0.19   &    $  144.2^ {+    6.09}_{   -5.90}$  &   26.31 $\pm$  0.09    &             \ldots        &           \ldots             &      \ldots             &  \ldots     &    101.7  $\pm$    27.05 & 25.50 $\pm$  0.50 & -24.22 $\pm$   0.12 & d \\ 
		A2247 &   6.31 $\pm$   1.29   &    $   72.43^{+   24.69}_{  -19.65}$  &   26.30 $\pm$  0.61    &             \ldots        &           \ldots             &      \ldots             &  \ldots     &     40.63 $\pm$     5.83 & 25.14 $\pm$  0.43 & -22.93 $\pm$   0.08 & \ldots \\ 
		A2248 &   4.39 $\pm$   0.27   &    $   39.20^{+    1.40}_{   -1.36}$  &   24.60 $\pm$  0.08    &             \ldots        &           \ldots             &      \ldots             &  \ldots     &     26.15 $\pm$     1.42 & 23.66 $\pm$  0.09 & -23.27 $\pm$   0.03 & \ldots \\ 
		A2255 &  23.77 $\pm$   6.61   &    $ 9338^  {+23985    }_{-7908    }$  &   34.64 $\pm$  2.92    &             \ldots        &           \ldots             &      \ldots             &  \ldots     &   5055     $\pm$  4844     & 30.86 $\pm$  3.90 & -25.53 $\pm$   0.75 & \ldots \\ 
		A2256 &  10.96 $\pm$   1.14   &    $  217.4^ {+   45.73}_{  -39.50}$  &   27.41 $\pm$  0.40    &             \ldots        &           \ldots             &      \ldots             &  \ldots     &    169.2  $\pm$    71.65 & 26.54 $\pm$  0.77 & -24.61 $\pm$   0.20 & a \\ 
		A2271 &   6.17 $\pm$   0.29   &    $  111.9^ {+    6.54}_{   -6.26}$  &   25.99 $\pm$  0.12    &             \ldots        &           \ldots             &      \ldots             &  \ldots     &     56.83 $\pm$     7.09 & 24.61 $\pm$  0.22 & -23.96 $\pm$   0.07 & c \\ 
		A2293 &   8.59 $\pm$   1.13   &    $   85.51^{+   14.85}_{  -13.14}$  &   26.10 $\pm$  0.34    &             \ldots        &           \ldots             &      \ldots             &  \ldots     &     89.57 $\pm$    27.84 & 26.16 $\pm$  0.64 & -23.81 $\pm$   0.12 & d \\ 
		A2308 &  10.92 $\pm$   1.49   &    $  270.9^ {+   83.32}_{  -67.68}$  &   27.92 $\pm$  0.56    &             \ldots        &           \ldots             &      \ldots             &  \ldots     &    130.0  $\pm$    66.38 & 26.34 $\pm$  1.32 & -24.39 $\pm$   0.21 & \ldots \\ 
		A2319 &   1.34 $\pm$   0.13   &    $   22.61^{+    2.12}_{   -1.98}$  &   22.69 $\pm$  0.13    &      1.18 $\pm$  0.25 &  $  168.6^ {+  11.88}_{ -11.29}$  &  26.39 $\pm$   0.14 & 0.64 $\pm$ 0.04 &     92.40 $\pm$     3.15 & 25.40 $\pm$  0.05 & -24.64 $\pm$   0.02 & d \\ 
		A2388 &  11.08 $\pm$   0.79   &    $  526.6^ {+  173.4 }_{ -139.0 }$  &   29.52 $\pm$  0.54    &      0.16 $\pm$  0.06 &  $  141.0^ {+1.7E11}_{-1.6E11}$  &  28.87 $\pm$   0.12 & 0.25 $\pm$ 0.04 &    327.4  $\pm$   123.1  & 27.79 $\pm$  0.51 & -24.06 $\pm$   0.19 & \ldots \\ 
		A2469 &   4.08 $\pm$   0.34   &    $   50.30^{+    3.65}_{   -3.46}$  &   25.99 $\pm$  0.15    &             \ldots        &           \ldots             &      \ldots             &  \ldots     &     31.83 $\pm$     2.92 & 25.14 $\pm$  0.24 & -22.47 $\pm$   0.06 & \ldots \\ 
		A2495 &   3.81 $\pm$   0.22   &    $  157.7^ {+   12.03}_{  -11.38}$  &   25.78 $\pm$  0.15    &             \ldots        &           \ldots             &      \ldots             &  \ldots     &    104.2  $\pm$     9.91 & 25.02 $\pm$  0.17 & -24.61 $\pm$   0.05 & \ldots \\ 
		A2506 &   4.06 $\pm$   0.17   &    $   23.76^{+    0.67}_{   -0.66}$  &   23.81 $\pm$  0.06    &             \ldots        &           \ldots             &      \ldots             &  \ldots     &     16.66 $\pm$     0.59 & 23.14 $\pm$  0.07 & -22.61 $\pm$   0.02 & d \\ 
		A2513 &   3.50 $\pm$   0.09   &    $   23.07^{+    0.41}_{   -0.40}$  &   23.47 $\pm$  0.04    &             \ldots        &           \ldots             &      \ldots             &  \ldots     &     20.22 $\pm$     0.43 & 23.19 $\pm$  0.04 & -23.20 $\pm$   0.01 & \ldots \\ 
		A2516 &   4.21 $\pm$   0.29   &    $   16.82^{+    0.61}_{   -0.59}$  &   22.91 $\pm$  0.08    &             \ldots        &           \ldots             &      \ldots             &  \ldots     &     17.31 $\pm$     0.53 & 23.02 $\pm$  0.05 & -23.38 $\pm$   0.02 & \ldots \\ 
		A2524 &   5.11 $\pm$   0.36   &    $   39.90^{+    1.74}_{   -1.69}$  &   24.37 $\pm$  0.09    &             \ldots        &           \ldots             &      \ldots             &  \ldots     &     32.46 $\pm$     2.48 & 23.96 $\pm$  0.18 & -23.73 $\pm$   0.04 & \ldots \\ 
		A2558 &   6.25 $\pm$   0.39   &    $   79.90^{+    5.94}_{   -5.63}$  &   26.25 $\pm$  0.15    &             \ldots        &           \ldots             &      \ldots             &  \ldots     &     55.86 $\pm$    12.99 & 25.46 $\pm$  0.59 & -23.32 $\pm$   0.11 & \ldots \\ 
		A2572 &   8.43 $\pm$   0.47   &    $  109.3^ {+   10.50}_{   -9.80}$  &   26.67 $\pm$  0.19    &             \ldots        &           \ldots             &      \ldots             &  \ldots     &     73.14 $\pm$    24.10 & 25.79 $\pm$  0.74 & -23.57 $\pm$   0.14 & d \\ 
		A2589 &   5.92 $\pm$   0.26   &    $  276.8^ {+   24.84}_{  -23.27}$  &   27.32 $\pm$  0.17    &             \ldots        &           \ldots             &      \ldots             &  \ldots     &    113.7  $\pm$    26.69 & 25.45 $\pm$  0.50 & -24.14 $\pm$   0.12 & c \\ 
		A2593 &   5.78 $\pm$   0.47   &    $   72.58^{+    6.14}_{   -5.77}$  &   25.41 $\pm$  0.17    &             \ldots        &           \ldots             &      \ldots             &  \ldots     &     45.63 $\pm$     6.32 & 24.52 $\pm$  0.36 & -23.56 $\pm$   0.07 & d \\ 
		A2618 &   6.97 $\pm$   0.48   &    $  129.4^ {+   12.31}_{  -11.49}$  &   26.17 $\pm$  0.19    &             \ldots        &           \ldots             &      \ldots             &  \ldots     &     88.33 $\pm$    19.27 & 25.32 $\pm$  0.49 & -24.50 $\pm$   0.10 & \ldots \\ 
		A2622 &   5.05 $\pm$   0.21   &    $   37.77^{+    3.01}_{   -2.84}$  &   24.35 $\pm$  0.15    &      0.67 $\pm$  0.19 &  $  410.8^ {+  38.95}_{ -36.36}$  &  28.76 $\pm$   0.18 & 0.32 $\pm$ 0.04 &     62.06 $\pm$     6.46 & 25.28 $\pm$  0.19 & -23.83 $\pm$   0.04 & \ldots \\ 
		A2625 &   2.82 $\pm$   0.13   &    $   13.16^{+    0.34}_{   -0.33}$  &   22.40 $\pm$  0.05    &             \ldots        &           \ldots             &      \ldots             &  \ldots     &     13.08 $\pm$     0.16 & 22.42 $\pm$  0.02 & -22.75 $\pm$   0.01 & \ldots \\ 
		A2626 &   4.97 $\pm$   0.33   &    $   61.03^{+    3.34}_{   -3.21}$  &   24.83 $\pm$  0.11    &             \ldots        &           \ldots             &      \ldots             &  \ldots     &     47.21 $\pm$     3.63 & 24.34 $\pm$  0.19 & -23.83 $\pm$   0.04 & d \\ 
		A2630 &   5.87 $\pm$   0.63   &    $    7.17^{+    0.52}_{   -0.49}$  &   22.18 $\pm$  0.15    &             \ldots        &           \ldots             &      \ldots             &  \ldots     &      7.16 $\pm$     0.18 & 22.08 $\pm$  0.04 & -22.16 $\pm$   0.02 & \ldots \\ 
		A2634 &   4.79 $\pm$   0.28   &    $   56.56^{+    3.38}_{   -3.23}$  &   24.73 $\pm$  0.13    &             \ldots        &           \ldots             &      \ldots             &  \ldots     &     42.75 $\pm$     3.31 & 24.05 $\pm$  0.17 & -23.67 $\pm$   0.04 & bd \\ 
		A2637 &   9.63 $\pm$   1.09   &    $   46.50^{+    3.79}_{   -3.57}$  &   24.82 $\pm$  0.17    &             \ldots        &           \ldots             &      \ldots             &  \ldots     &     36.44 $\pm$     7.56 & 24.21 $\pm$  0.35 & -23.76 $\pm$   0.09 & \ldots \\ 
		A2657 &   3.17 $\pm$   0.17   &    $   48.52^{+    2.05}_{   -1.99}$  &   24.86 $\pm$  0.09    &             \ldots        &           \ldots             &      \ldots             &  \ldots     &     32.94 $\pm$     1.36 & 23.98 $\pm$  0.08 & -23.04 $\pm$   0.03 & b \\ 
		A2665 &   4.29 $\pm$   0.15   &    $   61.63^{+    4.77}_{   -4.51}$  &   24.81 $\pm$  0.13    &      0.19 $\pm$  0.04 &  $  170.1^ {+  33.40}_{ -29.11}$  &  27.30 $\pm$   0.10 & 0.22 $\pm$ 0.04 &     75.08 $\pm$     4.43 & 24.97 $\pm$  0.17 & -24.25 $\pm$   0.03 & \ldots \\ 
		A2666 &   4.79 $\pm$   0.36   &    $   33.19^{+    1.90}_{   -1.82}$  &   24.06 $\pm$  0.13    &             \ldots        &           \ldots             &      \ldots             &  \ldots     &     27.40 $\pm$     1.89 & 23.80 $\pm$  0.14 & -23.25 $\pm$   0.04 & cd \\ 
		A2675 &   4.24 $\pm$   0.38   &    $   27.48^{+    5.41}_{   -4.71}$  &   23.85 $\pm$  0.32    &      1.87 $\pm$  0.31 &  $  751.8^ {+  96.65}_{ -88.15}$  &  28.28 $\pm$   0.22 & 0.80 $\pm$ 0.03 &    430.4  $\pm$    69.07 & 27.31 $\pm$  0.28 & -24.91 $\pm$   0.10 & \ldots \\ 
		A2678 &   5.76 $\pm$   0.40   &    $    9.59^{+    0.74}_{   -0.70}$  &   22.44 $\pm$  0.15    &      1.02 $\pm$  0.08 &  $  370.7^ {+  14.97}_{ -14.53}$  &  27.69 $\pm$   0.08 & 0.74 $\pm$ 0.02 &    235.4  $\pm$    19.22 & 26.89 $\pm$  0.07 & -24.04 $\pm$   0.05 & d \\ 
		AWM1  &   4.99 $\pm$   0.16   &    $   14.47^{+    0.33}_{   -0.32}$  &   23.00 $\pm$  0.05    &             \ldots        &           \ldots             &      \ldots             &  \ldots     &     16.51 $\pm$     0.69 & 23.28 $\pm$  0.11 & -22.73 $\pm$   0.02 & bc \\ 
		AWM5  &   6.53 $\pm$   0.16   &    $   58.72^{+    1.98}_{   -1.93}$  &   24.69 $\pm$  0.07    &             \ldots        &           \ldots             &      \ldots             &  \ldots     &     41.67 $\pm$     4.22 & 23.91 $\pm$  0.20 & -23.96 $\pm$   0.05 & \ldots \\ 
		AWM7  &   6.49 $\pm$   0.17   &    $  150.4^ {+    7.30}_{   -7.04}$  &   26.22 $\pm$  0.10    &             \ldots        &           \ldots             &      \ldots             &  \ldots     &     82.44 $\pm$    16.77 & 24.94 $\pm$  0.43 & -24.22 $\pm$   0.09 & c \\ 
		L2027 &  12.12 $\pm$   1.66   &    $  465.2^ {+  193.4 }_{ -147.3 }$  &   29.58 $\pm$  0.72    &             \ldots        &           \ldots             &      \ldots             &  \ldots     &    271.1  $\pm$   199.8  & 27.66 $\pm$  1.95 & -23.85 $\pm$   0.35 & \ldots \\ 
		L2030 &   2.47 $\pm$   0.11   &    $   34.97^{+    1.04}_{   -1.02}$  &   24.28 $\pm$  0.07    &             \ldots        &           \ldots             &      \ldots             &  \ldots     &     30.65 $\pm$     0.67 & 23.96 $\pm$  0.02 & -22.75 $\pm$   0.02 & \ldots \\ 
		L2069 &   5.04 $\pm$   0.29   &    $   64.08^{+    2.87}_{   -2.77}$  &   24.82 $\pm$  0.10    &             \ldots        &           \ldots             &      \ldots             &  \ldots     &     56.18 $\pm$     6.74 & 24.43 $\pm$  0.24 & -24.34 $\pm$   0.06 & d \\ 
		L2093 &   5.60 $\pm$   0.55   &    $   21.45^{+    1.50}_{   -1.42}$  &   24.00 $\pm$  0.15    &             \ldots        &           \ldots             &      \ldots             &  \ldots     &     15.73 $\pm$     1.15 & 23.50 $\pm$  0.15 & -22.49 $\pm$   0.04 & bd \\ 
		L2211 &   5.34 $\pm$   0.36   &    $   26.18^{+    1.03}_{   -1.00}$  &   23.73 $\pm$  0.09    &             \ldots        &           \ldots             &      \ldots             &  \ldots     &     21.23 $\pm$     1.21 & 23.32 $\pm$  0.11 & -23.28 $\pm$   0.03 & c \\ 
		L3009 &  77.14 $\pm$ 111.1    &    $ 3426^  {+177679    }_{ -805.4 }$  &   34.51 $\pm$ 14.30    &             \ldots        &           \ldots             &      \ldots             &  \ldots     &   3446     $\pm$  3405     & 30.51 $\pm$  5.51 & -23.87 $\pm$   0.67 & d \\ 
		L3055 &   5.39 $\pm$   0.60   &    $   57.26^{+    6.39}_{   -5.90}$  &   25.96 $\pm$  0.23    &             \ldots        &           \ldots             &      \ldots             &  \ldots     &     49.05 $\pm$     7.94 & 25.73 $\pm$  0.33 & -22.96 $\pm$   0.08 & \ldots \\ 
		L3152 &   5.39 $\pm$   0.22   &    $   30.84^{+    2.33}_{   -2.20}$  &   24.09 $\pm$  0.14    &      0.25 $\pm$  0.05 &  $  175.1^ {+   4.82}_{  -4.73}$  &  27.70 $\pm$   0.09 & 0.22 $\pm$ 0.03 &     39.91 $\pm$     1.59 & 24.52 $\pm$  0.08 & -23.41 $\pm$   0.02 & \ldots \\ 
		L3186 &   5.38 $\pm$   0.42   &    $   40.78^{+    2.01}_{   -1.94}$  &   24.59 $\pm$  0.10    &             \ldots        &           \ldots             &      \ldots             &  \ldots     &     32.27 $\pm$     3.29 & 24.02 $\pm$  0.19 & -23.49 $\pm$   0.05 & c \\ 
		MKW4  &   4.20 $\pm$   0.15   &    $   39.32^{+    1.00}_{   -0.98}$  &   23.76 $\pm$  0.06    &             \ldots        &           \ldots             &      \ldots             &  \ldots     &     31.81 $\pm$     1.12 & 23.29 $\pm$  0.07 & -23.46 $\pm$   0.02 & b \\ 
		\enddata
		\tablecomments{All parameters are corrected for PSF broadening, dust extinction, cosmic dimming and are K-corrected. The parameters of the single S\'ersic fits (see Equation (\ref{eq:ss})) or double S\'ersic fits (see Equation (\ref{eq:ds})) are given in columns (2) -- (7). The errors are calculated solely from the covariance matrices of the fits. The fraction of the integrated outer S\'ersic component (column 8) compared to the total galaxy light $S_2/(S_1+S_2)$ is calculated by integrating both semimajor-axis S\'ersic functions out to infinite radius. Both components are assumed to have the same ellipticity profiles. The parameters from 2-D profile integration are listed in columns (9) -- (11). They are corrected for undetected ICL (see Section \ref{sec:integrationcrop}) and the errors are estimated from that correction. Column (12) lists the found types of accretion signatures: a = 2 BCGs, b = shells, c = tidal streams and d = multiple nuclei.}
	\end{deluxetable}
\end{longrotatetable}

\clearpage
\bibliography{Paper1}
\bibliographystyle{aasjournal}

\end{document}